\documentclass[11pt,a4paper]{article}
\pdfoutput=0
\usepackage{jheppub}

\newcommand{\be}{\begin{equation}}
\newcommand{\ee}{\end{equation}}
\newcommand{\bea}{\begin{eqnarray}}
\newcommand{\eea}{\end{eqnarray}}
\newcommand{\half}{\frac{1}{2}}

\newcommand{\la}{\langle}
\newcommand{\ra}{\rangle}

\newcommand{\bq}{\mathbf{q}}

\numberwithin{equation}{section}

\makeatletter
\newcommand{\rmnum}[1]{\romannumeral #1}
\newcommand{\Rmnum}[1]{\expandafter\@slowromancap\romannumeral #1@}
\makeatother

\def\IR{\mathbb{R}}

\def\cH{{\cal H}}

\def\cO{{\cal O}}
\def\cP{{\cal P}}
\def\cQ{{\cal Q}}

\def\cT{{\cal T}}

\title{Split Flows in Bubbled Geometries}

\author{Chih-Wei Wang}

\affiliation{National Center for Theoretical Sciences, National Tsing-Hua University, \\
		Hsinchu 30013, Taiwan, R.O.C.}
		
\emailAdd{freeform1111@gmail.com}

\abstract{We propose a procedure to clarify part of the physical sector in the five dimensional bubble geometries based on ideas similar to the split attractor flow conjecture proposed by Denef. This procedure involves building some simple tree-like graphs that we call skeletons without referring to the moduli space. The skeleton (tree) exists if and only if it passes the existence conditions which are purely based on some local CTC's (closed timelike curves) checking. Then, we propose the conjecture similar to Denef's version which states that every existing skeleton (tree) should correspond to some solution in which the global absence of CTC's is ensured. Furthermore, we propose two pictures to identify this correspondence explicitly and use some numerical examples to show how this procedure works. We also analyze the physical sector of the simplest bubbled supertube and see how the existence conditions constrain the charge parameter space.}


\begin{document}

\maketitle

\flushbottom

\section{Introduction}

The black hole microstates counting \cite{Strominger:1996sh} is considered to be one of the most important achievements of string theory. However, the counting is done in the limit of weak string coupling and it is unclear what these microstates become in a strong coupling regime when the gravitational back-reaction can no longer be ignored. On the other hand, the very existence of the black holes brings another deep puzzle: the information paradox. To address this issue, Samir Mathur and his collaborators propose an interesting idea, the so-called fuzzball proposal (see \cite{Mathur:2005zp, Skenderis:2008qn} for a review). Roughly speaking, the proposal states that a classical black hole should be thought of as an effective description of the ensemble of many horizonless fuzzball states. In the stronger version of the proposal, the majority of fuzzball states can be realized as the solutions of supergravity. These solutions are completely smooth and horizonless and naturally can be considered as microstate geometries of a black hole.

This proposal has achieved some great successes in the two-charge BPS solutions (for a review see \cite{Mathur:2005zp}). However, to verify this proposal on the black holes with macroscopic horizons, one needs to extend the analysis to the three-charge solutions. There have been some developments in five dimensional BPS supergravity solutions \cite{Bena:2004de, Bena:2005va, Berglund:2005vb, Bena:2006is, Bena:2006kb, Bena:2007qc, Bena:2008wt, Bena:2008nh} in this direction and a review is in \cite{Bena:2007kg}. It is clear that there are a large number of bubbled  geometries which are smooth and horizonless. They are the natural candidates of microstate geometries of a given supersymmetric black hole or black ring. However, due to the difficulties of checking the conditions for the absence of closed time-like curves (CTC's), the physical sector of the bubbled geometries remains unclear. Without knowledge of this sector, it seems unlikely that a systematic study could be carried out or even to ``count" the number of these microstate geometries.

On the other hand, a similar problem is encountered in four dimensional multi-center BPS supergravity solutions but with the different physical origin. In \cite{Denef:2000nb,Denef:2007vg}, an interesting criterion of the well-behaved solutions was proposed. This so-called split attractor flow conjecture states that a multi-center solution exists if and only if there exist an corresponding attractor flow tree in the Calabi-Yau moduli space. There are also some studies of 4d-5d connection \cite{Gaiotto:2005gf, Gaiotto:2005xt, Bena:2005ni, Berglund:2005vb, Balasubramanian:2006gi, Cheng:2006yq} which suggest that the problem in four dimensional supergravity may relate to the CTC's checking in five dimensions. In \cite{Denef:2007vg, deBoer:2008fk}, it has been further suggested that this conjecture can be used in some five dimensional solutions as well by uplifting the four dimensional solutions. However, this uplifting procedure usually involves some rescaling of harmonic functions or coordinates and it is unclear whether there is always a unique way to connect four dimensional solutions to five dimensional solutions. Additionally, from the point of view of five dimensional theory, it is unclear why the solutions in five dimensional bubbled geometries should be related to the split attractor flow trees in four dimensional rather than five dimensional moduli space. And yet, the central charges and moduli space in five dimensional theory are all real and it seems unlikely to have the similar picture like the wall of marginal stability in the complex moduli space in four dimensional theory. 

To avoid the above subtleties and make the connection between the criterion in the conjecture and the solutions in bubble geometries more transparent, we take a more direct approach and propose two alternative forms of the conjecture based purely on CTC's conditions. These two forms are referred to as the ``static" and the ``dynamic" version respectively in this work. The ``static" version is based on locating a ``consistent" path in spacetime for a given solution while the ``dynamic" version is related to some adiabatic property of the solutions of the bubble equations. Both of them are developed by considering  the ``response" (i.e, decompose to two clusters) of the solutions under the tuning of the asymptotic constants similar to the original idea of the split attractor flow conjecture. However, to have a criterion that can be practically used to search CTC's free solutions, a systematic ``way" of tuning the asymptotic constants must be specified . For that, we at first propose a systematic method to build the tree-like graphics. These graphics are referred to as the ``skeletons" and are constructed by joining several single-center flows. Then, a criterion is proposed and it is amount to checking several CTC's conditions on these ``skeletons". For those ``skeletons" that pass the criterion, they will be one-to-one corresponding to some well-behaved components of the solution space of the bubble equations. In fact, this criterion is very similar to Denef's original split attractor flow conjecture even though it does not have the same physical picture related to the stability as we have explained. We will further explain the relation between the criterion, the two versions of the conjecture and Denef's original conjecture in section \ref{relation}. Although all of the different versions play some roles, it is the criterion based on the ``skeletons" that will be the main tool in this paper.

Unfortunately, the above picture based on the criterion does not cover all of the well-behaved components. Specifically, there are potentially large number of the components which simply resist breaking into two clusters even when the asymptotic constants hit the split point. These kinds of the solutions are referred to as the scaling solutions and for them, the picture will fail to distinguish different solutions or give an unambiguous answer as to whether the solutions are well-behaved or not. Nevertheless, through this split flow picture, one can explore the physical sector of the bubbled geometries at least partially or test the conjecture by checking the corresponding solutions explicitly. In this paper, we use the criterion to clarify the physical sector of the simplest bubbled supertube and also test the conjecture by examining several more complicated numerical examples. Additionally, we discover that in some systems there exists a bifurcation point between a scaling and a split flow component. In other words, the component of the solution space undergoes topological change and the split flow component emerges from the scaling component when the system is close to the split point. For more detail on this, see section \ref{bifurcation}.

From a different perspective, this picture can be treated as some kind of coarse graining  (or renormalization) scheme for solving bubble equations in which one reduces a large system of bubble equations to smaller and smaller subsets which become easier to solve. In other words, this basically collects some subset of the Gibbons-Hawing centers into a cluster and examines the CTC's conditions by treating this cluster as a single-center. Therefore, this might be precisely the scheme or positivity condition which are suspected to exist in \cite{Bena:2007kg}. However, as we have mentioned, the scaling solutions are like some kind of ``entanglement" states formed by several centers and cannot be captured by this coarse-graining scheme. They are not understood well and may yet play very important roles in microstate geometries. Nevertheless, improving our understanding of the split flow solutions may provide the first step to resolve the mystery of the scaling solutions.

The outline of this paper is the following. In section \ref{bubble}, we review the basic setup for bubbled geometries without fixing the asymptotic structures of the metrics. Then, we derive the bubble equations in general backgrounds. In section \ref{split}, we briefly review the wall-crossing and the split attractor flow conjecture. Then, we propose several alternative forms of the conjecture which are suitable to be used in bubbled geometries and explain the relation between them. In section \ref{split_components}, we explain the construction of the tree-like graphs (``skeleton") and their connection with the components of the solution space of the bubble equations. In section \ref{numerical}, we check several numerical examples. In section \ref{3center}, we use the existence conditions of the 3-center trees to identify the physical sector of bubbled supertubes and examining other exotic 3-center solutions. Finally, we present our conclusion and some discussion about potential future work in section \ref{conclusion}.

\section{Bubbled geometries with unfixed asymptotic backgrounds}
\label{bubble}

\subsection{Review of bubbled geometries}

It was shown in \cite{Bena:2004de} that the general three-charge five dimensional BPS supergravity solutions can be obtained from the compactification of the eleven dimensional M-theory. For $T^6$ compactification, the metric has the following form:
\bea
ds_{11}^2 &=&  - \left(\frac{1}{ Z_1 Z_2
Z_3}\right)^{2/3} (dt+k)^2 + \left( Z_1 Z_2 Z_3\right)^{1/3}
h_{mn}dx^m dx^n  + \left(\frac{Z_2 Z_3}{Z_1^2}\right)\,(dx_5^2+dx_6^2)  \nonumber \\  
&+& \left(\frac{Z_1 Z_3}{Z_2^2}\right)\,(dx_7^2+dx_8^2)  + \left(\frac{Z_1 Z_2}{Z_3^2}\right)\,(dx_9^2+dx_{10}^2) \\
& &  \nonumber \\
{\cal A} & =&
A^1 \wedge dx_5 \wedge dx_6 +A^2 \wedge dx_7 \wedge dx_8 + A^3
\wedge dx_9 \wedge dx_{10}~\,,
\label{background}
\eea
where $A^I$ and $k$ are one-forms in the five-dimensional spacetime. The base space metric $h_{mn}$ is a four-dimensional hyper-K\"ahler metric. The warp factors, $Z_I$, can be obtained from solving three linear BPS equations sequently:
\bea
\Theta^I  &=& \star_4 \Theta^I \nonumber \\ 
\nabla^2  Z_I & =&  \half C_{IJK} \star_4 (\Theta^J \wedge \Theta^K) \nonumber \\
dk + \star_4 dk &=& Z_I \Theta^I~ \,,
\label{BPS_eom}
\eea
where the self-dual ``dipole field strengths'' $\Theta^I$ are defined as:
\be
\Theta^I \equiv d A^I + d\left(  \frac{dt +k}{Z_I}\right)\,, 
\label{theta_def}
\ee
where $\star_4$ is the Hodge dual taken with respect to the base space metric, $h_{mn}$.

To construct an explicit solution, one needs to specify a hyper-K\"ahler metric for the base. In \cite{Bena:2005va, Berglund:2005vb}, the Gibbons-Hawking space was considered to be the base and it was shown that many completely smooth five dimensional solutions could be explicitly constructed. These are the so-called bubbled geometries. The Gibbons-Hawking space is a subclass of hyper-K\"ahler space for which the metric is explicitly known:
\be
h_{mn}dx^m dx^n  =V(dx_1^2+dx_2^2+dx_3^2) + \frac{1}{V} \big( d\psi + \vec{A} \cdot
d\vec{x}\big)^2\,,
\label{GHspace}
\ee
where $V$ is an arbitrary harmonic function in $\IR^3$ and $A$ is\footnote{One should not confuse $A$ with the three gauge fields, $ A^I$, which are coming from the reduction of the three-form fields, ${\cal A}$}:
\be
\vec \nabla \times \vec A ~=~ \vec \nabla V\,.
\ee
The solution of BPS equations (\ref{BPS_eom}) in this base space is specified by eight arbitrary harmonic functions $(V,K^1, K^2, K^3, L_1, L_2, L_3, M)$.  The warp factors $Z_I$ are:
\be
Z_I ~=~ \frac{1}{2}\, C_{IJK}\, \frac{K^J K^K}{V} ~+~ L_I \,,
\label{ZIdef}
\ee
and the one-from, $k$,  is:
\be
k ~=~ \mu\, ( d\psi + A   ) ~+~ \omega\,,
\ee
where $\mu$ is:
\be
\mu ~=~ \frac{1}{6}\, C_{IJK}\, \frac{K^I K^J K^K}{V^2} ~+~
\frac{ K^I L_I}{2 \,V} ~+~  M\,,
\label{mudef}
\ee
and $\omega$ satisfies
\be
\vec \nabla \times \vec \omega ~=~  V \vec \nabla M -
M \vec \nabla V +  \frac{1}{2}(K^I  \vec\nabla L_I - L_I \vec
\nabla K^I )\,.
\label{omegadef}
\ee
For $T^6$ compactification, $C_{IJK}\equiv  |\epsilon_{IJK}|$. To have smooth geometries, we need to further overlap the sources of these harmonic functions. In other words, for a system with N-centers, we have the following harmonic functions:
\bea
V~=~ q_0 ~+~ \sum_{j=1}^N\, \frac{q_j}{r_j}\,, ~&&~ K^I ~=~ k_0^I ~+~ \sum_{j=1}^N\,\frac{k_j^I}{r_j}\,,\nonumber \\  
L_I ~=~ (l_0)_I ~+~ \sum_{j=1}^N\,\frac{(l_j)_I}{r_j}\,, ~&&~  M~=~ m_0 ~+~ \sum_{j=1}^N\,\frac{m_j}{r_j}\,, 
\label{8harm}
\eea
where $r_j$ is the distance to the center `$j$'. The regularity  of $Z_I$ and $\mu$  at $r_j=0$ implies $(l_j)_I$ and $m_j$ should be:
\be
 (l_j)_I ~=~ -\frac{1}{2}\,C_{IJK}\,\frac{k_j^J\,k_j^K}{q_j}\,,  ~~~ m_j ~=~ \frac{1}{12}\,C_{IJK}\,\frac{k_j^I\,k_j^J\,k_j^K}{q_j^2}\,. 
 \label{reg_const}
 \ee
The constants of the harmonic functions, $(q_0, k_0^I, (l_0)_I, m_0)$, will determine the asymptotic structure of the metric. We will leave them unfixed at this moment.

\subsection{Bubble equations in general asymptotic background}

With the asymptotic metric unfixed, the parameters of the bubbled geometries with N centers will include the eight asymptotic constants, the charge parameters (i.e., $(q_j\,,k_j^1\,,k_j^2\,,k_j^3)$) and the positions of all centers. However, not every solution with arbitrary parameters will be physical. In general, these solutions will contain CTC's (closed-time-like curves) and require one to constrain these parameters carefully to avoid their appearance. Specifically, it was shown that the necessary conditions to be free of CTC's are:
\be
\cQ ~\geq~ 0\,\,, \qquad V\,Z_I ~\geq~
0\,\,, ~\quad~ I=1,2,3,
\label{CTC_nc}
\ee
where,
\be
\cQ ~\equiv~ Z_1\,Z_2\,Z_3\,V ~-~ \mu^2\,V^2\,.
\label{Qdef}
\ee
These conditions are the main sources of difficulty to understanding the physical sector of the bubbled geometries because one needs to check these conditions globally in order to make sure the solutions are well-behaved. However, there are some equations that can be obtained by verifying the first condition in (\ref{CTC_nc}) at the positions of the centers. These so-called bubble equations are relatively easier to check and are proven to be quite useful even though they alone cannot guarantee the well-behavedness of the solutions.

To derive the bubble equation, recall that $Z_I$ are regular while $V$ diverges at the centers and therefore it is clear that $\mu$ must vanish at the position of every center in order for the first condition in (\ref{CTC_nc}) to be satisfied. Using the definition of $\mu$ in (\ref{mudef}) and harmonic functions in (\ref{8harm}), the constant part of $\mu$ at each center is:
\bea
\mu(r_j=0) &= & \frac{k_j^1\, k_j^2\, \bar{k}^3  +  k_j^2\, k_j^3\, \bar{k}^1 + k_j^3\, k_j^1\, \bar{k}^2}{q_j^2} ~-~ 2\,\frac{k_j^1\, k_j^2\, k_j^3}{q_j^3}\,\bar{q} ~+~ \bar{m} \nonumber \\ 
~&&-~ \frac{1}{2\,q_j}\,\sum_{I=1}^3\,\Big(\frac{\bar{q}}{q_j}\,k_j^I (l_j)_I   ~-~ k_j^I \bar{(l)}_I - (l_j)_I \bar{k}^I \Big) \,,
\label{muconst}
\eea
where $(\bar{q}, \bar{k}^I, \bar{(l)}_I, \bar{m})$ are constants including the contributions from the asymptotic constants and other centers. By using the regularity constraints in (\ref{reg_const}), after some algebra, one can show that the conditions, $\mu(r_j=0)=0$, can be concisely written as:
\be
-\sum_{i \neq j}^N\, \frac{\la\, \Gamma_j, \Gamma_i \,\ra}{r_{ij}} ~=~ \la\, \Gamma_j, h \,\ra\,,~\hspace{1cm}~ j=1, 2, ...N,
\label{bubeq}
\ee
where $N$ is the number of centers , $r_{ij}$ is the distance between center `i' and `j' and $\Gamma_j$ is a charge vector\footnote{In bubble geometries, the charge parameters of the center can be specified by its magnetic part alone, for example, $\Gamma_j^{(m)}=(q_j,\, k_j^I)$.}:
\be
\Gamma_j ~\equiv~  \Big(\, q_j,\, k_j^I,\, \frac{1}{2} \,C_{IJK}\,\frac{k_j^J k_j^K}{q_j},\, \frac{1}{6}\,C_{IJK}\,\frac{k_j^I k_j^J k_j^K}{q_j^2}\Big)\,.
\label{chg_vec}
\ee
Similar to charge vectors, $h$ is an eight-component vector formed by the asymptotic constants:
\be
h ~\equiv~ \Big(\, q_0,\, k_0^I,\, -(l_0)_I,\, 2\,m_0 \Big),
\label{hdef}
\ee
and generally we will refer to it by the asymptotic vector. Finally, the symplectic intersection product is defined as:
\be
\Big\langle (A^0,\,A^I,\,A_I,\,A_0)\,,\,   (B^0,\,B^I,\,B_I,\,B_0) \Big\rangle  ~:=~ -A^0 B_0 ~+~ A^I B_I ~-~ A_I B^I ~+~ A_0 B^0\,.
\label{simp_prod}
\ee
Notice that for the bubbled geometries, this symplectic product of the two charges is actually equivalent to the flux between them as defined in \cite{Bena:2005va}:
\be
\la \Gamma_i\,, \Gamma_j \ra ~=~ q_i\,q_j\,\Pi^{(1)}_{ij}\,\Pi^{(2)}_{ij}\,\Pi^{(3)}_{ij}\,,
\label{prod_flux}
\ee
with $\Pi_{ij}^{(I)} \equiv \left(\frac{k_i^I}{q_i}-\frac{k_j^I}{q_j}\right)$. Additionally, for  convenience, we define the vector, $\cH$, which is composed of the eight harmonic functions as the following:
\be
\cH ~=~ \Big(\,V\,,\,K^I\,,\,-L_I\,,\,2\,M\,\Big)\,.
\ee
To fix the geometry to be asymptotically flat, $h$ must be chosen appropriately. If we normalize the constants of $Z_I$ to be one, $h$ should be set to:
\be
 h_0 ~=~ (0,\,0,\,0,\,0,\, -1,\, -1,\, -1,\,-\sum_{I=1}^3\,\sum_j^N\,k_j^I)\,\,.
 \label{h_flat}
 \ee
By fixing $h$ at $h_0$ and using (\ref{prod_flux}), the bubble equations in (\ref{bubeq}) reduce to the familiar form as defined in \cite{Bena:2005va}. However, one notices that for general $h$, the summation of all $N$ equations will not be satisfied because the left-hand side of (\ref{bubeq}) sums to zero due to the antisymmetric property of the symplectic product (\ref{simp_prod}) while the right-hand side is in general not zero. Therefore the self-consistency requires $h$ to satisfy the following condition:
\be 
\Big\langle \Gamma_t \equiv \sum_j^N\,\Gamma_j \,,\, h \Big\rangle ~=~ 0\,.
\label{hflow}
\ee
This consistent condition of $h$ and the general bubble equations (\ref{bubeq}) will play an important role in section \ref{split}.

\subsection{Large gauge transformation}

It was noticed in \cite{Bena:2005ni} that there is a symmetry of the solutions associated with the shifting of the harmonic functions in the following way:
\bea
V &\rightarrow& V\,, \nonumber \\
K^I &\rightarrow& K^I ~+~ c^I\,V\,, \nonumber \\
L_I &\rightarrow& L_I ~-~ C_{IJK}\,c^J\,K^K ~-~ \frac{1}{2}\,C_{IJK}\,c^J\,c^K\,V\,, \nonumber \\
M &\rightarrow& M~-~ \frac{1}{2}\,c^I\,L_I ~+~ \frac{1}{12}\,C_{IJK}\,(V\,c^I\,c^J\,c^K+3\,c^I\,c^J\,K^K)\,.
\eea
This transformation does not change the metric, CTC's conditions or bubble equations but shifts the gauge fields by some constant forms:
\be
A^I ~\rightarrow~ A^I ~+~ c^I\,(d\psi\,+\,\vec{A})\,.
\ee
Since it changes the boundary value of the gauge field, the symmetry associated with this transformation can be regarded as some kind of large gauge symmetry.

For a multi-center solution, this transformation is equivalent to shift every charge parameter and asymptotic constant as the following:
\bea
k_j^I &\rightarrow& k_j^I ~+~ c^I\,q_j\,, \nonumber\\
h &\rightarrow& \Big(q_0\,,\, k_0^I + c^I\,q_0\,,\, -(l_0)_I + C_{IJK}\,c^J\,k_0^K  + \frac{1}{2}\,C_{IJK}\,c^J\,c^K\,q_0\,, \nonumber \\
&& ~ 2\,m_0 - c^I\,(l_0)_I +\frac{1}{6}\,C_{IJK}\,\left(q_0\,c^I\,c^J\,c^K + 3\,c^I\,c^J\,k_0^K\right)\Big)\,.
\label{gauge_inv}
\eea
Because all centers in bubble geometries have nonzero Gibbons-Hawking (GH) charges, $q_i$, every $k_i$ will be shifted by some constant under this gauge transformation. Therefore, for every center, there exists a particular gauge choice such that the center contains only the GH charge (i.e., $\Gamma=(q,0,0,0,0,0,0,0)$). This implies that $k_i$ may not be considered as physical quantities since they are not gauge invariant. However, some combinations\footnote{Typical examples are the total asymptotic charges and angular momenta of the system.} or relations of $k_i$ for several centers may become gauge invariant. Therefore, to define the physical sector, one needs to have some constraints on this combinations or relations between centers rather than the exact values of $k_i$. This is probably one of the reasons why it is difficult to define the physical sector in bubbled geometries.

\section{Split (attractor) flow conjecture}
\label{split}

We have explained the difficulty of checking CTC's condition globally in five dimensional bubbled geometries and it turns out that a similar difficulty is shared by four dimensional BPS multi-center solutions. In four dimensions, the metric is determined by the entropic function, $\Sigma$, and it can potentially become imaginary somewhere in the space. To ensure the well-behavedness of the solutions, one needs to check if $\Sigma$ is real globally which is similarly difficult to the CTC's checking. To overcome this difficulty, an interesting conjecture was proposed in \cite{Denef:2000nb, Denef:2007vg}. The conjecture states that a multi-center solution exists if and only if a split attractor flow tree exists in the moduli space. A split attractor flow tree is a tree-like graph in the moduli space and it starts from the moduli at infinity and terminates at attractor points of each of the centers. The flow can only split when it crosses the wall of marginal stability which is some codimensional one surface in moduli space. On the wall, the phases of the central charges of two constituents are aligned  and can be separated infinitely far away. 

In order to use this conjecture to explore the physical sector of the bubbled geometries or verify the conjecture, we need to identify the split attractor flow trees in five dimensional moduli space and find the explicit connection between the trees and the solutions. However, the central charges and the moduli space in five dimensional theory are not complex and it seems unlikey to have a similar picture as with the wall of marginal stability in four dimensional moduli space. On the other hand, it was suggested in \cite{Denef:2007vg, deBoer:2008fk} that one can study five dimensional solutions in a similar manner through the uplifting of 4d to 5d. However, this uplifting procedure usually involves the rescaling of the coordinates or harmonic functions which seems to make the connection to five dimensional solutions become ambiguous. Also, the lacking of the understanding of how this conjecture works from the point view of the five dimensional theory is unsatisfactory. Here, we will take a more direct approach. We identify some tree-like structures from the five dimensional solutions without referring to the moduli space and propose the criterion of the existence of the trees and solutions solely based on CTC's conditions. We will still refer to these tree-like structures as the split flow trees due to their similarity with the trees in the four dimensional moduli space. We propose two methods\footnote{We use ``static" and ``dynamic" to distinguish two pictures. Sometimes, we ignore the quotation marks but it should be understood that they should not be interpreted literally.} to construct the pictures of  the split flow trees. Then from these two pictures, we propose two forms of the split flow conjecture. Finally, we will compare these two versions of the conjecture with Denef's original conjecture.

\subsection{``Static" split flow trees in space and the conjecture}
\label{static}

\begin{figure}[t]
\centering
\includegraphics[width=12cm]{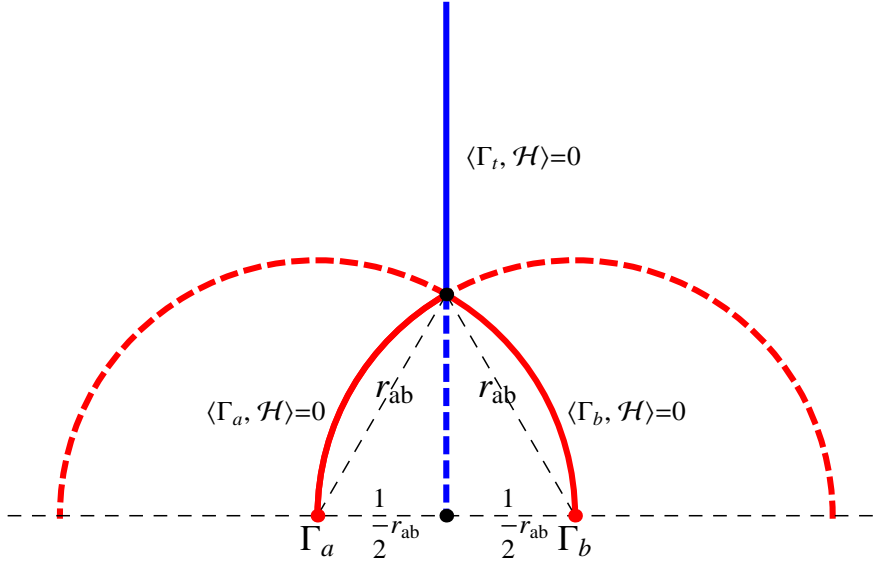}
  \caption{\it\small This picture represents the two-center split flow in space and was obtained by cutting a slice along the axis of the two charges at a particular azimuthal angle. The solid blue and red lines represent the actual flows while the whole lines included the dashed ones are defined by (\ref{Gt_flow}-\ref{Gb_flow}).}
\label{twocenter}
\end{figure}

In this section, we present a method to ``locate" the flows in spacetime once a solution is given. For bubbled geometries, the space where the centers are located is $\IR^3$ inside the Gibbon-Hawking space. To see the image of the flow in this space, one just needs a simple observation that when changing the asymptotic moduli to any point along the flows, the system should become another valid solution with a different asymptotic metric.   For a single-center solution with spherical symmetry, the flow is a single line in the moduli space. Any point on this line map to a sphere in $\IR^3$. Then, taking the asymptotic moduli to any point down the flow will result in another valid solution with a truncated flow. However, for a two-center solution, the situation is different because we don't have spherical symmetry but only an axial symmetry. So, what should be the trace the flow follows in $\IR^3$? Specifically what is the orientation it should take when flowing? To answer this question, one notices that for an arbitrary path, the values of the harmonic functions at the points along the path will not satisfy the consistent condition (\ref{hflow}). This means if we assign the asymptotic constants by the values of the harmonic functions at these points, there will be no solution for the bubble equations. This, of course, implies that the ``image" of the flows in space cannot take any arbitrary path presuming that assigning the asymptotic moduli to any point down the flow in moduli space should result in another valid solution. Therefore, the correct path should be confined by the consistent condition. For a two-center solution, we have the following condition:
\be
\langle \Gamma_t,\, \cH(\vec{r}) \rangle ~=~  \Big\langle \Gamma_t\,,\, \frac{\Gamma_a}{r_a} + \frac{\Gamma_b}{r_b} +h_{\infty} \Big\rangle ~=~ \langle \Gamma_a,\, \Gamma_b \rangle\,(\frac{1}{r_b}-\frac{1}{r_a}) ~=~0\,\,,
\label{Gt_flow}
\ee
where $h_{\infty}$ is the asymptotic vector and we have used the fact that $\langle \Gamma_t,\, h_{\infty} \rangle=0$. One can see that if $\langle \Gamma_a, \Gamma_b \rangle \neq 0$, this condition defines the equal-distance plane between two centers. Therefore, the main flow in the moduli space maps to a plane in $\IR^3$ while a point on the flow maps to a circle on this plane due to the axial symmetry. Similarly, to find the flows which terminate at $\Gamma_a$ and $\Gamma_b$, we need to consider the following conditions:
\bea
\langle \Gamma_a,\, \cH(\vec{r}) \rangle ~&=&~ \frac{\langle \Gamma_a,\,\Gamma_b \rangle}{r_b} ~+~ \langle \Gamma_a,\,h_{\infty} \rangle ~=~0\,\,,  \label{Ga_flow} \\
\langle \Gamma_b,\, \cH(\vec{r}) \rangle ~&=&~ \frac{\langle \Gamma_b,\,\Gamma_a \rangle}{r_a} ~+~ \langle \Gamma_b,\,h_{\infty} \rangle ~=~0\,\,. \label{Gb_flow}
\eea
One may notice how they are very similar to the two-center bubble equations. In fact, to solve these equations is equivalent to solving the bubble equation and it gives $r_a=r_b=r_{ab}$. Therefore, these conditions define two spheres centering at $\Gamma_a$ and $\Gamma_b$ and with the radius equal to the separation of the two centers. To see this more clearly, we can look at a slice at a particular azimuthal angle and the flows will be look like Figure \ref{twocenter}. The blue line represents the main flow while the red lines represent the split flows. If one parametrizes the blue line by the distance to one of the charges (i.e., $r=r_a=r_b$), then the domain of this line is $r\in [\infty, \frac{r_{ab}}{2}]$. The end of the line is at $r=\frac{r_{ab}}{2}$ and we will call it the ``half-split point". The point where the blue and red lines cross (i.e., $r=r_{ab}$) is called the split point. 

For the solution with just two centers, the tree-like pattern is clear. However, if we include more centers, we need to consider many possible lines with different combinations of charges. The resulting pattern may be quite confusing.  The best way to extract the tree from the pattern is probably tracing the lines starting from the centers. When the two lines cross, remove the remaining lines and join them to the line corresponding to the summation of the two charges. Eventually, one trace it back to the line that goes to infinity and the result is the tree that corresponds to this solution.

\begin{figure}
\centering
\begin{tabular}{ccc}
\includegraphics[width=0.3\linewidth]{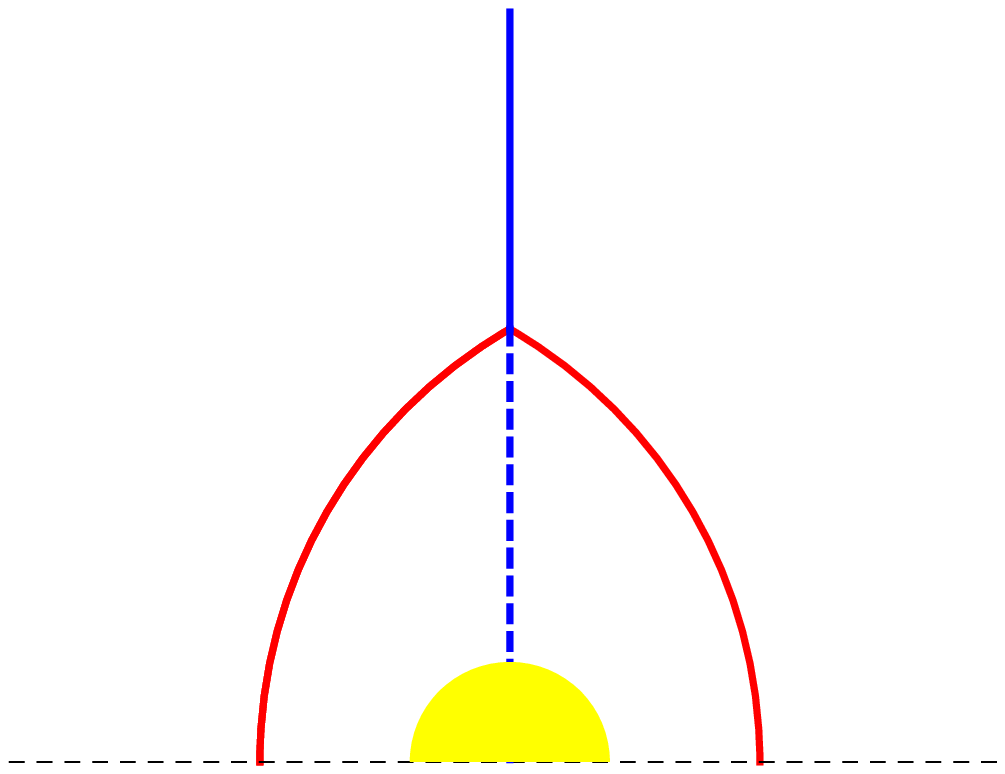} &
\includegraphics[width=0.3\linewidth]{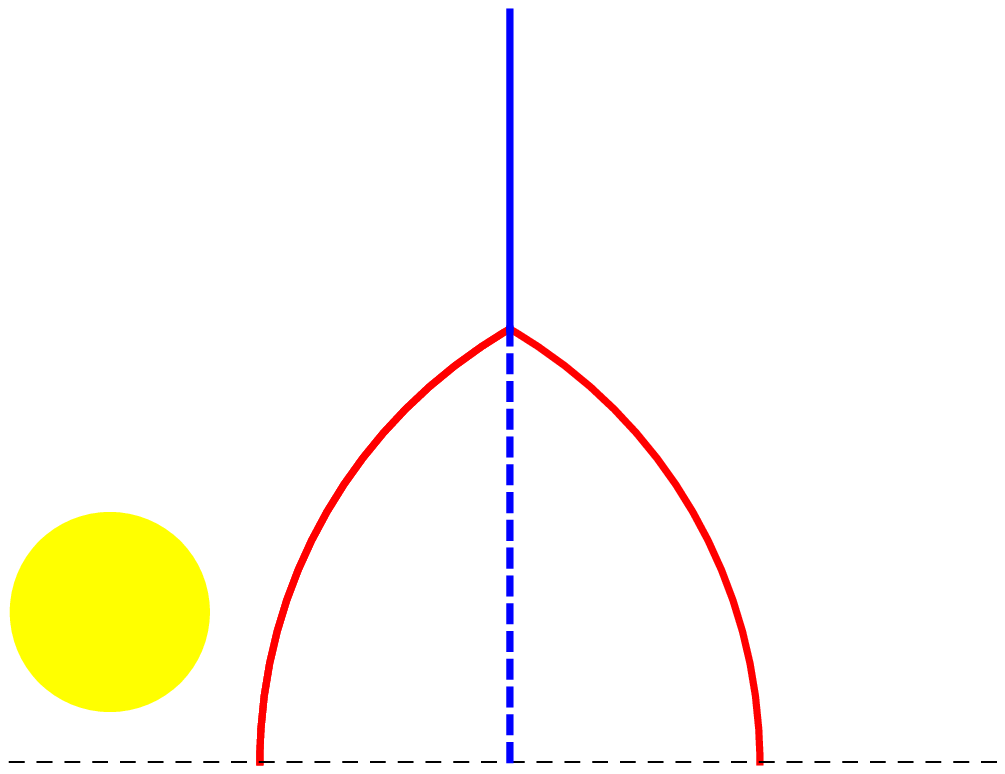} &
\includegraphics[width=0.3\linewidth]{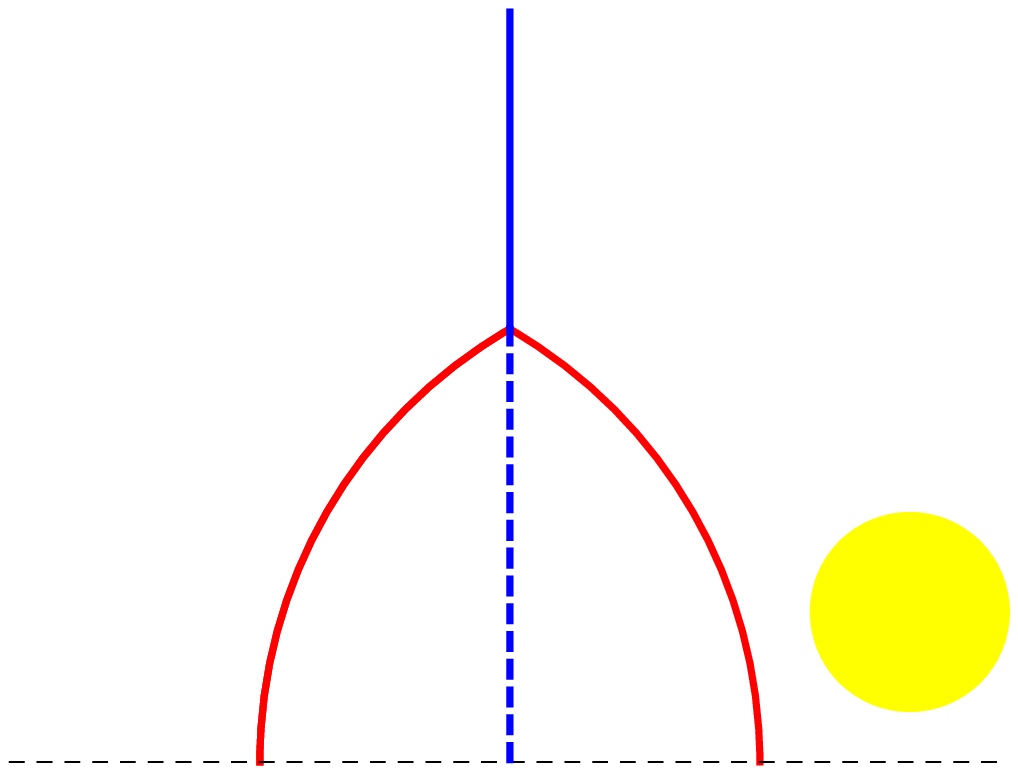} 
 \end{tabular}
 \caption{\it\small These are several simple examples of the configurations which will violate the split flow conjecture if they exist. The yellow circle represents the region full of CTC's.}
\label{ctcvs}
\end{figure}

With this picture of the split flows in $\IR^3$, we can state the static version of the split flow conjecture: {\it for a solution in which a clear split-flow pattern can be identified by the method described above, the absence of the CTC's along the paths of the flows guarantees that the whole space is CTC's free.} In other words, this version of split flow conjecture implies that the situations in Figure \ref{ctcvs} cannot exist. Among them, the first one is particularly interesting because we know for some multi-center charge configuration, the attractor point of the total charge is not well-behaved. This usually means there will be CTC's near this attractor point and since the half-split point is closer to the attractor point than the split point, it seems possible that the half-split point dips into CTC's region while the split point remains free of CTC's. Therefore, testing CTC's near the half-split points for those solutions with clear split-flow patterns provides an interesting way to test the conjecture.

This ``static" version and the corresponding conjecture provide the representation of the trees in spacetime and a natural resolution if one wants to find out the tree that a particular solution corresponds to. However, there exists several drawbacks for this approach. First, from several numerical examples, we found out that if the fluxes that are involved in the split points of the tree are very different from each other in scale or very close to zero, some lines may be broken even though the solution itself is indeed a split flow solution based on the ``dynamic" analysis in the next subsection. This not only makes it difficult to extract the tree from the solution unambiguously but also makes it hard to distinguish this solution from the scaling solutions in such cases. Furthermore, in order to use the ``static" version of the conjecture, one needs to construct solutions first. Therefore, it is clear that this version is not suitable to be used in any systematic search. Due to these drawbacks, we will mostly rely on the ``dynamic" version combined with the ``skeleton" construction which will be introduced in sections \ref{dynamic} and \ref{skeleton}. However, due to its close connection with spacetime, the ``static" version is useful for testing the conjecture directly. Therefore, we will mostly use it to verify the conjecture by looking for any counter example. Specifically, we will check the CTC's conditions at the half-split points when we verify the existence of three-center trees in section \ref{3center}.

\subsection{``Dynamic" version of the split flow trees and the conjecture}
\label{dynamic}

Corresponding to the wall-crossing in the moduli space, what one will see in the four dimensional supergravity picture is that when the asymptotic moduli (vector) approaches the wall of marginal stability, the system will be divided into two parts which are separated further from each other  under the constraint of the integrability conditions. Since the bubble equations (\ref{bubeq}) in bubbled geometries have exactly the same form as the integrability conditions in four dimensional supergravity, it seems quite natural to follow this thinking and consider how a solution deforms under the constraint of the bubble equations when we change the asymptotic vector, $h$.

To be more specific, consider a certain number of centers with the total charge equal to $\Gamma_t$. Then we start with some asymptotic vector, $h_{\infty}$, and begin to change it ``continuously" by following the flow:
\be
h_f ~=~ h_{\infty} ~+~ \eta\, \Gamma_t\,, ~~\qquad~~  \eta: 0 \rightarrow \infty\,.
\label{hflow_dyn}
\ee
Then we ``continuosly" trace how this solution changes under the constraint of the bubble equations. At some point, one may find the solution is divided into two parts which are separated further away when $\eta$ is approaching a particular value, $\hat{\eta}$. After $\eta$ passes this value, the solution no longer exists because it decays to two parts separated infinitely far away. However, one can trace the two parts separately starting from $\hat{\eta}$ and go on until eventually the whole solution reduces to several independent single centered solutions. If one find a solution can be completely deconstructed in this way, one can conclude it is a split flow solution and the order of the splitting determines the tree it corresponds to. However, in some cases, one may find some or all of the centers form some composite element which does not expand or separate to two parts no matter how one changes the asymptotic vector and on the contrary, it usually decreases in size when one follows the flow in (\ref{hflow_dyn}). This kind of behavior is the signature of the scaling solutions and unfortunately, we cannot use the split flow techniques to search them or classify them.

Notice that the solution is not uniquely determined by the bubble equations. What we trace here is just a point in one of the components of the solution space which will be explained more in section \ref{split_components}. However, this behavior of splitting should not be considered as the feature or property of a particular point in the solution space but the whole component. In other words, it does not matter which point in the component you are tracing\footnote{There are some subtleties if the scaling and split flow solutions coexist. In this situation, it is possible to have a bifurcation point between scaling and split flow components. If such a point exists, it will be crucial to decide which part of the component to trace. For more detail, see the numerical example in section \ref{bifurcation} } as long as you do it continuously. The reason we should do it ``continuously" is to prevent the solution jumping to the potentially many other components nearby. This indeed causes some difficulty in implementing this picture on the solutions practically. However, in general, it seems to provide a more accurate and unambiguous classification of the solutions compared to the ``static" version.
  
From this picture, we propose another form of the split flow conjecture as the following: {\it The fact that if a solution contains CTC's or not cannot be changed by any continuous deformation under the constraint of the bubble equations. This includes those deformations induced by tuning the asymptotic vector continuously as long as one keeps it inside of the physical region}.  More precisely, what we mean is that the asymptotic vector is free to move around or even split when it hits the split point but it must be bounded inside of the region where the corresponding self-consistent condition in (\ref{hflow}) is satisfied and the metric at infinity determined from it must be CTC's free. While we have been considering this ``dynamic" procedure involving tuning the asymptotic vector as some kind of operational definition of saying one solution belongs to a particular tree, it may have a physical interpretation as well. The nature guess is that it corresponds to changing the background of the whole system adiabatically. For example, the Taub-Nut interpolation in \cite{Bena:2005ni} that connects five dimensional black rings to four dimensional black holes can be regarded as some special flow in this ``dynamic" picture. From this point of view, the ``dynamic" version of the conjecture seems quite natural. Because it is very hard to believe that one can render\footnote{Naturally, when we take the asymptotic vector to the split point, the system will be divided into two parts. However, the resulting two independent systems are still both physically well-defined.} a well-defined solution unphysical by changing the background adiabatically presuming, of course, the background itself is physical.

Notice that in the ``dynamic" version of the conjecture, how one should tune the asymptotic vector was never specified. However, in order to use the conjecture practically, one needs to construct a systematic method to tune the asymptotic vector. In the beginning of this subsection, we have introduced a simple choice like the one in (\ref{hflow_dyn}) which is a single centered flow. In fact, one can construct a complete split flow tree solely from joining these single centered flows. The result is a tree-like graph referred to as the ``skeleton" and its construction will be explained more in section \ref{skeleton}. Moreover, we will also introduce a criterion for identifying the well-behaved trees based on checking CTC's conditions on  these skeletons. Unlike the ``static" version of the conjecture, we do not need to solve the bubble equations in order to construct a ``skeleton" and therefore, this criterion is useful for a systematic search.

\subsection{The relation between different versions of the conjecture}
\label{relation}

We have introduced several versions of the conjecture and here we clarify the relation between them and Denef's original split attractor flow conjecture.

As we have mentioned in the previous subsection, the ``dynamic" version never specify a way to tune the asymptotic vector. From the perspective of the ``dynamic" version, to identify a split flow solution is equivalent to ask if there exist a way or path to tune the asymptotic vector such that the solution can be completely deconstructed. However, this path by no means needs to be unique. For example, in the ``static" version, once the split-flow pattern is clearly identified and no CTC is discovered on the pattern, it immediately provide a valid choice of the path for the ``dynamic" version since the harmonic functions are analytic and they satisfy the constraint in (\ref{hflow}). Furthermore, the precise path of this pattern will be different depending on the internal moduli of the solution and all of them are valid choices of the path. Additionally, a ``skeleton" that pass the criterion in the next section is also a valid choice. Therefore, it is clear that the ``dynamic" version is stronger than any criterion based on a specific choice of path including the ``static" version and the ``skeleton". On the other hand, it is also clear that the ``dynamic" version is more general because it is supposed to be applicable to any solution including a scaling solution. However, in scaling solutions, there may be multiple components including good or bad ones connected to the same scaling point at which all of the centers are overlapping with each other. Clearly, if the scaling point exists, this will violate the ``dynamic" version because through this point one can connect a good component to a bad one. However, from the speculations in \cite{Bena:2007qc} and the detailed analysis in \cite{deBoer:2008zn}, it seems quite likely that the scaling point will be removed by some quantum effects. If this is true, the ``dynamic" version can still be true even for scaling solutions. Even so, the ``dynamic" version is not exactly a criterion but more like some kind of adiabatic property of the solutions. Therefore, It cannot help us sort out the scaling components because we cannot ``disentangle" the scaling solutions by tuning the asymptotic vector.

Even though there exist several differences between the different versions, it is still possible that they are all equivalent when all of them are applicable. One can image that the several choices of path including those in the ``static" version and the ``skeleton" form a representative  ``bundle" of paths. That is if any path in this bundle hit any pathology, there is no other choice of path can do better job. If this is true, all of the different versions will give same verdict to any solution when all of them are applicable. This is the point of view we take in this paper. On the other hand, it is also possible that the ``dynamic"  version is too strong and only the criterion based on the certain path choice is true. In that case, the difference between the ``static" version and the criterion based on the ``skeletons" will be an interesting open question.

If we accept the above ``bundle" picture, it is quite natural to select the ``skeleton" as the representation of the whole ``bundle" because it can be systematically constructed. Therefore, the criterion based on the ``skeleton" can be used to search the healthy ``bundles" and the corresponding trees. As we have mentioned, the ``skeletons" are constructed by joining several single centered flows. If we change the setting to four dimensional theory and map the asymptotic vector to a point in moduli space, a ``skeleton" will look like a split attractor flow tree. If we change the physical conditions accordingly, for example, instead of checking CTC's conditions, we make sure the central charge never hit zero in the middle of the flows, etc..., the criterion in the next section will be exactly like Denef's original conjecture. This suggests that the criterion should be considered as an equivalent\footnote{However, this will imply the causality (e.g. CTC's conditions) in five dimensions are connected to the stability issues or the wall-crossing phenomena in four dimensions. How that is so may be an intertesting question worth to explore.} or a natural extension of Denef's original conjecture to five dimensional theory.

\section{Split flow trees $\Longleftrightarrow$ split flow components}
\label{split_components}

\subsection{Solution space of bubble equations}

Since a solution must satisfy the bubble equations in order to be a candidate for well-behaved solutions, it is essential to understand the solutions of bubble equations. To be more precise, we should call them a {\it solution space}. The reason is that the bubble equations cannot fix the solutions completely because the number of equations is smaller than the degrees of freedom from the positions of the centers. Therefore, the solutions have some moduli or the ``sliding" degrees of freedom left to deform themselves. In \cite{deBoer:2008zn}, the space spanned by these moduli was referred to as the solution space\footnote{In this paper, it is the solution space of the integrability conditions that was studied but the conditions are practically the same as the bubble equations in (\ref{bubeq}).} and was studied in some simple cases. In the paper, the symplectic structure of the space was identified and the solution space was promoted to the phase space and quantized. By doing so, the wall-crossing formula \cite{Denef:2007vg} was recovered. This suggested that the wall-crossing or split attractor flows may have a very close relation with the solution space of bubble equations. However, it is still difficult to say something concrete about this space especially when the effective number of pairs with nontrivial fluxes between them is not so small. Nevertheless, through numerical studies, it is clear that in general the solution space is not even connected. This means that in most situations, the solution space has multiple components and each component is spanned by its own ``sliding" degrees of freedom.  Here, we roughly classify the components of solution space into three categories:  
\begin{itemize}
\item {\bf Split flow components:}  Any solution in these components can be deconstructed to several single-center solutions by tuning the asymptotic vector continuously along the split flow trees. There exists a one-to-one correspondence between these kinds of components and split flow trees. They are the main subjects of the split flow conjecture and their absence of CTC's globally is guaranteed by the conjecture.
\item {\bf Scaling components:} the solution of this kind has some or all of its centers forming scaling solutions. Roughly speaking, these centers trap in some kind of ``entanglement" state and cannot be torn apart by tuning the asymptotic vector. The whole solution may still have tree-like structure with some flows terminating at the scaling solutions. From the perspective of the split flows, these scaling solutions should be treated as single-center points or attractor points and should only exist if the corresponding attractor points are well-behaved. Because there are potentially many scaling components corresponding to the same attractor point in the bubbled geometries, the one-to-one correspondence is no longer true in these cases. Furthermore, if a solution shows scaling behavior, the conjecture can no longer help us determine whether this solution is well-behaved or not. Depending on whether the solution is truly absent of CTC's globally or not, they will belong to the scaling components or the bad components.
\item {\bf Bad components:} the solutions of this kind contain CTC's even though they satisfy the bubble equations and should be regard as unphysical.
\end{itemize}
From the perspective of microstate geometries, the scaling components are quite interesting and probably very important. Unfortunately, even though there are some numerical examples showing their existence, there is no systematic method to search or even distinguish the scaling from the bad components. On the other hand, the split flow solutions which belong to the split flow components are under better controls and we will show how to find the explicit connection between them and split flow trees.

\subsection{Building the ``skeletons" of the split flow solutions}
\label{skeleton}

In this section, we demonstrate the procedure to identify all split flow components given a set of centers with charges, $\Gamma_i$, and the asymptotic vector, $h_{\infty}$. At first, we exhaust every possible split flow tree and build the ``skeleton" for each of them. A ``skeleton" is a tree-like graph that contains the complete information of all split points, how the harmonic functions change along the flows, and finally, the topological data of the tree. In other words, a skeleton can be seen as the blueprint of a solution. Then, we propose the existence conditions for the skeletons and use them as a criterion to pick out the physical trees.

\begin{figure}[t]
\centering
\includegraphics[width=10cm]{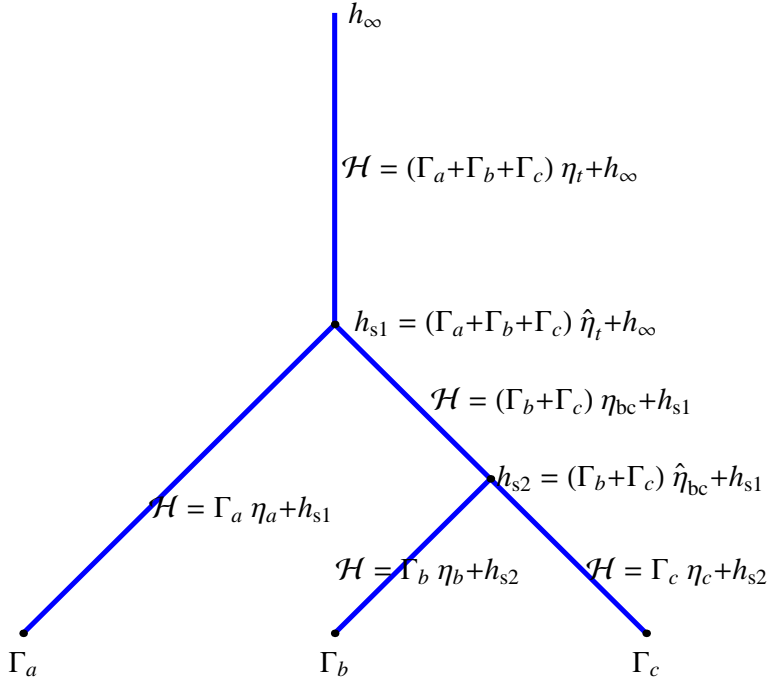}
\caption{\it\small  This figure shows the ``skeleton" of the tree, $\{\Gamma_a\,,\{\Gamma_b\,,\Gamma_c\}\}$. The edges are like the singe-center flows parametrized by $\eta$. How the harmonic functions change along the flows is completely determined once the asymptotic vector, $h_{\infty}$, is chosen. The two split points are present and labeled as $h_{s_1}$ and $h_{s_2}$.}
\label{skeleton_fig}
\end{figure}

As an example, we will use a simple three-center system and will adopt the convention in \cite{Denef:2007vg} to use a nested list for identifying a tree. To start, consider a set of three centers, $\{\Gamma_a\,, \Gamma_b\,, \Gamma_c\}$, there are three possible trees:
\be
\{\,\Gamma_a\,, \,\{\,\Gamma_b\,,\, \Gamma_c\,\}\}\,, ~\quad~ \{\,\Gamma_b\,,\, \{\,\Gamma_a\,,\, \Gamma_c\}\}\,, ~\quad~ \{\,\Gamma_c\,,\, \{\,\Gamma_a\,,\, \Gamma_b\,\}\}\,.
\ee
From a given asymptotic vector, $h_{\infty}$, the structure of the tree and the charge of each center, we can construct the ``skeleton" for each tree. For example, the skeleton of the first tree is shown in Figure \ref{skeleton_fig}. One can see that every edge in the skeleton looks like an attractor flow of a single-center solution and every split is treated like a two-center split. How the harmonic functions change along the flow and the positions of the split points are therefore determined accordingly. Every flow is parametrized by some positive variable, $\eta_i$, which is related with the inverse of the radius of the corresponding single-center solution. The parameter, $\eta_i$, starts from zero to infinity if the flow terminates at a center or stops at some particular positive value, $\hat{\eta}_i$, if it hits a split point. The exact value of $\hat{\eta}_i$ for a particular split is determined by the flux between two centers and the value of the asymptotic vector at the beginning of the flow. For example, $\hat{\eta}_t$ and $\hat{\eta}_{bc}$ are:
\bea
\hat{\eta_t} ~=&~ \frac{1}{r_{a, bc}} ~&=~ - \frac{\la \Gamma_a\,,\,h_{\infty} \ra}{\la \Gamma_a\,,\,\Gamma_b+\Gamma_c\ra}\,, \\
\hat{\eta}_{bc} ~=&~ \frac{1}{r_{bc}} ~&=~ - \frac{\la \Gamma_b\,,\,h_{s_1} \ra}{\la \Gamma_b\,,\,\Gamma_c\ra}\,.
\eea
It is immediately clear that not every skeleton exists because $\hat{\eta_{i}}$ must be positive while it is not necessarily true for arbitrary charges and $h$. Also, in order to use the skeleton as the path for tuning the asymptotic vector just like the ``dynamic" procedure in section \ref{dynamic}, we need to ensure there is no CTC along every edge in the skeleton. Therefore, we have the following existence conditions:
\begin{itemize}
\item The positivity of the separations: $\hat{\eta_{i}} > 0$ for every split point.
\item Absence of CTC's along the flows: $\cQ\,(\eta_i)>0\,, ~ V Z_I\,(\eta_i)>0$\,.
\end{itemize}
The conditions above technically reduce the CTC's checking from three dimensional space to several one-dimensional lines. However, it is still difficult to check the second condition in practical and we adopt a weaker form of the existence conditions which reduce the checking to several points:
\begin{samepage}
\begin{itemize}
\label{weaker_exist}
\item $\hat{\eta_{i}} > 0$ for every split point\footnote{One finds that this condition is related to the existence condition of the wall of marginal stability in \cite{Denef:2007vg}. The subtle difference arises when the wall of anti-marginal stability is involved. However, it is possible that the anti-marginal stability will be ruled out if one includes the second condition.}.
\item Absence of CTC's at every vertex and end point, namely: 
\be
 \cQ\,(\Gamma_i)>0\,,~\quad ~ V Z_I\,(\Gamma_i)>0\,, ~\quad~ \cQ\,(h_i)>0\,,~\quad ~ V Z_I\,(h_i)>0\,. \nonumber 
\ee   
\end{itemize}
\end{samepage}
Here we would like to explain more about $\cQ\,(\Gamma_i)>0$ and  $V Z_I\,(\Gamma_i)>0$. In bubbled geometries, for a true single-center flow that starts from some $h$ and terminates at a GH center, the CTC's conditions are greatly simplified. For example, for a flow starting from $h_s$ and terminating at $\Gamma^{(m)}=(q,\,k^I)$, the harmonic functions change according to the following:
\be
\cH ~=~ \Gamma\,\eta ~+~ h_s\,.
\ee
As $\Gamma$ is a single GH center, we can always choose the gauge such that $\Gamma^{(m)}=(q,0,0,0)$. Supposedly, $h_s$ becomes $h'_s$ in this gauge:
\be
h'_s ~=~ (q_0,\, k_0^I,\,-(l_0)_I,\,0)\,,
\ee
where $m_0$ is required to be zero by the self consistent condition (i.e., $\la \Gamma^{(m)},\,h'_s \ra=0$). Then $V Z_I$ and $\cQ$ are greatly simplified:
\bea
V\,Z_I (\eta) ~&=&~ q\,(l_0)_I\,\eta ~+~ \cO(1)\,, ~~\quad~~ I=1,2,3, \label{sc_VZI} \\
\cQ (\eta) ~&=&~ q\,(l_0)_1\,(l_0)_2\,(l_0)_3\,\eta ~+~ \cO(1)\,.
\eea
Since all of these functions change linearly with $\eta$, we only need to check the beginning and the end to ensure the whole flow is free of CTC's. The starting point is either at infinity or some split point and should already be verified in the other existence conditions. The conditions at the end can be imposed simply by requiring that the slopes in (\ref{sc_VZI}) are positive:
\be
q\,(l_0)_1 > 0\,, ~\quad~ q\,(l_0)_2 > 0\,,~\quad~ q\,(l_0)_3 > 0\,.
\label{scf_CTC}
\ee
Although these conditions are expressed in a particular gauge, it should be understood that CTC's conditions themselves are gauge invariant. Also, it may seem quite strange that whether the center is well-behaved or not should depend on the starting point of the flow. This is probably because the centers in bubble geometries are not singular and contain no source.

We have shown the simplest example when the weak conditions are enough to imply the satisfaction of the strong conditions. However, it is unclear if this is true or not in more general situations. Nevertheless, the weaker form is apparently far easier to implement because we only need to check several points and will be very useful in the numerical studies. For the examples we present in section \ref{numerical}, the weak conditions are actually enough.

With more centers included in the set, the number of possible trees increases very quickly\footnote{Assuming every center is distinguishable, the number of trees is $\frac{(2N-2)!}{(N-1)!\,2^{N-1}}$.}. However, the procedure to build the ``skeletons" and pick out only those trees that satisfy the existence conditions is essentially the same. Therefore, given a set of a certain number of centers with some assigned charges and an asymptotic vector, one can write a simple computer program to list all of the trees that satisfy at least the weaker form of the existence conditions. Furthermore, one can use the ``skeletons" of these trees to find the corresponding solutions of the bubble equations through assembling the centers which will be explained in section \ref{assemble}.

\subsection{Disassembling and assembling the solutions}
\label{assemble}


As we have explained in section \ref{relation}, one can use the skeleton as the specific path for tuning the asymptotic vector. By doing so, one can disassembling the corresponding solution. For example, one change the asymptotic vector according to the flow depicted in the skeleton and continuously traces the deformation of the solution under the constraint of the bubble equations. When the asymptotic vector approaches the split point, the solution will divide into two independent clusters as the distance between them gradually increase. After it hit the split point, the system is divided into two subsystems and one continues to follow the split flows for them individually. Eventually, one can disassemble this solution pairwise completely and it serves as an unambiguous way to classify the solution as a particular tree. 

On the other hand, the skeletons can also be used as the blueprints to assemble the solutions. To find a solution explicitly, we need to fix its sliding degrees of freedom and this means to pick a particular point from the whole component. Which point should we choose? Although this is not proven, it is quite natural to suspect every split flow component should contain at least one solution that respects the axial symmetry due to their nature of pairwise decay. Therefore, for simplicity, we should try to assemble this particular axi-symmetric solution. For example, one can try to assemble this particular solution for the tree in Figure \ref{skeleton_fig}. To do that, we solve the bubble equations of only $\Gamma_b$ and $\Gamma_c$ with $h_{\infty}$ set to $h_{s_1}$.  Then we bring in the $\Gamma_a$ by pushing $h_{\infty}$ up flow a little bit and probing the axi-symetric solution which has $\Gamma_a$ separated very far away from $\Gamma_b$ and $\Gamma_c$. After we have this solution with three centers, we can change $h_{\infty}$ to $h_0$ slowly and trace the solution. Of course, for this case with just small number of centers, this procedure seems redundant. However, in principle, this procedure can assemble all split flow solutions with any number of centers. The only subtleties are the orientations of the different subsets. We find that sometimes, these orientations are important and one needs to change the orientations of some subsets in order to complete the flow.

\section{Numerical examples}
\label{numerical}

\subsection{The merger of two bubbled supertubes}

In this subsection, we review some numerical examples in \cite{Bena:2007qc} but look at them from the new perspective we have developed. The first numerical example was originally used to study the merger of two bubbled supertubes and it turns out to be a good example to demonstrate the different aspects of the split flow picture. The basic setup is the following: there are two bubbled supertubes and each contains two GH centers with the opposite GH charges. Therefore, in total we have five GH centers including the isolated GH center at the origin. The charges of a bubbled supertubes can be specified by their GH charge, $Q$ and two other parameters: $d^I$ and $X^I$ which are related to the charge parameters of the individual GH center. If the magnetic part of the charge parameters of the two GH centers in the bubbled supertube are:
\be
\Gamma_-^{(m)} ~=~ (\,-Q\,,\, k_{(-)}^1\,,\, k_{(-)}^2\,,\,k_{(-)}^3\,)\,, ~\quad~
\Gamma_+^{(m)} ~=~ (\,Q\,,\, k_{(+)}^1\,,\, k_{(+)}^2\,,\,k_{(+)}^3\,)\,,
\ee
where we assume $Q$ is a positive number and then $d^I$, $X^I$ are defined as:
\be
d^I ~\equiv~  2\,(k_{(-)}^I ~+~ k_{(+)}^I) ~\quad~ X^I ~\equiv~ 2\,\frac{k_{(-)}^I-k_{(+)}^I}{Q}\,.
\ee
In this example, the charge assignment of the two rings is the following:
\bea
Q_a=105\,,~&\quad& ~d_a^I ~=~ (\, 50\,,\, 60\,,\, 40 \,) \,, ~\quad~  X_a^I ~=~ (\, 110\,,\, 560\,,\, 50 \,) \,,\\ 
Q_b=105\,,~&\quad& ~d_b^I ~=~ (\, 80\,,\, 50\,,\, 45 \,) \,, ~\quad~  X_b^I ~=~ (\, x \,,\, 270\,,\, 280 \,) \,,
\eea
where $x$ was varied from about $64$ to the merger value of $x \approx 90.3$. We  are interested in axi-symmetric solutions and therefore we put all of five centers on the z-axis and fixed their order. After solving the bubble equations numerically, the separations between the center can be obtained. Because the number of the center is not so large, we can find all of the solutions. In the range of $x$ which goes from about $71.7$ to $90.3$, we found there are three branches of solutions.  In the following, we will see how these different branches correspond to the different components. We will show two cases with $x=85$ and $x=90$ in detail.

\subsubsection{$x=85$}

For $x=85$, three solutions written with the centers' $z$ coordinates are:
\begin{center}
\begin{tabular}{ c | c | c | c | c | c}
 & $\Gamma_0$ & $\Gamma_{+a}$ & $\Gamma_{-a}$ & $\Gamma_{+b}$ & $\Gamma_{-b}$  \\ \hline
 (\rmnum{1}) & 0 & 1774.222 & 1774.225 & 2916.575 & 2916.581 \\ \hline
 (\rmnum{2}) & 0 & 2346.036 & 2346.037 & 2423.872 & 2423.873 \\  \hline
 (\rmnum{3}) & 0 & 2397.797 & 2397.816 & 2398.288 & 2398.318 \\
 \end{tabular}
 \end{center}
The subscript, $\pm a$ and $\pm b$, indicate they have a positive or negative GH charge and belong to the ring $a$ or $b$. In \cite{Bena:2007qc}, the solution (\rmnum{3}) has been discovered with a large region of CTC's while the other two are free of CTC's. Moreover, the solution (\rmnum{2}) belongs to the branch which will eventually reach the merger of the two rings when $x \approx 90.3$. 

At first, we use the procedure described in section \ref{skeleton} to find how many split flow trees we have for this setting. For the system with five centers, we have $105$ possible trees. After going through the weak existence conditions, only two of them still remain.
\bea
\cT^{(1)} ~&\equiv&~\{\{\Gamma_0\,,\,\{\,\Gamma_{+a}\,,\, \Gamma_{-a}\}\}\,,\,\{\Gamma_{+b}\,, \,\Gamma_{-b}\}\}\,, \nonumber \\
\cT^{(2)} ~&\equiv&~  \{\Gamma_0\,, \,\{\{\,\Gamma_{+a}\,,\, \Gamma_{-a}\}\,,\,\{\Gamma_{+b}\,,\, \Gamma_{-b}\}\}\} \,.
\eea

\begin{figure}[t]
\centering
 \begin{tabular}{ccc}
\includegraphics[width=0.3\linewidth]{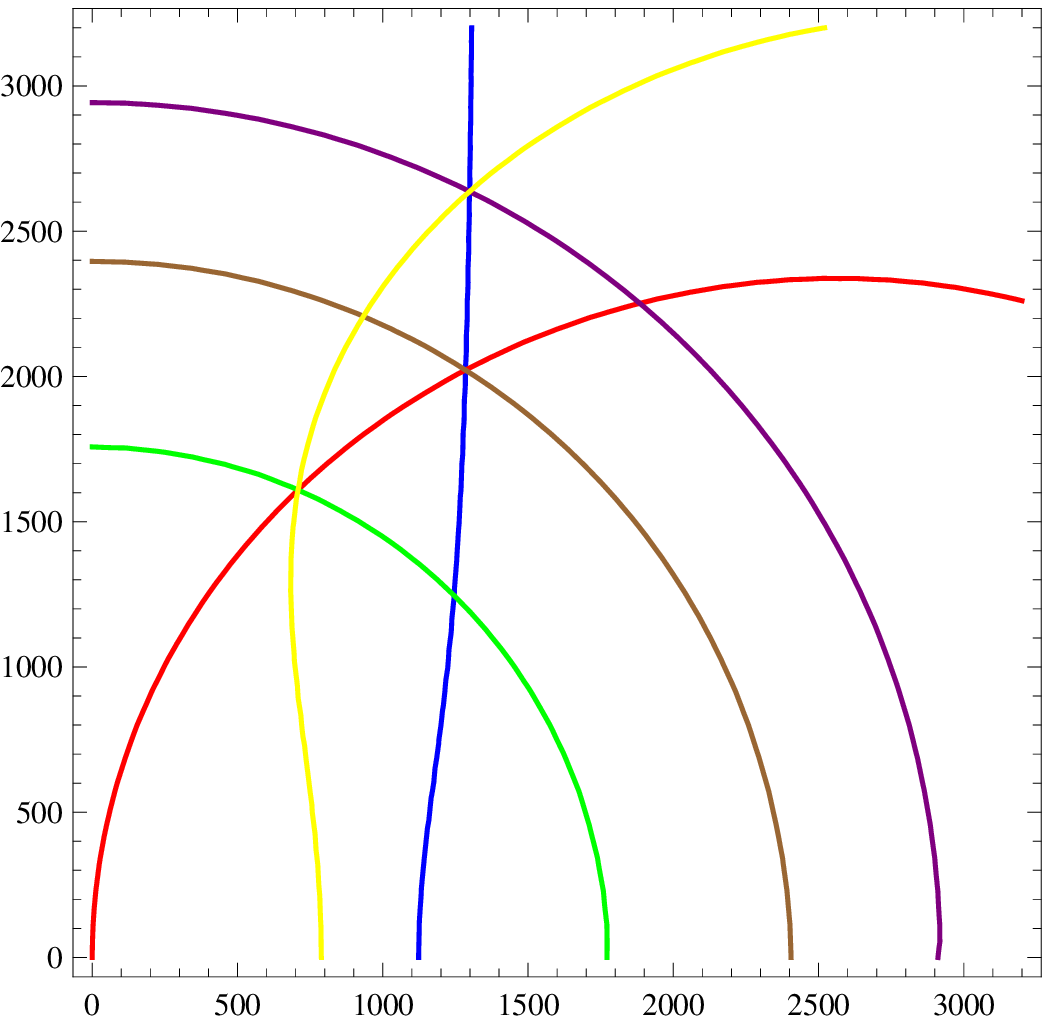} &
\includegraphics[width=0.3\linewidth]{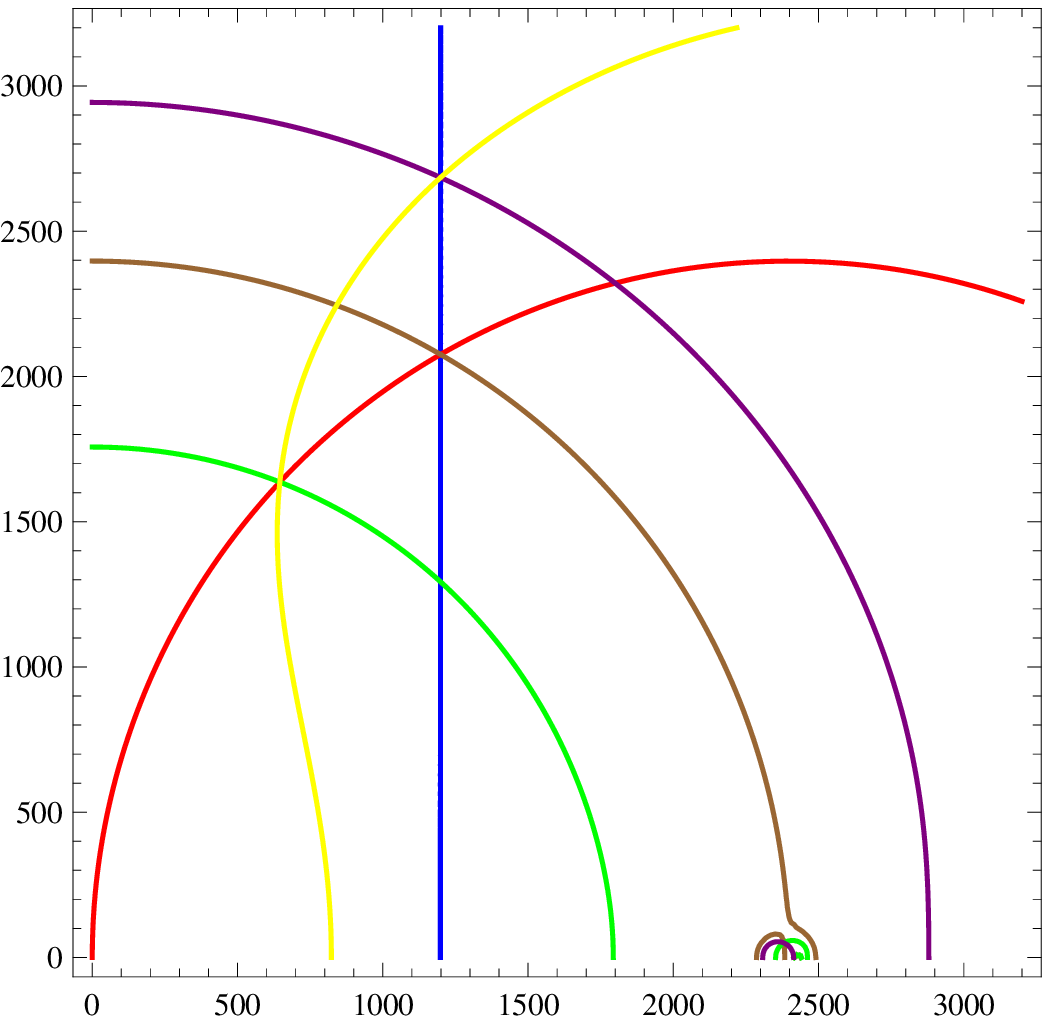} &
\includegraphics[width=0.3\linewidth]{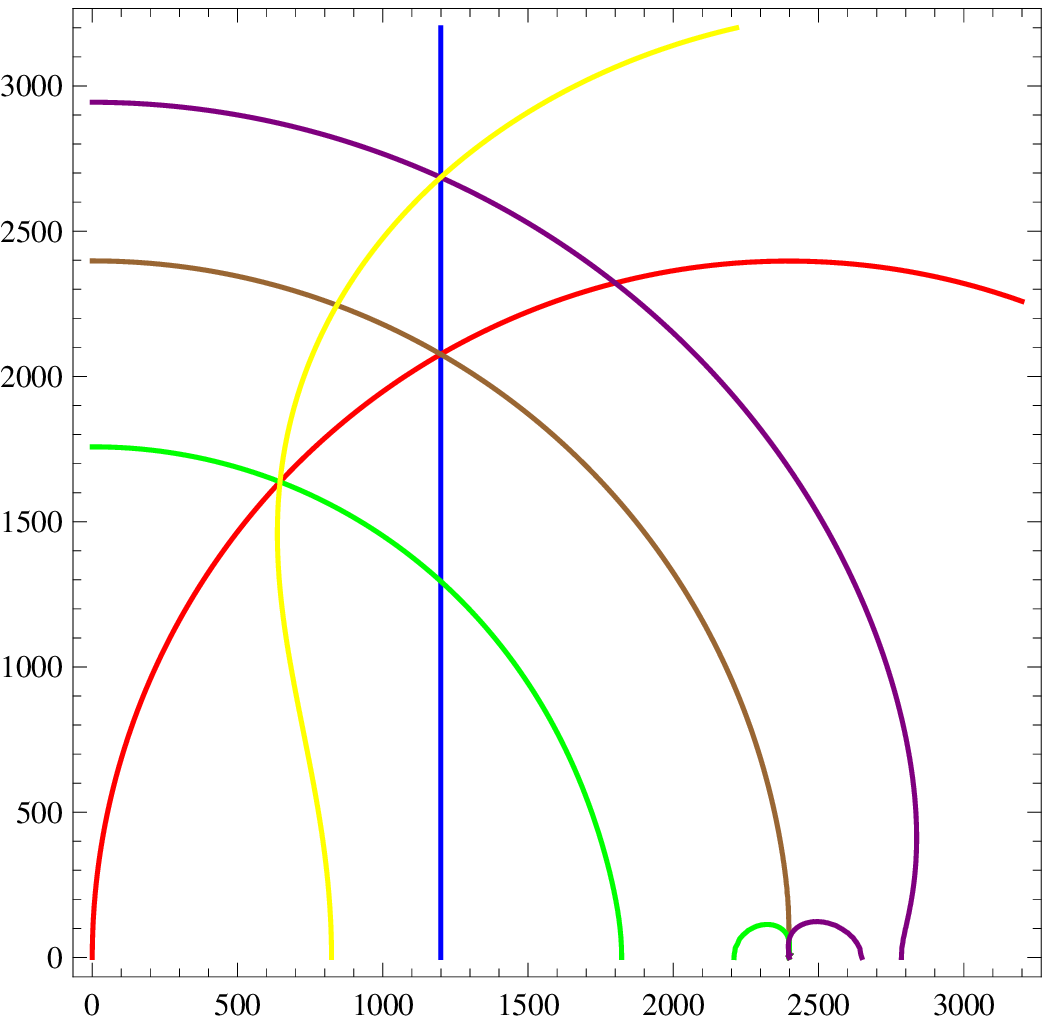}
 \end{tabular}
 \caption{\it\small From left to right, these graphs correspond to the solutions (\rmnum{1}-\rmnum{3}). The lines in different colors are drawn by several different conditions: $\la \Gamma_t\,, \cH \ra=0$ (blue), $\la \Gamma_0\,, \cH \ra =0$\,(red)\,, $\la \Gamma_a\,, \cH \ra =0$\,(green)\,, $\la \Gamma_b\,, \cH \ra =0$\,(purple)\,,$\la \Gamma_0 + \Gamma_a\,, \cH \ra =0$\,(yellow) and $\la \Gamma_a + \Gamma_b\,, \cH \ra =0$ (brown).}
\label{sfg}
\end{figure}

\begin{figure}[t]
\centering
\begin{tabular}{ccc}
 \includegraphics[width=0.3\textwidth]{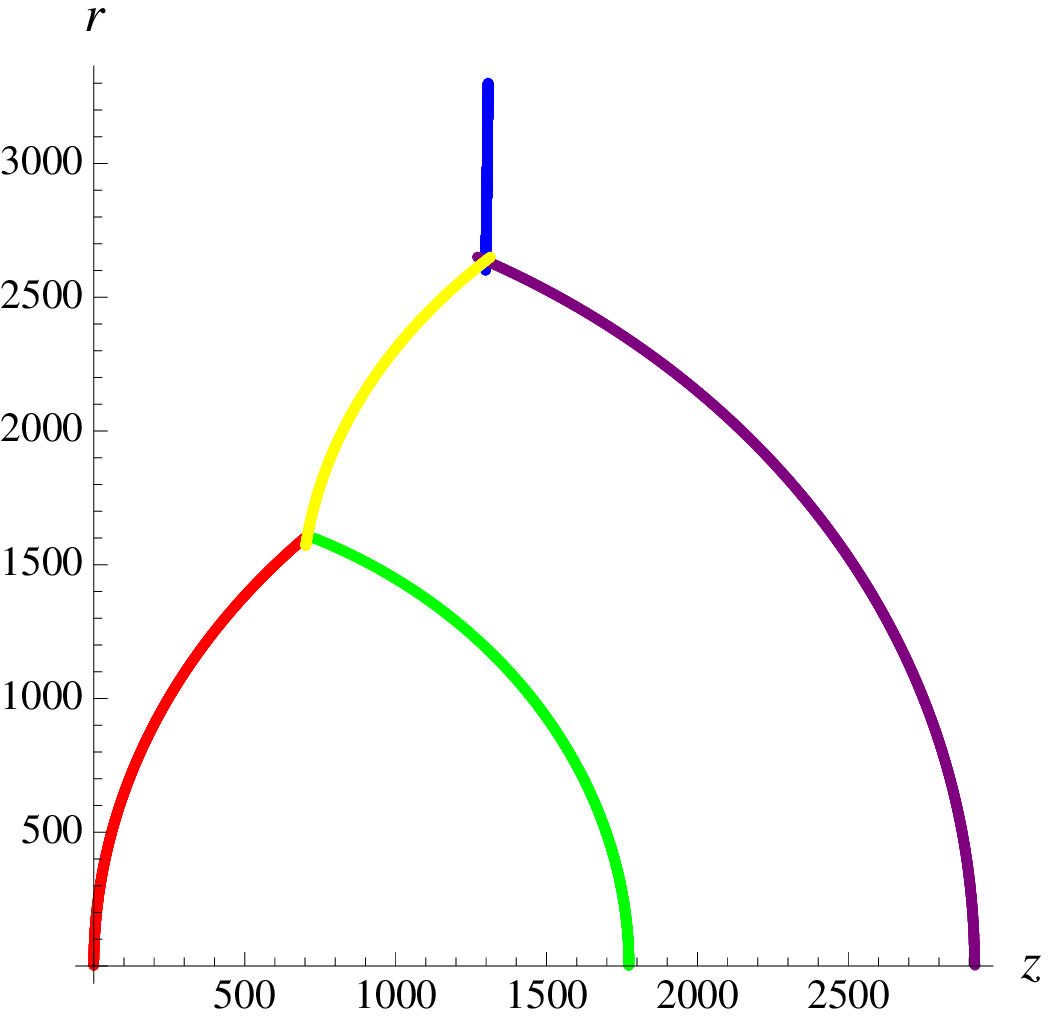} &
 \includegraphics[width=0.3\textwidth]{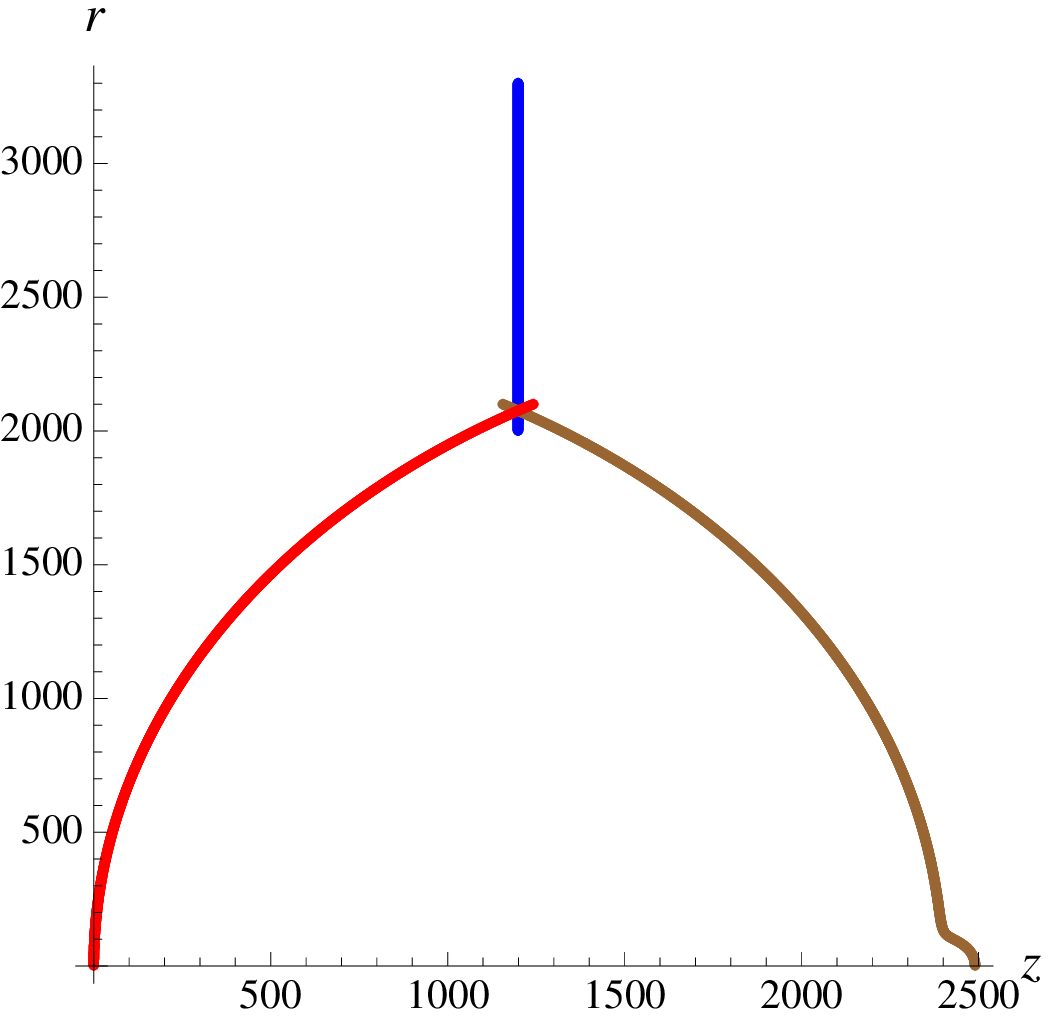} &
 \includegraphics[width=0.3\textwidth]{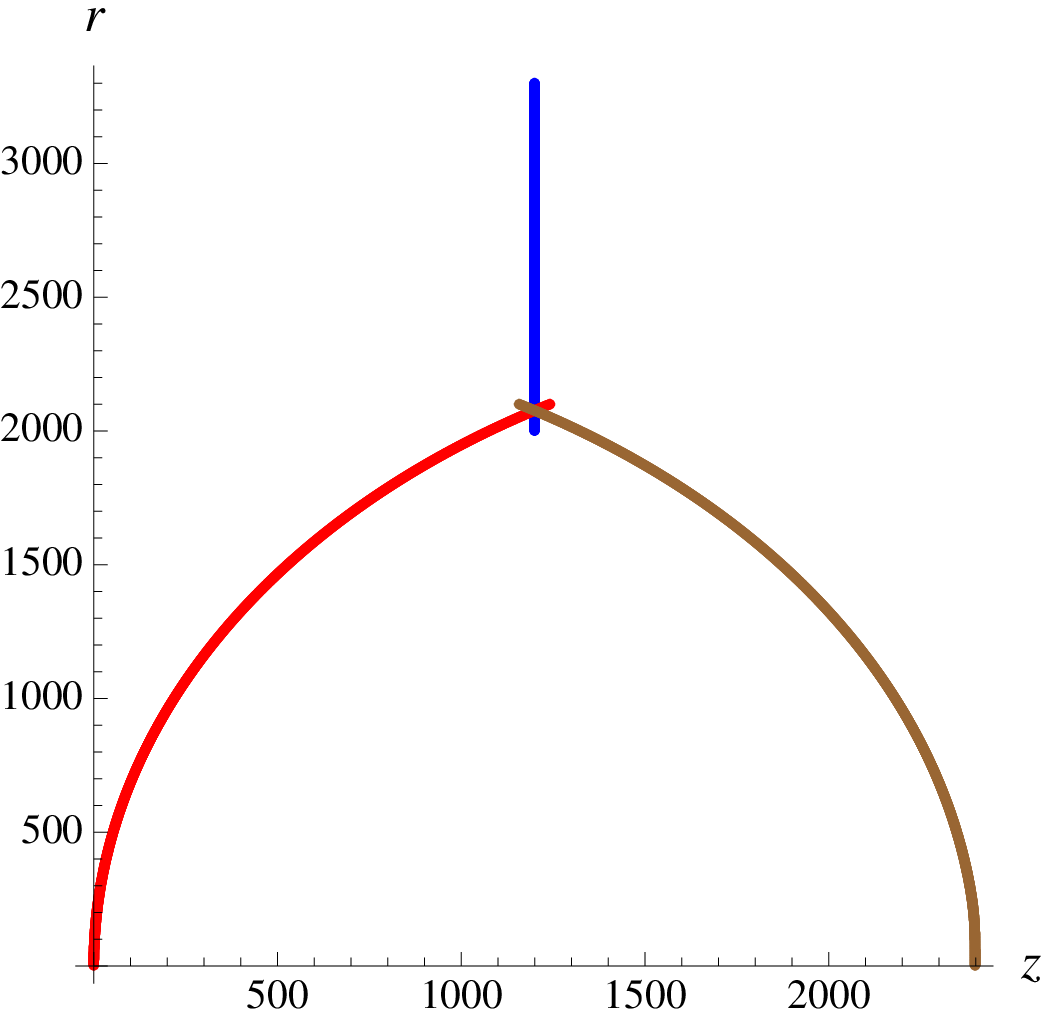}
 \end{tabular}
 \caption{\it\small From left to right, these graphs correspond to solutions (\rmnum{1}-\rmnum{3}). They are what remains after removing the irrelevant lines and disconnected lines. They clearly show the flow patterns of the three solutions at least up to the first split. The solution (\rmnum{1}) undergoes the first split $\Gamma_t \rightarrow  \Gamma_{0+a} +\Gamma_b$ and then the second split $\Gamma_{0 + a} \rightarrow \Gamma_0+\Gamma_a$. The solution (\rmnum{2}) and (\rmnum{3}) are similar and only their first splits, $\Gamma_t \rightarrow \Gamma_0 + \Gamma_{a+b}$, are visible in the figures.}
\label{sfg_ps}
\end{figure}

To link these two trees to the solutions, at first, we can check the split flow graphs of these solutions. We use the normal cylindrical coordinate, $(r, \theta, z)$ and plot several lines defined by $\langle \Gamma_i\,, \cH \rangle =0$ on the $(r,z)$ plane at some fixed $\theta$. The results are shown in Figure \ref{sfg}.
One may notice that they are very similar even for the very different solutions. This may be due to the similar underlying structures coming from the same asymptotic vector and charge assignments. As the  figures contain several irrelevant lines, it is difficult to see the pattern of the flows. If we remove the irrelevant and the disconnected lines, we have  Figure \ref{sfg_ps}. One can see the patterns of the flows become clearer in these graphs. For solution (\rmnum{1}), the flow starts from $\Gamma_t$ and splits to $\Gamma_{0+a}$ and $\Gamma_{b}$ and then $\Gamma_{0+a}$ further splits to $\Gamma_0$ and $\Gamma_a$. Finally, each ring undergoes the splits, $\Gamma_{a,b} \rightarrow \Gamma_{+(a,b)} + \Gamma_{-(a,b)}$, which are not visible in the figure. Therefore, we can conclude that solution (\rmnum{1}) belongs to the split flow component with the tree, $\cT^{(1)}$.

\begin{figure}[t]
\centering
 \begin{tabular}{cc}
\includegraphics[width=0.35\linewidth]{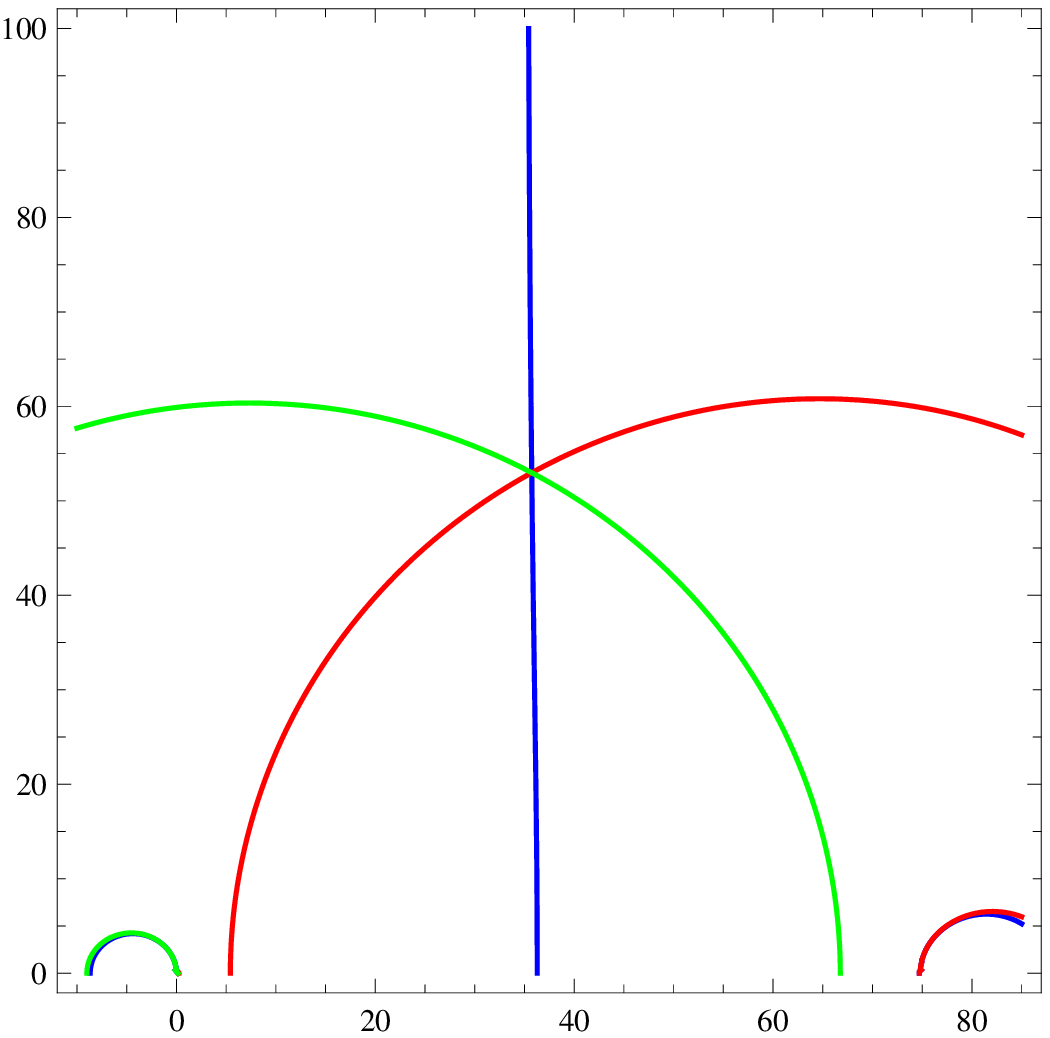} &
\includegraphics[width=0.35\linewidth]{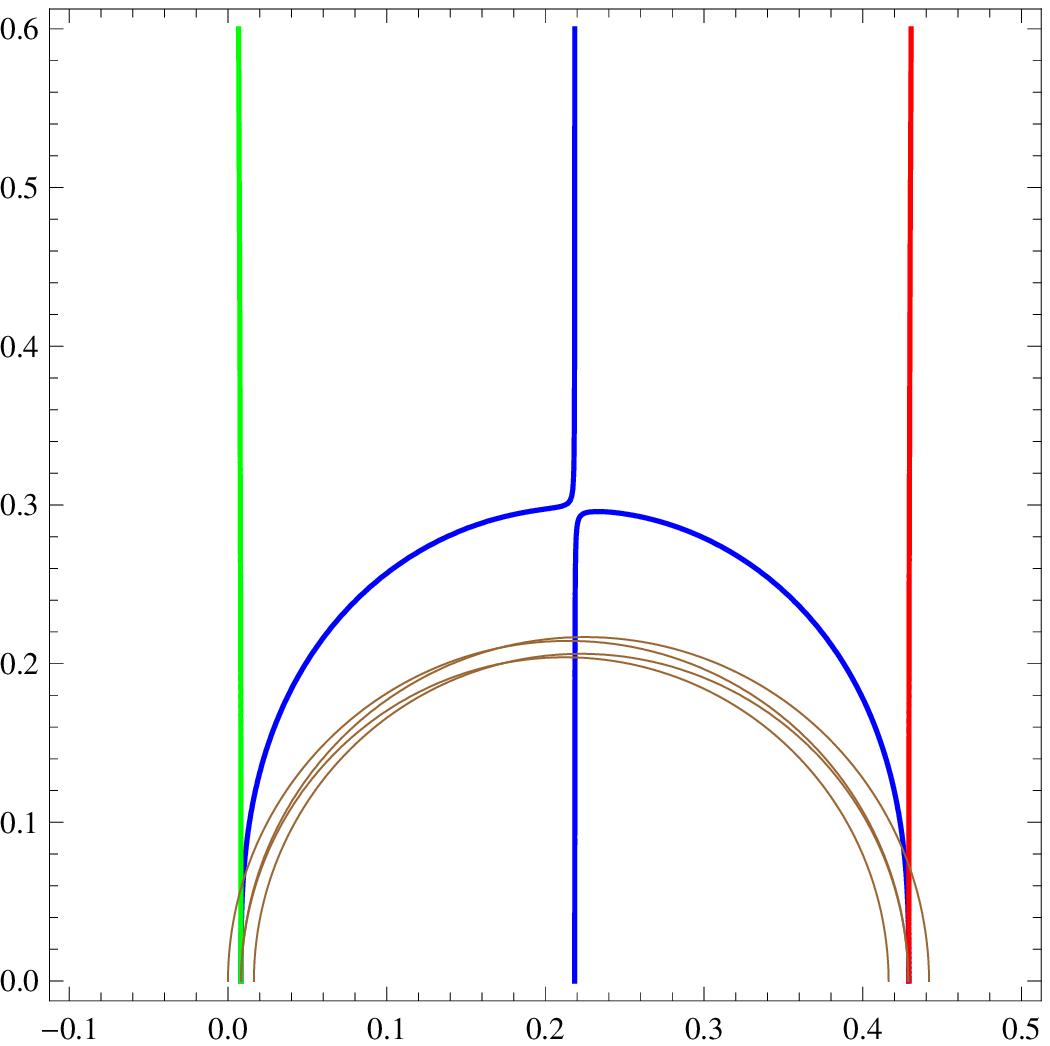} 
 \end{tabular}
 \caption{\it\small From left to right, these graphs correspond to solutions (\rmnum{2}-\rmnum{3}). There are several lines including: $\la \Gamma_a+\Gamma_b\,,\,\cH \ra = 0$ (blue),  $\la \Gamma_a\,,\,\cH \ra = 0$ (red),  $\la \Gamma_b\,,\,\cH \ra = 0$ (green) and all other individual flows,  $\la \Gamma_{\pm a,b}\,,\,\cH \ra = 0$ (brown).}
\label{sfg_ss}
\end{figure}

For solutions (\rmnum{2}) and (\rmnum{3}), only their first splits, $\Gamma_t \rightarrow \Gamma_0 +\Gamma_{a+b}$, are visible in the graphs. In fact, even we zoom in the near ring region, the next split point cannot be recognized due to the bending and disconnection of some lines. To see if the  $\Gamma_{a+b}$ flows end with scaling solutions or they actually have further split, we can decouple the isolated GH center by taking $h_{\infty}$ to the first split point. This means we fix $h_{\infty}$ at\footnote{This is the value that we get from the skeleton. One can use the precise value at the split point in the split flow graph but the quantitative difference is not significant}:
\be
h_s \approx \Big( 0.000417\,,\,0.0271\,,\,0.0229\,,\,0.0177\,,\,-3.688\,,\,-2.727\,,\,-4.152\,,\,0\Big)\,,
\ee
%
%
and then we solve bubble equations for four centers with this new $h$. We found there are two solutions this time:
\begin{center}
\begin{tabular}{ c | c | c | c | c }
 & $\Gamma_{+a}$ & $\Gamma_{-a}$ & $\Gamma_{+b}$ & $\Gamma_{-b}$  \\ \hline
 (\rmnum{2}) & 0 & 0.00052 & 74.8067 & 74.8079\\ \hline
 (\rmnum{3}) & 0 & 0.0163 & 0.416 & 0.442 \\ 
 \end{tabular}
\end{center}
Why they are corresponding to solution (\rmnum{2}) and (\rmnum{3}) will become clear once we study the dynamical picture. For now, we can speculate the connection by the ratios of several distances. Then the split flow graphs that we get for the solutions are shown in Figure \ref{sfg_ss}.
We can see the first split is clear in solution (\rmnum{2}) but again we have several lines torn apart and it is probably because the distance of two rings is too close. However, we can proceed similarly and then we see there are indeed further splits: $\Gamma_{a,b} \rightarrow \Gamma_{+(a,b)} + \Gamma_{-(a,b)}$. Therefore, we conclude that solution (\rmnum{2}) again belongs to a split flow component but with the other tree, $\cT^{(2)}$. On the other hand, the graph of solution (\rmnum{3}) has no clear first split. This is consistent that it should not belong to a split flow component since the two possible trees have been identified. In fact, since there is a large CTC's region surrounding the two rings,  solution (\rmnum{3}) belongs to a bad component.

\begin{figure}[t]
\centering
\includegraphics[width=0.7\linewidth]{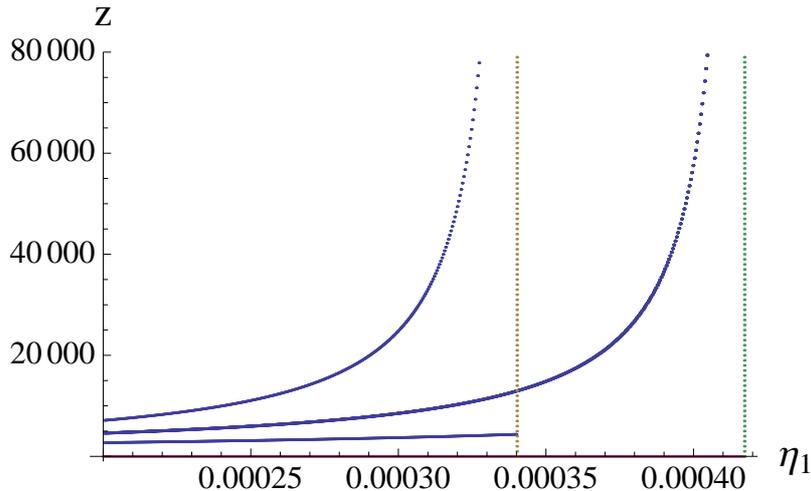} 
 \caption{\it\small This graph traces the z-coordinate of all centers for all solutions under the change of $\eta_1$. As there is not much change when $\eta_1=0\sim 0.0002$, we show the graph starting from $\eta_1=0.0002$. The red line along $z=0$ represents $\Gamma_0$ for all solutions. The positions of the other centers are shown with blue dots. The remaining centers of solutions (\rmnum{2}) and (\rmnum{3}) are very close to each others and are represented by the middle blue line. The top blue line is $\Gamma_b$ and the bottom blue line is $\Gamma_a$ for solution (\rmnum{1}). The brown dashed line marks the exact value of $\eta_{s_1}^{(1)}$ and the green dashed line marks $\eta_{s_1}^{(2)}$.}
\label{1st_spt_dyn}
\end{figure}

To verify the above identifications, we employ the dynamical picture. As we have mentioned, it does not matter which point in the component we trace. For simplicity, we focus on the solution which preserves the axial symmetry. In other words, we keep the same setup but we change $h_{\infty}$ following the flow described by the skeleton:
\be
h_{\infty}^{(f)} ~=~ \eta_1\,\Gamma_t   ~+~ h_0\,.
\ee
We start from $\eta_1=0$, increase it by a small value at a time and solve the bubble equations for each new $\eta_1$. The increment should be small enough to trace all of the solutions. The first splits of three solutions are shown in Figure \ref{1st_spt_dyn}. There are two split points corresponding to the two different split flow trees at:
\be
\eta^{(1)}_{s_1} ~=~ -\frac{\la\, \Gamma_0 + \Gamma_a\,,\, h_0\, \ra}{\la\, \Gamma_0 + \Gamma_a\,,\, \Gamma_b \,\ra} ~\approx~ 0.00034\,, ~~\quad~~  \eta^{(2)}_{s_1}  ~=~ -\frac{\la\, \Gamma_0 \,, h_0 \,\ra}{\la\, \Gamma_0\,,\,  \Gamma_a +\Gamma_b \,\ra} ~\approx~ 0.00042\,.
\ee
The subscript $``s_1"$ denotes the first split and the superscript label the tree it belongs to.
 We can see these two split points clearly in the figure. When the  value of $\eta$ is close to $\eta_{s_1}^{(1)}$, the two rings in solution (\rmnum{1}) indicated by the top and the bottom blue lines are quickly separated far apart and after $\eta_1$ crosses $\eta_{s_1}^{(1)}$, solution (\rmnum{1}) disappear and we left with only two solutions. Notice that all of the rings in solutions (\rmnum{2}) and (\rmnum{3}) are very close to each other and accumulate at the middle blue line. When $\eta_1$ increases further and is close to $\eta_{s_1}^{(2)}$, the separation between rings and $\Gamma_0$ for the remaining two solutions begins to increase very rapidly. After $\eta_1$ passes $\eta_{s_1}^{(2)}$, no solution exists. This analysis confirms the existence of the first split for all of the solutions.
 
\begin{figure}[t]
\centering
 \begin{tabular}{cc}
  \includegraphics[width=0.35\linewidth]{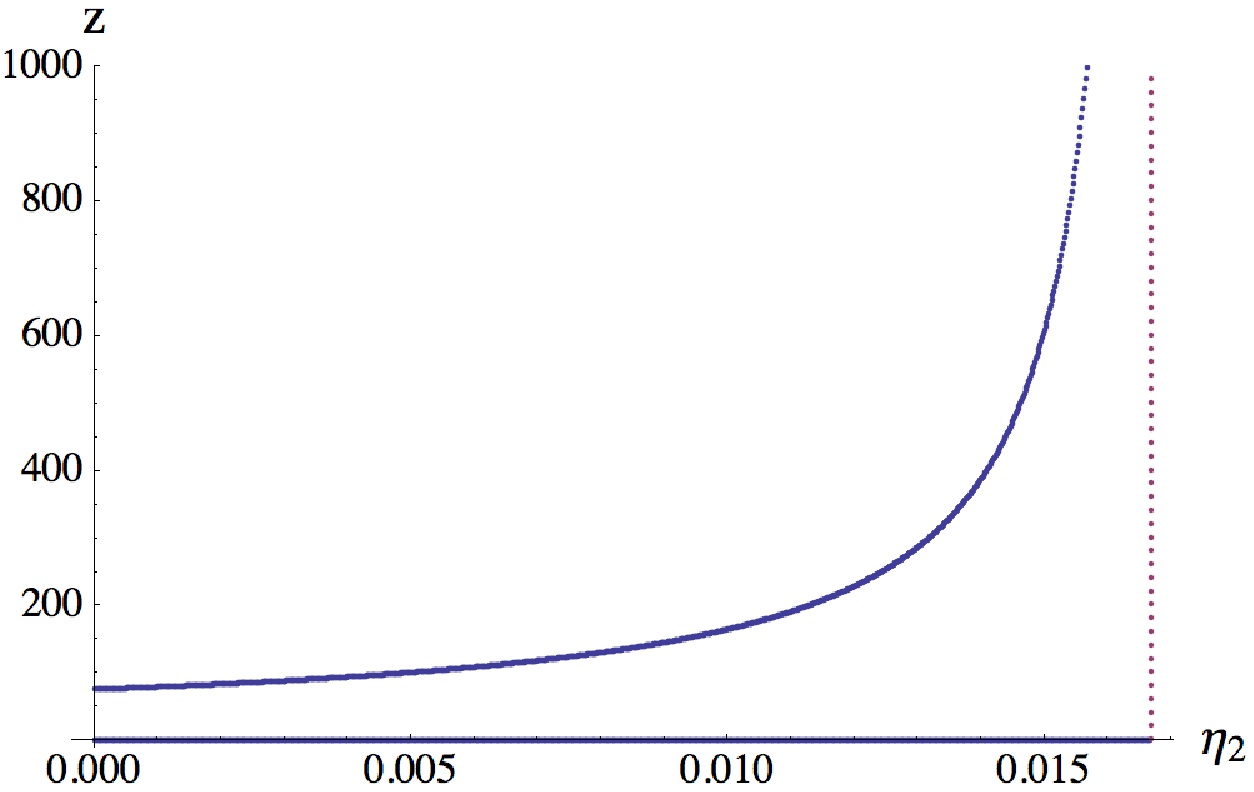} &
  \includegraphics[width=0.35\linewidth]{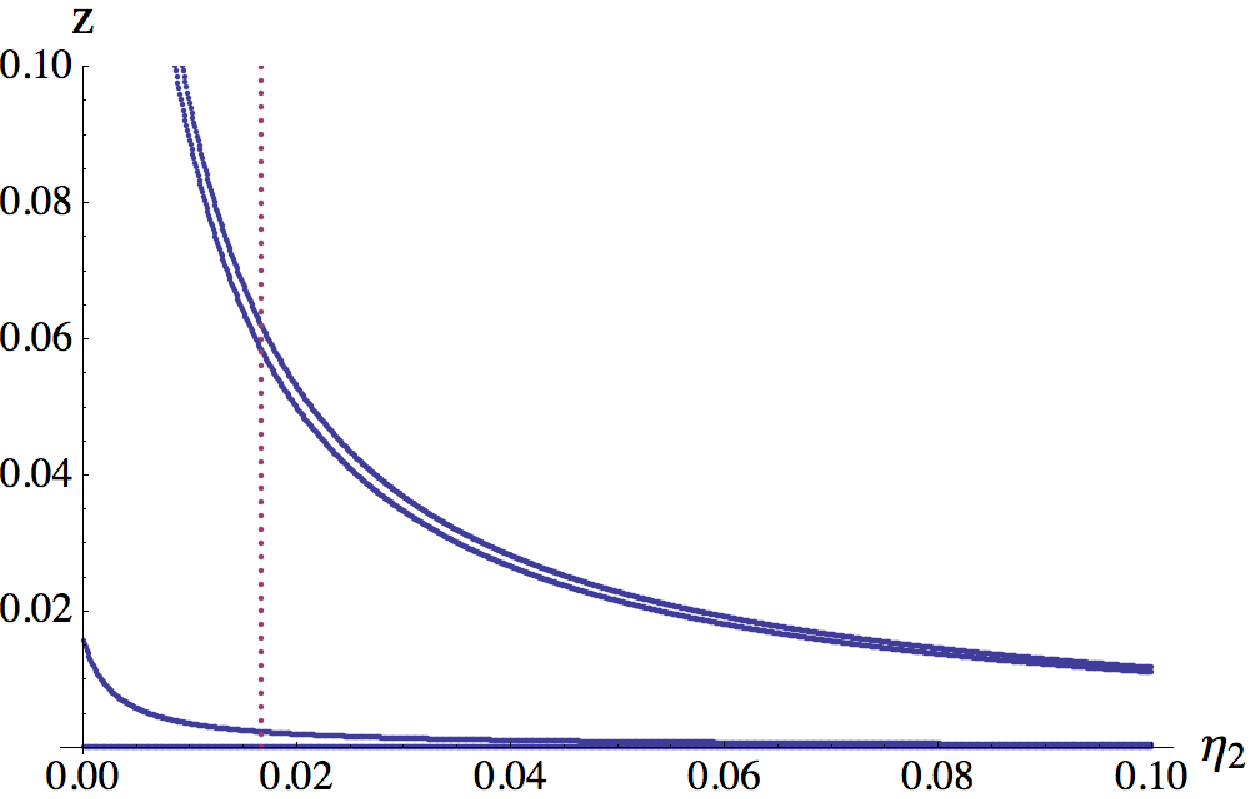}
 \end{tabular}
 \caption{\it\small From left to right, these graphs correspond to solutions (\rmnum{2}-\rmnum{3}) and they show how the solutions change under the flow of $\eta_2$. The top (double) blue indicates $\Gamma_b$ while the bottom (double) blue line represents $\Gamma_a$. The red dashed line marks the exact value of $\eta_{s_2}^{(2)}$.}
\label{2nd_spt_dyn}
\end{figure}

For the second split, we focus on solutions (\rmnum{2}) and (\rmnum{3}) because the split-flow pattern of  solution (\rmnum{1}) is already quite clear in Figure \ref{sfg_ps}. We consider that $h_{\infty}$ starts from $\eta_1=\eta_{s_1}^{(2)}$ and follows the flow of $\Gamma_{a+b}$:
 \be
 h_{\infty}^{(f)} ~=~ \eta_2\, (\Gamma_a+\Gamma_b) ~+~ h_{s_1} ~=~\eta_2\, (\Gamma_a+\Gamma_b) ~+~ \eta_{s_1}^{(2)}\,\Gamma_t  ~+~ h_{\infty}\,,
 \ee
 where we parametrize this flow with $\eta_2$ and start the flow from $\eta_2=0$. Then we solve the bubble equations for four centers and the result is shown in Figure \ref{2nd_spt_dyn}.
The next split point of $\cT^{(2)}$ can be easily computed:
\be
\eta_{s_2}^{(2)} ~=~ -\frac{\la \Gamma_a\,,\, h_{s_1}\ra}{\la \Gamma_a\,,\,\Gamma_b \ra} ~\approx~ 0.0167\,.
\ee
We can clearly see that for solution (\rmnum{2}), $\Gamma_a$ and $\Gamma_b$ are separated further away when $\eta$ approach $\eta_{s_2}^{(2)}$ and after $\eta_2$ crosses $\eta_{s_2}^{(2)}$, the solution disappears. The following split points of solution (\rmnum{2}) can also be clearly identified and we again conclude that it belongs to $\cT^{(2)}$. On the other hand, the behavior of solution (\rmnum{3}) is completely different. The overall size of solution (\rmnum{3}) decreases as $\eta_2$ increases and even after $\eta_2$ crosses $\eta_{s_2}^{(2)}$, it continues to exist with decreasing size. This is the typical behavior of the scaling solution and this part of solution (\rmnum{3}) should be regarded as a single flow from the point view of the split flow picture . However, as we have already mentioned, solution (\rmnum{3}) has CTC's and should belong to a bad component.  This concludes that the dynamic picture is consistent with the static one in this case.

\subsubsection{$x=90$}

\begin{figure}[t]
\centering
 \begin{tabular}{cc}
   \includegraphics[width=0.35\linewidth]{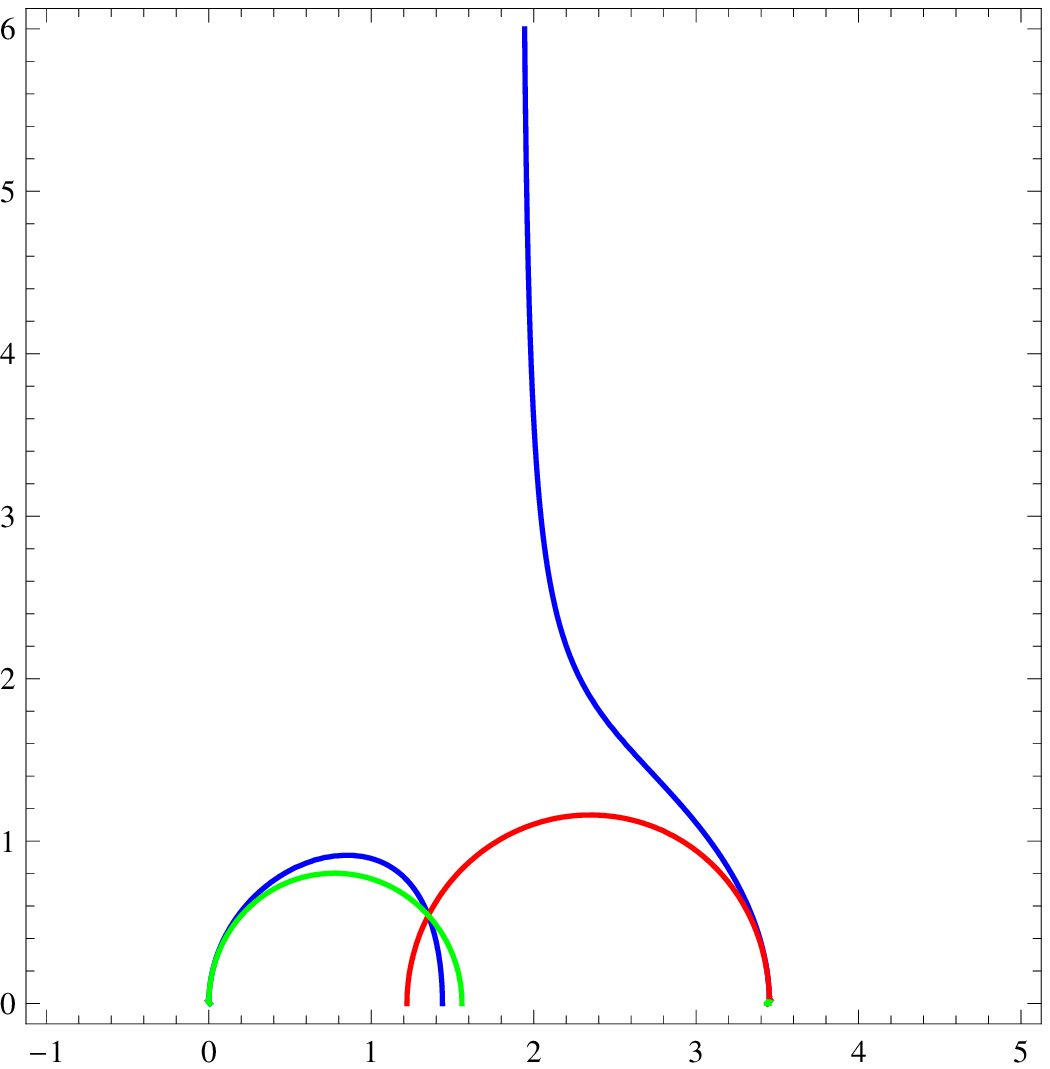} &
   \includegraphics[width=0.35\linewidth]{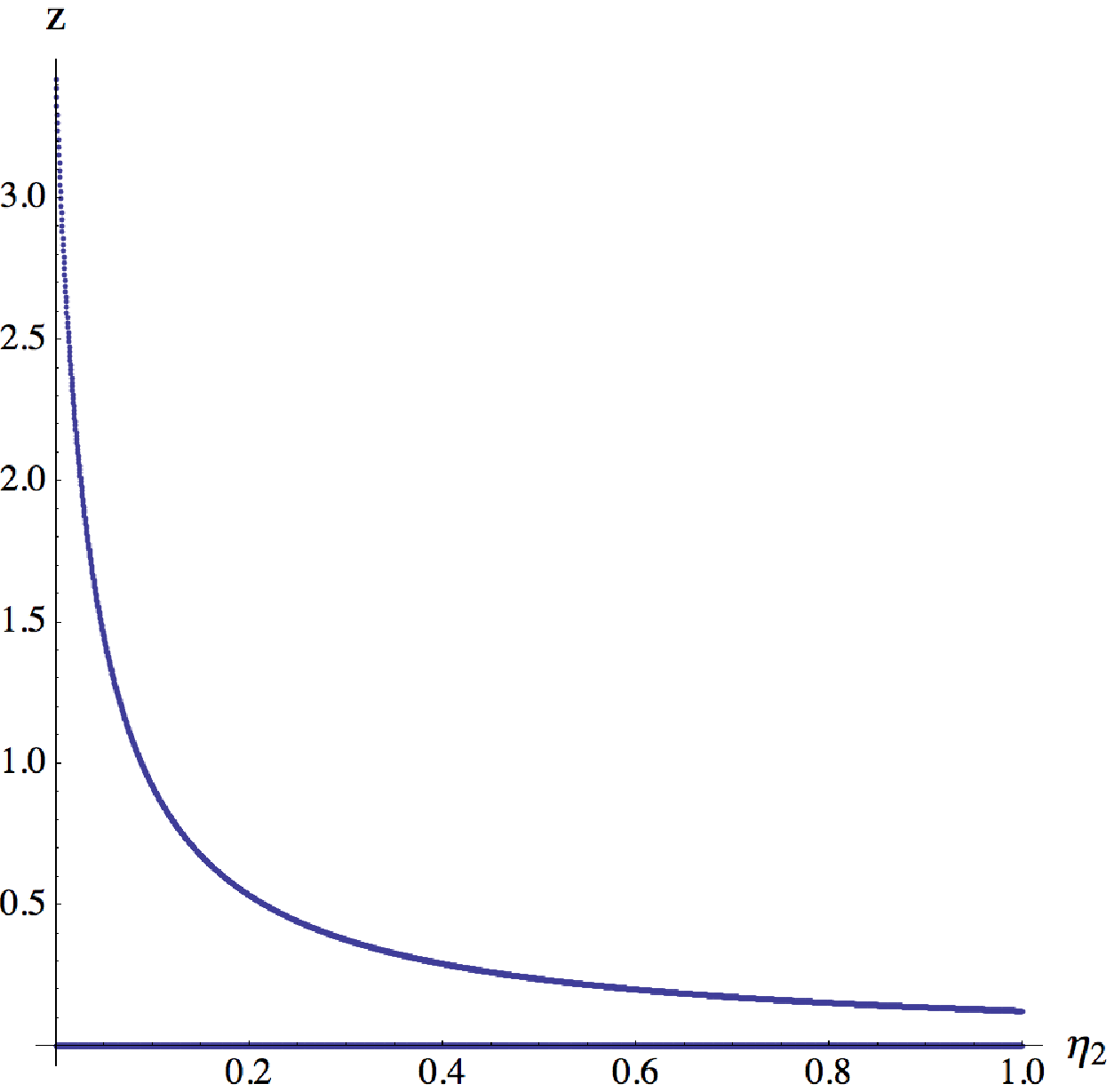}
 \end{tabular}
 \caption{\it\small On the left, we have the split flow graph for solution (\rmnum{2}) after the first split. The blue line is $\la \Gamma_a+\Gamma_b,\, \cH \ra =0$ and the green and red lines are $\la \Gamma_{a,b},\, \cH \ra =0$.  On the right, the graph shows how solution (\rmnum{2}) changes under the flow with the top blue line indicating the positions of  $\Gamma_b$ and the bottom blue line marking the positions of $\Gamma_a$.}
\label{2nd_spt_dynx2}
\end{figure}

For $x=90$, we still have three solutions:
\begin{center}
\begin{tabular}{ c | c | c | c | c | c}
 & $\Gamma_0$ & $\Gamma_{+a}$ & $\Gamma_{-a}$ & $\Gamma_{+b}$ & $\Gamma_{-b}$  \\ \hline
 (\rmnum{1}) & 0 & 1774.10378 & 1774.10683 & 3005.96801 & 3005.97352 \\ \hline
 (\rmnum{2}) & 0 & 2419.88471 & 2419.88474 & 2423.38597 & 2423.38604 \\  \hline
 (\rmnum{3}) & 0 & 2422.25748 & 2422.25842 & 2422.27682 & 2422.27830 \\
 \end{tabular}
 \end{center}
However, if we again look at the physical trees, we find that only one remains: $\cT^{(1)}$. The reason that $\cT^{(2)}$ vanishes is because $\la \Gamma_a\,,\Gamma_b \ra$ change sign and consequently $\eta_{s_2}^{(2)}$ become negative. To understand why we still have three solutions, we again employ the static and dynamic analysis. We find that the behavior of solution (\rmnum{1}) does not change much and it still match with $\cT^{(1)}$. On the other hand, solution (\rmnum{3}) still has CTC's and should belong to a bad component. Therefore, we focus on solution (\rmnum{2}). Furthermore, since the first split point still exists just like the previous case, we focus on the analysis after the isolated GH center is decoupled. Proceeding similarly, we have the result shown in Figure \ref{2nd_spt_dynx2}. We see a curious fact that even though the solution itself does not change much, its behavior under the flow  dramatically changes. From Figure \ref{2nd_spt_dynx2}, it is clear that now solution (\rmnum{2}) shows the scaling behavior. Since solution (\rmnum{2}) is free of CTC's unlike solution (\rmnum{3}), it truly belongs to a scaling component.
To put it more precisely, the $\cT^{(2)}$ vanishes at exactly $x=\frac{4185}{47}$ when $\la \Gamma_a\,,\Gamma_b \ra$ flips the sign and that is exactly\footnote{For the case with $x$ only slightly smaller than the critical value, $\frac{4185}{47}$, it will be difficult to determine the split point in the split flow graph. However, the solution continues to show the splitting behavior in the dynamic analysis.} the point when solution (\rmnum{2}) begins to show the scaling behavior. This concludes that when $\frac{4185}{47}\leq x \lesssim 90.3$, solution (\rmnum{2}) belongs to a scaling component. 

Another interesting configuration is $x\approx 71.7$ when solutions (\rmnum{1}) and (\rmnum{2}) become very similar. However, there are still two trees when $\frac{4185}{47}>x\gtrsim 14.156$\footnote{When $x \lesssim 14.156$, $\la \Gamma_a, h_{s_1} \ra$ changes sign and again the $\eta_{s_2}^{(2)}$ becomes negative and $\cT^{(2)}$ vanishes.}. If these two solutions somehow merge to one solution, the split flow picture will not be correct. However, interestingly, no matter how much  we fine tune the value of $x$, we always get either both of the solutions or neither of them. With the help of flux quantization, these two components may be separated without touching each other. Also, it may seems peculiar why both of them vanish when $x \lesssim 71.7$ while the corresponding trees still exist. The reason is simply that the order of the centers sometimes becomes important and when $x\lesssim 71.7$ the solutions corresponding to these two trees exist with the different order of the centers or more general orientations without axial symmetry.

\subsection{Triangular solution}

\begin{figure}[t]
\centering
\begin{tabular}{cc}
 \includegraphics[width=0.45\linewidth]{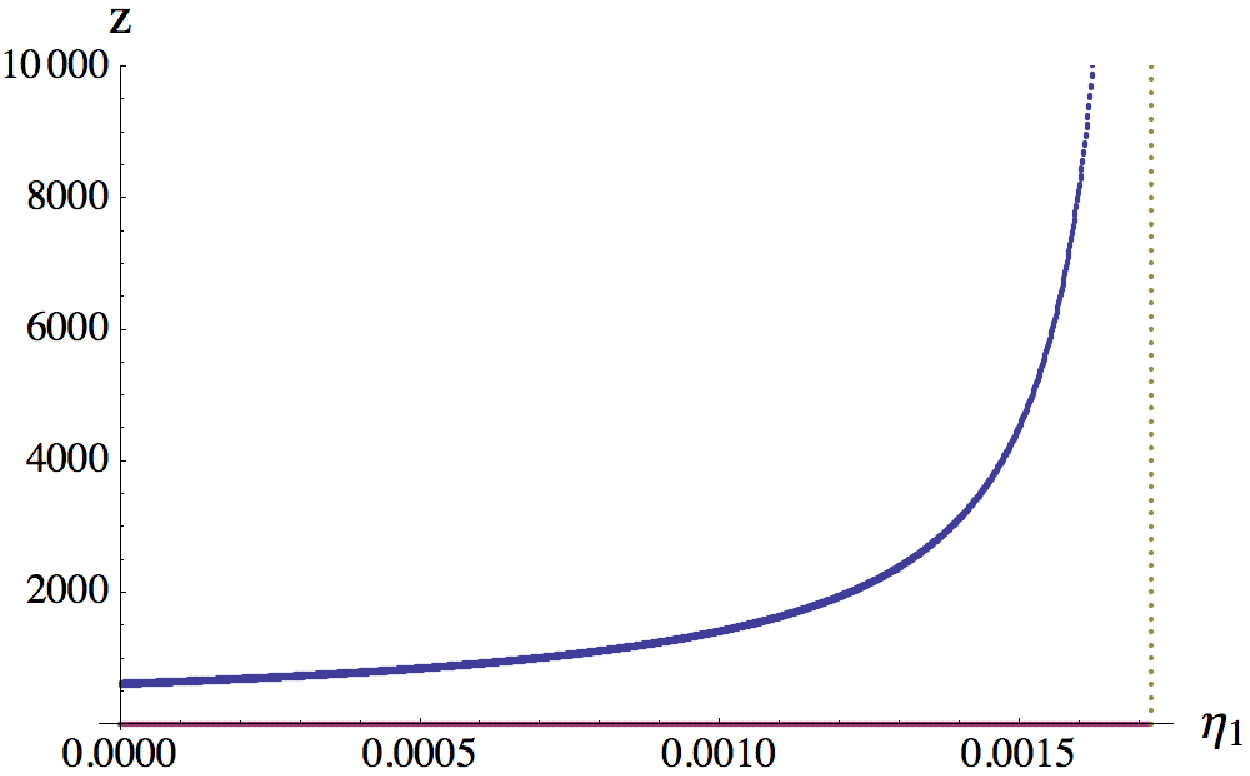} &
 \includegraphics[width=0.45\linewidth]{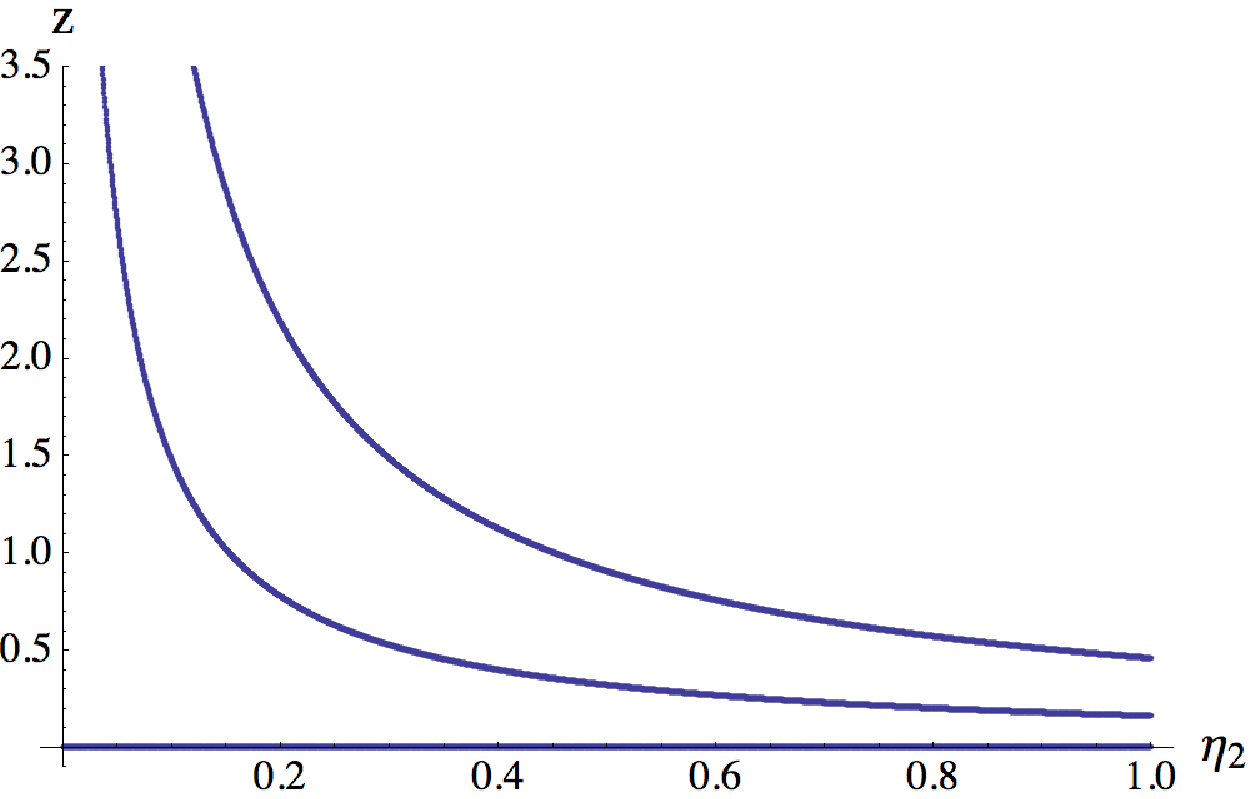}
 \end{tabular}
  \caption{\it\small On the left, the graph show how the triangular solution changes under the flow before the split. The red line indicates $\Gamma_0$ while the other centers accumulate at the blue line. The yellow dashed line marks the critical value, $\eta_{s}\approx 0.00172$, for the split.  On the right, the graph traces the solution with the remaining three centers after the split. }
  \label{tri_scal}
\end{figure}

In \cite{Bena:2007qc}, it was found that if the fluxes between the three centers form a closed triangle, the scaling solution exists. Specifically, one can adjust the angles between three centers to make the distances between them shrink and eventually merge into a point in the coordinate space. Physically, this produces a very long throat and makes the solution become very similar to a classical black ring with a finite horizon area when the three centers are very close to each others. Also, an example of this kind of scaling solution was given. We would like to see if this example shows the scaling behavior from the perspective of the split flows. The setup is the following:
\newline
\begin{tabular}{ll}
& \\
$\Gamma_0^{(m)} = (1,\,0,\,0,\,0)\,, $ & $\Gamma_1^{(m)} = ( -105,\,2210,\,7887,\,7800)\,, $\\
$\Gamma_2^{(m)} = (-105,\,1200,\,16000,\,1613)\,,$ & $\Gamma_3^{(m)} = (210,\,525,\,-20000,\,6400)\,.$ \\
&
\end{tabular}
\newline
For this setting, there is an axisymmetric solution. However, no tree exists for this set of charges. This implies that this solution should belong to either a scaling or a bad component. Since it has been verified that no pathology is present, this solution indeed should belong to a scaling component. We can implement the dynamic picture to confirm this. Since the scaling solution is formed only by $\Gamma_1^{(m)}$, $\Gamma_2^{(m)}$ and $\Gamma_3^{(m)}$, the split between them and $\Gamma_0^{(m)}$ still exists. This split is shown clearly in Figure \ref{tri_scal}. After the isolated GH center is decoupled, we continue the flow for the three centers in Figure \ref{tri_scal}. The scaling behavior appears at this stage and one finds that it cannot be torn apart by tuning the asymptotic vector along the flow. This confirms that this triangular solution indeed belongs to a scaling component.
%
%

\subsection{Three rings}
\label{3ring}

\begin{figure}[t]
\centering
 \begin{tabular}{ccc}
  \includegraphics[width=0.3\linewidth]{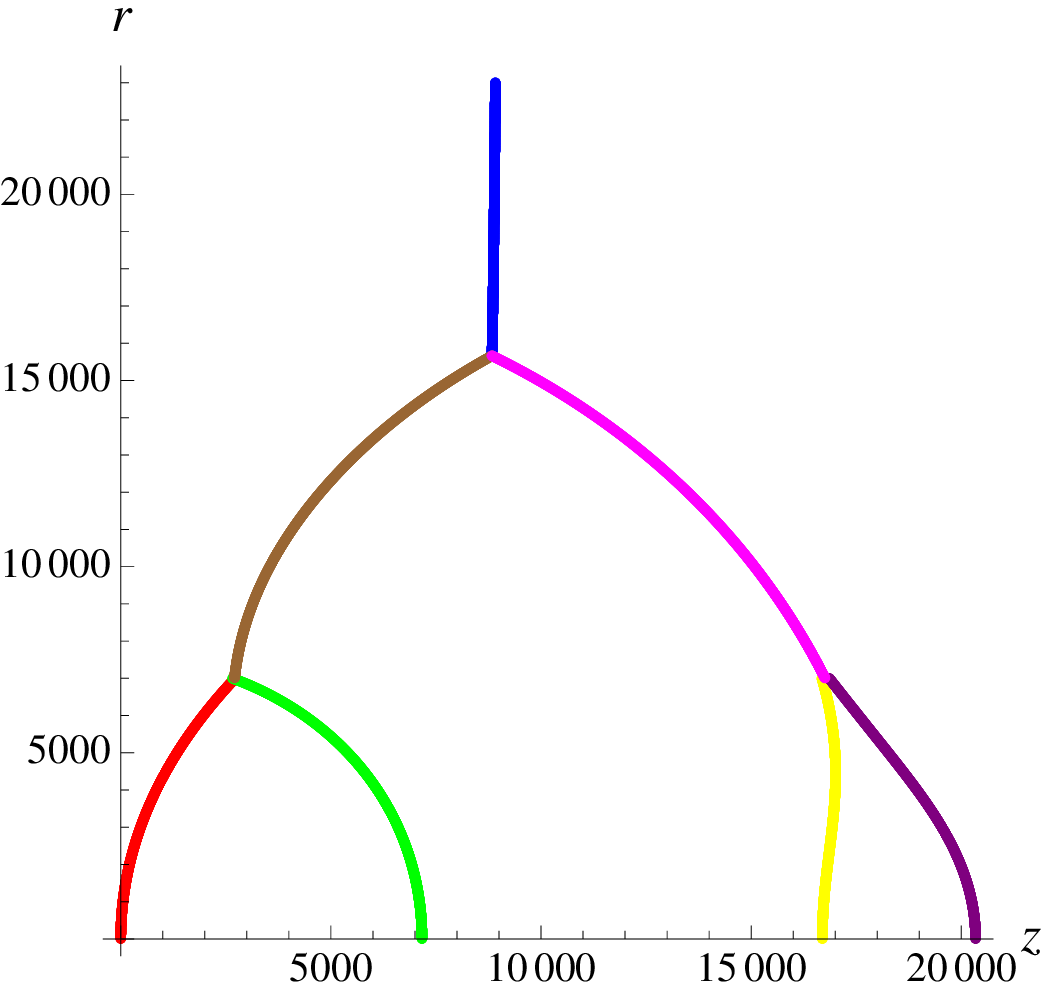} &
  \includegraphics[width=0.3\linewidth]{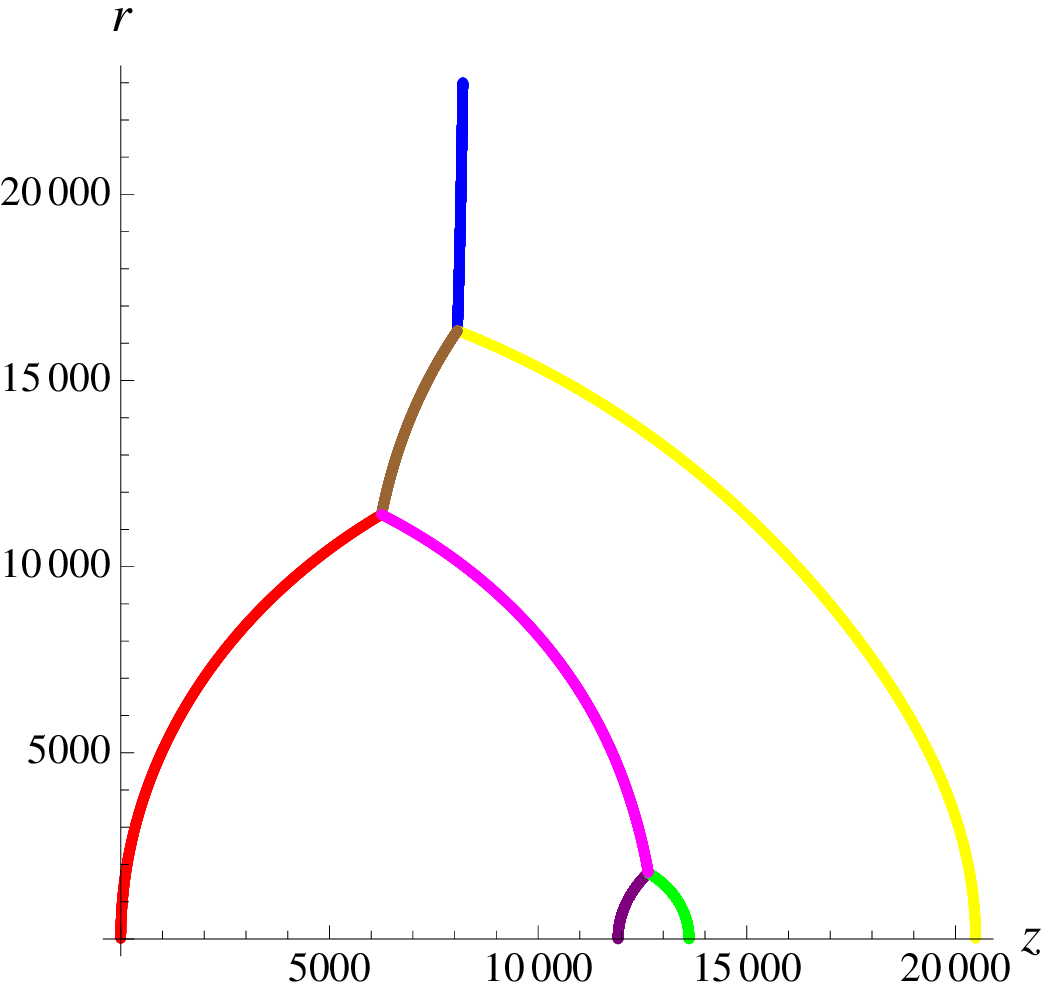} &
  \includegraphics[width=0.3\linewidth]{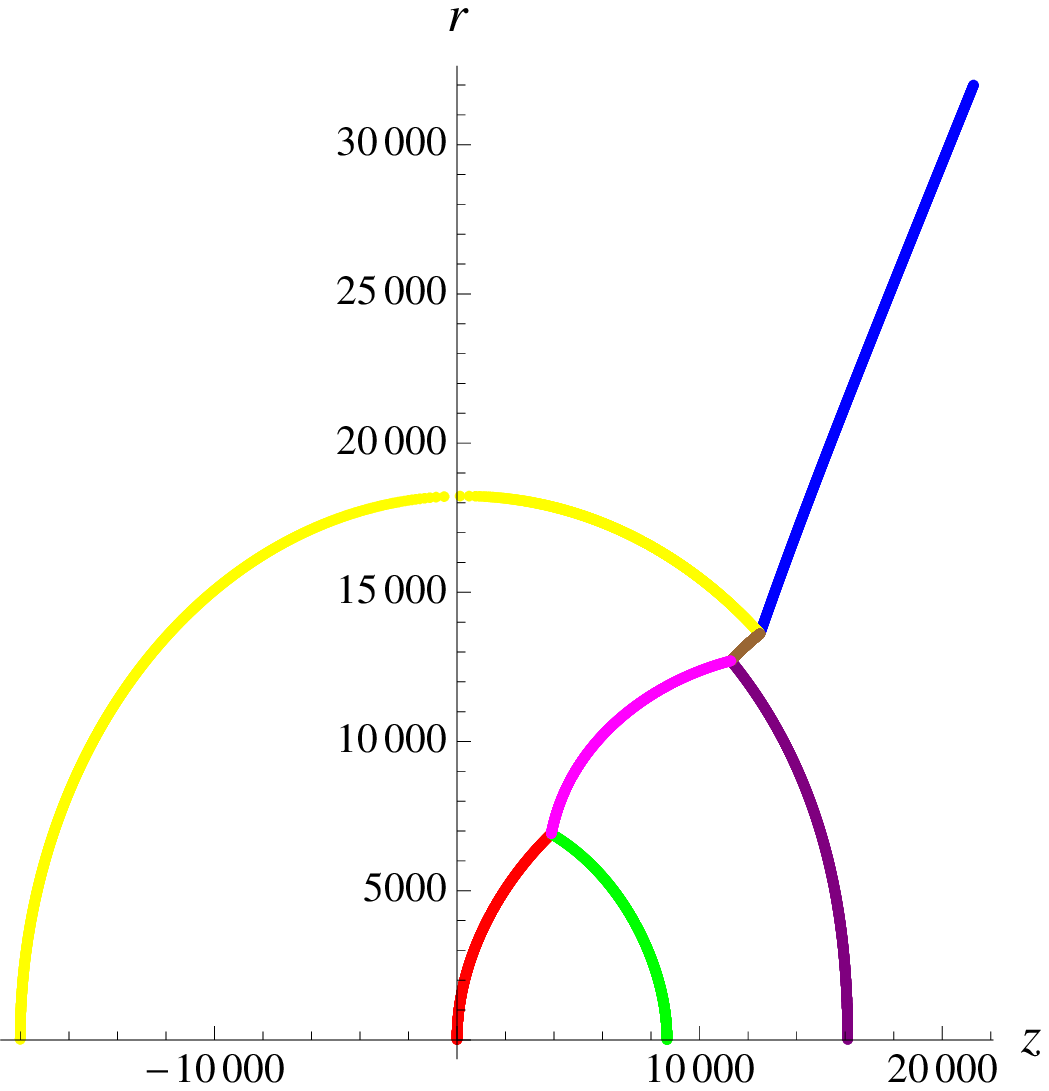}
 \end{tabular}
 \caption{\it\small From left to right are the split flow graphs for solutions (1) to (3). The red line is $\la \Gamma_0,\,\cH \ra =0$, the green line is $\la \Gamma_a,\,\cH \ra =0$, the yellow line is $\la \Gamma_b,\,\cH \ra =0$ and the purple line is $\la \Gamma_c,\,\cH \ra =0$. The other color lines are for some combinations of the charges. The splits between two centers in the rings are not visible in the graphs.}
 \label{sfg_tr}
 \end{figure}

To increase our confidence in this picture, we would like to study more complicated solutions. For example, we can have three rings and this means in total we have seven centers. We take the charge assignment of the three rings as the following:
\bea
Q_a=105\,,~&\quad d_a^I ~=~ (\, 50\,,\, 70\,,\, 65 \,) \,, \quad& X_a^I ~=~ (\, 920\,,\, 880\,,\, -270 \,) \,, \nonumber\\ 
Q_b=105\,,~&\quad d_b^I ~=~ (\, 106\,,\, 55\,,\, 35 \,) \,, \quad& X_b^I ~=~ (\, 310 \,,\, 590\,,\, 600 \,) \,, \nonumber\\
Q_c=105\,,~&\quad d_c^I ~=~  (\, 35\,,\, 80\,,\, 100 \,) \,, \quad& X_c^I ~=~ (\, 520 \,,\, 880\,,\, 220 \,) \,. 
\label{3ring_chg}
\eea
For this case, we find that only three trees pass the weaker existence conditions:
\bea
\cT^{(1)} ~&\equiv&~ \{\{\Gamma_0\,,\,\{\Gamma_{+a}\,,\,\Gamma_{-a}\}\}\,,\,\{\{\Gamma_{+b}\,,\,\Gamma_{-b}\}\,,\,\{\Gamma_{+c}\,,\,\Gamma_{-c}\}\}\}\,,\nonumber \\
\cT^{(2)} ~&\equiv&~\{\{\Gamma_0\,,\,\{\{\Gamma_{+a}\,,\, \Gamma_{-a}\}\,,\,\{\Gamma_{+c}\,, \,\Gamma_{-c}\}\}\}\,,\,\{\Gamma_{+b}\,,\,\Gamma_{-b}\}\} \,, \nonumber \\
\cT^{(3)} ~&\equiv&~  \{\{\{\Gamma_0\,,\,\{\Gamma_{+a}\,,\, \Gamma_{-a}\}\}\,,\,\{\Gamma_{+c}\,, \,\Gamma_{-c}\}\}\,,\,\{\Gamma_{+b}\,,\,\Gamma_{-b}\}\} \,. \nonumber 
\eea
For this many centers, one will have difficulty finding all of the solutions even numerically. However, by the method described in section \ref{assemble}, one can assemble all of the solutions corresponding to these trees. For simplicity, we still look for the axisymmetric solutions. The corresponding solutions we find are the following:
\begin{center}
\begin{tabular}{ c | c | c | c | c | c | c | c}
 & $\Gamma_0$ & $\Gamma_{-a}$ & $\Gamma_{+a}$ & $\Gamma_{-b}$ & $\Gamma_{+b}$ & $\Gamma_{-c}$ & $\Gamma_{+c}$   \\ \hline
 (1) & 0 & 7166.711 & 7166.712 & 16693.003 & 16693.006 & 20337.983 & 20337.987\\ \hline
 (2) & 0 & 13615.977 & 13615.979 & 20481.737 & 20481.739 & 11916.918 & 11916.922 \\ \hline
 (3) & 0 & 8654.738 & 8654.741 & -18007.984 & -18007.988 & 16105.593 & 16105.598 
 \end{tabular}
 \end{center}
 As usual, we express the solution with the $z$-coordinate of each center and fix $\Gamma_0$ at the origin. Notice that for solution (2) and (3), we need to rearrange the order of the centers in order to find the solutions. Their split flow graphs are shown in Figure \ref{sfg_tr}.

\begin{figure}[t]
\centering
\includegraphics[width=0.6\linewidth]{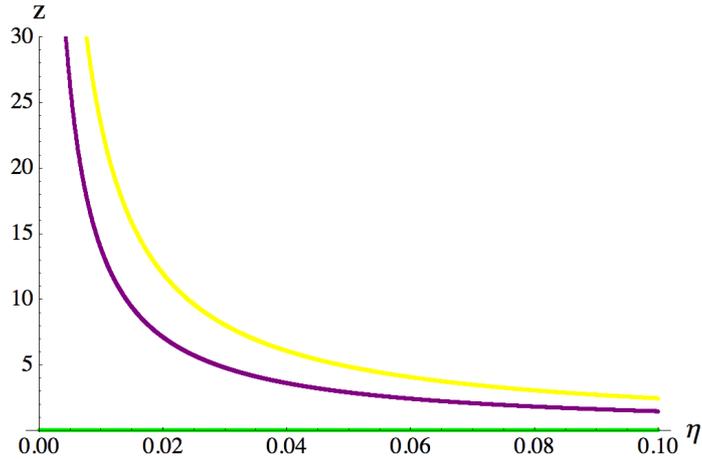}
\caption{\it\small This graph traces three rings under the flow after $\Gamma_0$ is decoupled. The green line is $\Gamma_a$, the yellow line is $\Gamma_b$ and the purple line is $\Gamma_c$.}
\label{dyn_tri}
\end{figure}

Additionally, we find another axisymmetric solution for this set of charges:
\begin{center}
\begin{tabular}{ c | c | c | c | c | c | c | c}
 & $\Gamma_0$ & $\Gamma_{-a}$ & $\Gamma_{+a}$ & $\Gamma_{-b}$ & $\Gamma_{+b}$ & $\Gamma_{-c}$ & $\Gamma_{+c}$   \\ \hline
 (s) & 0 & 14228.9671 & 14228.9676 & 14809.0745 & 14809.0752 & 14557.1942 & 14557.1951
 \end{tabular}
 \end{center}
This solution does not belong to any tree and in fact, should be classified as a scaling solution. To see this clearly, one can bring the asymptotic vector to the only split point and decouple $\Gamma_0$. One can then carry out the dynamic analysis for the remaining six centers. The result is shown in Figure \ref{dyn_tri}. Their scaling behavior is quite clear.

The structure of this scaling solution is actually similar to the triangle solution introduced in previous subsection. One can see that the fluxes between three rings indeed form a closed triangle. Therefore, the existence of this solution implies that the scaling solutions may have multi-layer structures just like the split flow solutions. It also show that the scaling and split flow components can coexist and a set of centers can have multiple split flow and scaling components at the same time.

\subsection{Bifurcation points: splittings of the components}
\label{bifurcation}

In this subsection, we will show the possibility that a connected component can be split into two components under the flow. We will use the example in section \ref{3ring}  with only some minor changes. We change only $d_c^I$ to the following and keep the other settings the same as (\ref{3ring_chg}):
\be
d_c^I ~=~ (35,\,20,\,100)\,.
\ee
For this setting, we have the similar scaling solution but different trees that pass the weak existence conditions:
\bea
\cT^{(1)} ~&\equiv&~ \{\Gamma_0\,,\,\{\{\Gamma_{+b}\,,\,\Gamma_{-b}\}\,,\,\{\{\Gamma_{+a}\,,\,\Gamma_{-a}\}\,,\,\{\Gamma_{+c}\,,\,\Gamma_{-c}\}\}\}\}\,,\nonumber \\
\cT^{(2)} ~&\equiv&~\{\{\Gamma_0\,,\,\{\{\Gamma_{+a}\,,\, \Gamma_{-a}\}\,,\,\{\Gamma_{+c}\,, \,\Gamma_{-c}\}\}\}\,,\,\{\Gamma_{+b}\,,\,\Gamma_{-b}\}\} \,, \nonumber \\
\cT^{(3)} ~&\equiv&~  \{\{\{\Gamma_0\,,\,\{\Gamma_{+a}\,,\, \Gamma_{-a}\}\}\,,\,\{\Gamma_{+b}\,, \,\Gamma_{-b}\}\}\,,\,\{\Gamma_{+c}\,,\,\Gamma_{-c}\}\} \,. \nonumber 
\eea
Among these three, $\cT^{(1)}$ is the most interesting because it has the same first split point as the scaling solution. This means that after we decouple $\Gamma_0$, we should have two components corresponding to $\cT^{(1)}$ and the scaling solution respectively. However, we find that there is only one component when we take $h_{\infty}$ to the split point:
\be
h_{s_1} ~=~ \Gamma_t\,\eta_{s_1} ~+~ h_0\,,
\ee
where $\Gamma_t$ is the total charge and $\eta_{s_1}$ is equal to:
\be
 \eta_{s_1} ~\equiv~ -\frac{\la \Gamma_0,\,h_0 \ra}{\la \Gamma_0,\,\Gamma_t \ra} ~\approx~ 0.000064805\,.
\ee
  If we change $h_{\infty}$ to continue the flow:
 \be
 h_{\infty} ~=~ (\Gamma_a+\Gamma_b+\Gamma_c)\,\eta_2 ~+~ h_{s_1}\,,
 \ee
 we find that at some point, the component breaks into two parts. We show the axisymmetric solutions in the following table:
 \begin{center}
\begin{tabular}{ c | c | c | c | c | c | c}
  & $\Gamma_{-a}$ & $\Gamma_{+a}$ & $\Gamma_{-b}$ & $\Gamma_{+b}$ & $\Gamma_{-c}$ & $\Gamma_{+c}$   \\ \hline
$ \eta_2=0 $ & 0 & 0.0021 & 1454.3822 & 1454.3830 & 8066.9990 & 8067.0021 \\ \hline
$ \eta_2 = \eta_2^{\ast}$,\, (\rmnum{1}) & 0 & 0.00158 & 12269.368 & 12269.370 & -1280.6451 & -1280.6444 \\ \hline
$ \eta_2 = \eta_2^{\ast}$,\, (\rmnum{2})& 0 & 0.00158 & 12271.433 & 12271.435 & -1280.6819 & -1280.6812 \\ \hline
$ \eta_2\sim 2.65\times 10^{-4}$,\, (\rmnum{1}) & 0 & 0.00147 & 7186.637 & 7186.639 & -1134.0126 & -1134.0119 \\ \hline
$ \eta_2\sim 2.65\times 10^{-4}$,\, (\rmnum{2}) & 0 & 0.00165 & 29530.173 & 29530.176 & -1409.4583 & -1409.4576 \\ \hline
 \end{tabular}
 \end{center}
From the table, one can see that at the beginning of the flow ($\eta_2=0$), only one component exists. However, starting from $\eta_2 \sim \eta_2^{\ast}= 2.5459751757 \times 10^{-4}$, we have two components, solution (\rmnum{1}) and (\rmnum{2}).

\begin{figure}[t]
\centering
  \includegraphics[width=0.6\linewidth]{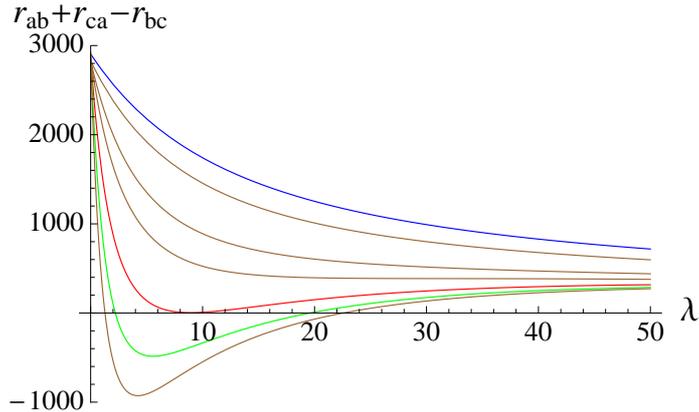}
 \caption{\it\small This figure shows how $r_{ab}+r_{ca}-r_{bc}$ changes with $\lambda$ for several different values of $\eta_2$. The blue curve corresponds to $\eta_2=0$, the red curve corresponds to $\eta_2 = \eta_2^{\ast}$ and the green curve corresponds to $\eta_2 = 2.65\times 10^{-4}$. When one increases the value of $\eta_2$, the curve becomes gradually lower and eventually, some part of it becomes negative which violates the triangular inequality.}
 \label{bp_trieq}
 \end{figure}

To understand what is happening, we need to study the whole solution space. In this situation, this space can be effectively captured by a three-center system. In other words, we will analyze the solution space of  the coarse grained system by replacing each ring with only one center. For $\eta_2=0$, the independent part of the bubble equations can be written as:
\bea
\frac{\la \Gamma_a,\, \Gamma_c \ra}{r_{ac}} ~&-&~ \frac{\la \Gamma_b,\, \Gamma_a \ra}{r_{ab}} ~=~ - \la \Gamma_a,\, h_{s_1} \ra\,, \\
\frac{\la \Gamma_b,\, \Gamma_a \ra}{r_{ab}} ~&-&~ \frac{\la \Gamma_c,\, \Gamma_b \ra}{r_{bc}} ~=~ - \la \Gamma_b,\, h_{s_1} \ra\,.
\eea
As $\la \Gamma_a,\, \Gamma_c \ra$, $\la \Gamma_b,\, \Gamma_a \ra$ and $\la \Gamma_c,\, \Gamma_b \ra$ are all positive and satisfy triangular inequality, the scaling solution of these three centers exists. Furthermore, there are at most two components in the solution space (for more detail, see appendix A in \cite{deBoer:2008zn}).  To know how the components change along the flow, we can parametrize the component with one parameter, $\lambda$, as the following:
\be
r_{ab} = \frac{1}{\frac{\lambda}{\la \Gamma_b,\,\Gamma_a \ra} + \frac{1}{x}}\,, ~\quad~ r_{bc} = \frac{1}{\frac{\lambda}{\la \Gamma_c,\,\Gamma_b \ra} + \frac{1}{y}}\,, ~\quad~ r_{ca} = \frac{1}{\frac{\lambda}{\la \Gamma_a,\,\Gamma_c \ra} + \frac{1}{z}}\,,
\label{r_solsp}
\ee
where $x$, $y$ and $z$ are the three distances of a particular solution in the component. For example, one can use axisymmteric one as this particular solution. Then, for any value of $\lambda$ such that all distances are positive and the triangular inequalities are satisfied will correspond to a solution. Therefore, the solution space in this case is the collection of some (at most two) intervals on a one dimensional line parametrized by $\lambda$. To check how this space changes along the flow, we take the axisymmetric solution with $y+z=x$ and monitor $r_{ab}+r_{ca}-r_{bc}$ for several points along the flow. The result is shown in Figure \ref{bp_trieq}. For $\lambda<0$, either some distances will become negative or some triangular inequality (i.e., $r_{bc}+r_{ca}>r_{ab}$) will be violated and the only possible domain is positive $\lambda$. In Figure \ref{bp_trieq}, the valid region must have the positive value of $r_{ab}+r_{ca}-r_{bc}$, it is clear that the interval (i.e., $[0,\infty)$) splits into two intervals (i.e., $[0,\lambda_-]$ and $[\lambda_+, \infty)$) when $\eta_2$ is larger than $\eta_2^{\ast}$.  From (\ref{r_solsp}), it is clear that $\lambda \rightarrow \infty$ corresponds to the scaling point. Therefore, we clearly see that along the flow, the scaling component splits into two in which one of them becomes the split flow component, $\cT^{(1)}$.

Since we can continuously change the background, $h_{\infty}$, there should be a precise bifurcation point when the component just splits into two. This precise value of $\eta_2$ for this bifurcation point naturally must be smaller than the next split point of $\cT^{(1)}$ (i.e., $\eta_{s_2}\sim 2.9\times 10^{-4}$). From the point view of the split attractor flow, when the flow passes the wall of marginal stability, it can split or continue the flow toward the single attractor point if this point is well-defined. What we show here indicates the system must decide its fate precisely at this bifurcation point when the component just splits into two. Interestingly, at this moment, if the solution is exactly at the joint point of the two components, it cannot decide whether it should fall into the scaling component and have all of the centers merging into the scaling point or it should fall into the split flow component and completely reduce to several decoupled GH centers. The physical implication of this is unclear at this moment. However, one thing is clear that one cannot distinguish the split flow from the scaling components unambiguously before the appearance of this bifurcation point.

\section{The existence of three-center trees}
\label{3center}

In bubbled geometries, the solutions are composed of several centers and each center is basically parametrized by a three dimensional vector, $k_i^I$, and its GH charge. In order to have solutions which are free of CTC's, all of these charge parameters must be constrained in some way. How they should be constrained in general is not clear. However, for the solutions that belongs to split flow components, the existence conditions of the tree should immediately provide these constraints assuming the conjecture is true. In this section, we will see how these constraints emerge from the existence conditions by using the next-to-simplest tree with three centers as an example. 

\begin{figure}[t]
\centering
 \includegraphics[width=9cm]{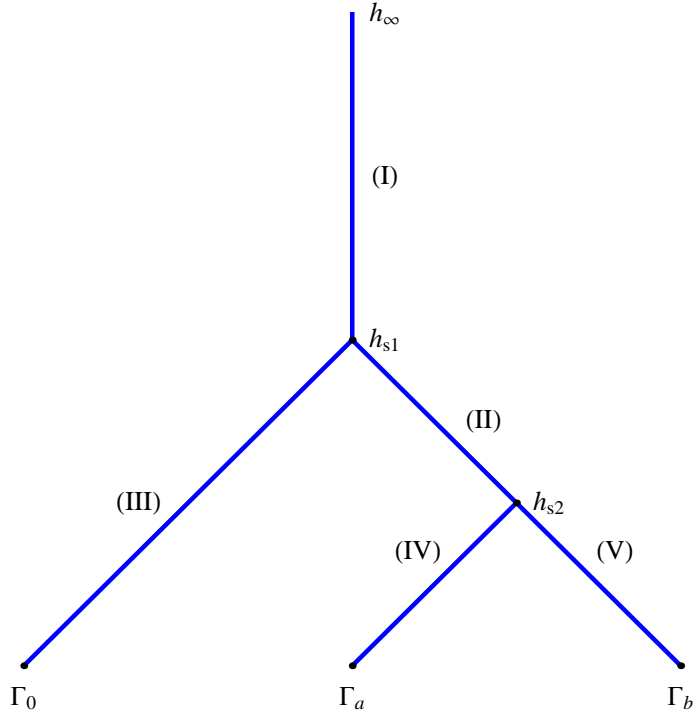}
  \caption{\it\small The structure of a general 3-center tree. There are in total five edges labeled from (\Rmnum{1}-\Rmnum{5}). There are two split points labeled with $h_{s_1}$ and $h_{s_2}$.}
\label{3c_split}
\end{figure}

For simplicity,  we assume every center has a GH charge of either $+1$ or $-1$. Also, we focus on the case in which the metric is asymptotically flat and this means that the GH charges sum to one and $h_{\infty}$ is always set to $h_0$ defined in (\ref{h_flat}). Given a set of three centers, $(\Gamma_0,\,\Gamma_a,\,\Gamma_b)$, the possible trees are $\{\Gamma_0,\,\{\Gamma_a,\,\Gamma_b\}\}$  with two other cyclic ones (see Figure \ref{3c_split}). Naively, we will have nine parameters coming from the remaining charge parameters but we can remove three of them by using the gauge symmetry in (\ref{gauge_inv}). We can always choose to fix the gauge such that the isolated charge becomes trivial and the resulting charges are:
\be
\Gamma^{(m)}_0=\pm\,(1,0,0,0)\,, ~\quad~ \Gamma_a^{(m)} =  (\,1,\, k_a^1,\, k_a^2,\, k_a^3)\,, ~\quad~ \Gamma_b^{(m)} = \mp\,(\,1,\, k_b^1,\, k_b^2,\, k_b^3\,)\,,
\ee
If the isolated charge is positive, the solution will have the same split flow pattern as the simplest bubbled supertubes. Therefore, it is convenient to implement the ring charge parameterization and define $\Gamma_r$ as:
\be
\Gamma_r ~\equiv~ \Gamma_a+\Gamma_b ~=~  (\,0,\, p^1,\, p^2,\, p^3,\, -s_1\, p^2 p^3,\,  -s_2\, p^1 p^3,\, -s_3\, p^1 p^2,\, 2\, s_m\, p^1 p^2 p^3)\,,
\label{gr}
\ee
where $(p^I, s_I)$ are related to $(k_a, k_b)$ by:
\bea
k_a^I-k_b^I ~=~ p^I\,, ~&~&~k_a^2\,k_a^3-k_b^2\,k_b^3 ~=~-s_1\,p^2\,p^3\,, \nonumber \\
k_a^1\,k_a^3-k_b^1\,k_b^3 ~=~-s_2\,p^1\,p^3\,,~&~&~k_a^1\,k_a^2-k_b^1\,k_b^2 ~=~-s_3\,p^1\,p^2\,,
\label{kprel}
\eea
and $s_m$ is not free but fixed by $s_I$:
\be
 s_m ~=~ \frac{1}{8}\,(1 - s_1^2 - s_2^2 - s_3^2 + 2\,s_2\, s_3 + 2\, s_1\, s_3 + 2\,s_1\, s_2)\,.
 \label{smdef}
 \ee
Notice that $s_I$ and $s_m$ are all dimensionless quantities.
To further discuss the constraints, we will divide the analysis into three sectors:
\begin{itemize}
\item{ Sector (\Rmnum{1}): The isolated charge is positive (i.e., $\Gamma_0^{(m)} = (1,0,0,0)$) and $p^I$ are either all positive or all negative.}
\item{ Sector (\Rmnum{2}): The isolated charge is positive and one or two of $p^I$ are negative while the remaining are positve.}
\item{ Sector (\Rmnum{3}): The isolated charge is negative (i.e., $\Gamma_0^{(m)} = -(1,0,0,0)$).}
\end{itemize}
These three sectors cover all possibilities for the three-center trees in bubbled geometries. We will analyze them in the following subsections. 

%




\subsection{Sector (\Rmnum{1})}

%

In this sector, $\Gamma_0$ is positive and $p^I$ are either all positive or all negative. The tree is $\cT=\{\Gamma_0\,,\,\{\Gamma_a\,,\,\Gamma_b\}\}$ and there are in total five edges in the tree as shown in Figure \ref{3c_split}. The strong existence conditions of the tree require there is no CTC along any of these edges. However, we will show that the weak conditions are enough. In fact, the sufficient conditions are the even smaller subset of that for this sector:
\begin{samepage}
\begin{itemize}
 \item{The separation of the first split is positive: $\eta_{s_1} = \frac{1}{r_{s_1}} > 0$}.
 \item{The separation of the second split is positive: $\eta_{s_2} = \frac{1}{r_{s_2}} > 0$}.
 \item{$V Z_I\,(\Gamma_0) >0\,, ~\quad~ I=1,2,3$}.
\end{itemize}
\end{samepage}
The first condition is:
\be
\eta_{s_1} ~=~ \frac{1}{r_{s_1}} ~=~ - \frac{\la \Gamma_0,\, h_\infty \ra}{\la \Gamma_0,\, \Gamma_r \ra} ~=~ \frac{\sum_I\, p^I}{2\,s_m\,p^1 p^2 p^3} ~>~ 0 \,.
\label{rs1st}
\ee
In this sector, this implies $s_m$ must be positive. The second condition is:
\be
\eta_{s_2} ~=~ \frac{1}{r_{s_2}} ~=~ - \frac{\langle \Gamma_a,\, h_{s_1}\rangle}{\langle \Gamma_a,\, \Gamma_b \rangle} ~=~ \frac{\langle \Gamma_a,\, \Gamma_0 + \Gamma_b \rangle + \langle \Gamma_a,\, h_\infty \rangle\, r_{s_1}}{p^1 p^2 p^3}\,\eta_{s_1} ~=~ \Delta_s\,\eta_{s_1}\,,
\label{rs2nd}
\ee
%
where $h_{s_1}$ is:
\be
h_{s_1} ~=~ (\Gamma_0  + \Gamma_r)\,\eta_{s_1}~+~ h_\infty\,,
\label{hs1st}
\ee
and $\Delta_s$ is defined as:
\bea
\Delta_s ~&\equiv&~2\,s_m\,(1-\mu_1\,s_1-\mu_2\,s_2-\mu_3\,s_3) -1+s_1\,s_2\,s_3\,,
\label{delta_s}
\eea
with $\mu_I ~\equiv~ \frac{p^I}{ \sum_I\,p^I}$. The positivity of $\eta_{s_2}$ and $\eta_{s_1}$ implies that $\Delta_s$ must be positive. The last condition requires $V Z_I$ to be positive at $\Gamma_0$. As we have explained in section \ref{skeleton}, this is equivalent to the conditions in (\ref{scf_CTC}). Using the explicit form of $h_{s_1}$ in (\ref{hs1st}), we have the following conditions:
\bea
1~+~ \frac{s_1\, p^2\, p^3}{r_{s_1}} ~>~ 0 ~&\Longrightarrow&~ 2\,\mu_1\,s_m ~+~ s_1 ~>~ 0\,, \label{g0vz1} \\
1~+~ \frac{s_2\, p^1\, p^3}{r_{s_1}} ~>~ 0 ~&\Longrightarrow&~ 2\,\mu_2\,s_m ~+~ s_2 ~>~ 0\,, \label{g0vz2} \\
1~+~ \frac{s_3\, p^1\, p^2}{r_{s_1}} ~>~ 0 ~&\Longrightarrow&~ 2\,\mu_3\,s_m ~+~ s_3 ~>~ 0 \,.
\label{g0vz3}
\eea

To summarize, we have the following conditions:  $s_m>0$\,, $\Delta_s>0$ and (\ref{g0vz1}-\ref{g0vz3}) in this sector. These conditions impose the constraints on the charges. However, before we can claim these constraints define the physical region, we need to verify that the strong existence conditions of the trees are satisfied. Due to the complexity of the CTC's conditions, we start with the parallel charge case to get a better idea about the basic features of these constraints and then move on to the general case. Additionally, we also check the dangerous half-split point to verify the conjecture.

%

\subsubsection{Parallel charges}
\label{parallel}

For the parallel charge case, we simply set $s_I$ to be equal (i.e., $s_1=s_2=s_3=s$). Immediately, we have $s_m = \frac{1}{8}  (1+3\,s^2)$ and the positivity of it is given. The remaining conditions become:
\be
s^3 + 3\,s^2 - s - 3 ~ >~  0\,, ~~\qquad~~  3\,s^2  +  12\, s  + 1 ~>~ 0\,,
\ee
where the second inequality comes from the summation of the conditions (\ref{g0vz1}-\ref{g0vz3}). Combining the above two conditions yields $s>1$. Therefore, $s>1$ defines the physical region for the parallel charge case. This also implies that the total charges of harmonic function $L_I$ must be positive. Consequently, $L_I$ is positive along edges (\Rmnum{1}-\Rmnum{3}) and $VZ_I>0$ is trivially satisfied along these edges. In the following, we will check the remaining unverified CTC's conditions for all of the five edges in the physical region.

\subparagraph{ Edge (\Rmnum{1}):}  

Since $V Z_I>0$ is always true for $s>1$, we only need to check the other condition, $\cQ>0$, along this edge. To monitor how $\cQ$ changes along the flow, we expand $\cQ$ with the power of $\frac{\eta_1}{\eta_{s_1}}$:
\be
\cQ|_{\eta_1} ~=  \bq_1^{(0)} ~+~ \bq_1^{(1)}\,\frac{\eta_1}{\eta_{s_1}}+\bq_1^{(2)}\,\left(\frac{\eta_1}{\eta_{s_1}}\right)^2 + \bq_1^{(3)}\,\left(\frac{\eta_1}{\eta_{s_1}}\right)^3+\bq_1^{(4)}\,\left(\frac{\eta_1}{\eta_{s_1}}\right)^4 \,,
\ee
and the coefficients are:
\bea
&& \bq_1^{(0)}~=~ 0\,, ~\quad~ \bq_1^{(1)} ~=~ \eta_{s_1}\,, ~\quad~ \bq_1^{(2)} ~=~  (p^1 p^2\,+\,p^2 p^3\,+p^1 p^3)\,(1+s)\,\eta_{s_1}^2\,, \nonumber \\
&& \bq_1^{(3)} ~=~ p^1 p^2 p^3\, (\sum_I\,p^I)\,(1+s)^2\,\eta_{s_1}^3\,, \nonumber \\
&& \bq_1^{(4)} ~=~ -\frac{1}{64}\,(p^1 p^2 p^3)^2\,(1+s)^2\,(17-10s+9s^2)\,\eta_{s_1}^4\,.
\eea
For $s>1$,  $\bq_1^{(4)}$ is always negative while the other coefficients are all positive. This indicates that if we put all of the charges into a single center, it will not be a well-defined attractor point and implies the single center solution or the scaling solution does not exist and the first split must happen. To check $\cQ$ remains positive along edge (\Rmnum{1}) even with the negative $\bq_1^{(4)}$, we will show that $\bq_1^{(3)} \geq 4 |\bq_1^{(4)}|$ in the physical region. For the parallel charge case, this is immediately followed by:
\be 
(1+3s^2) ~\geq~ \frac{1}{4}\, (17-10s+9s^2)\,.
\ee
This inequality is always true for $s>1$ and saturates at the boundary, $s=1$. Because of $\bq_1^{(3)} \geq 4 |\bq_1^{(4)}|$, only the third and fourth order terms are enough to guarantee the positivity of $\cQ$ at the split point where $\eta_1=\eta_{s_1}$. In fact, this is still true even when we go deeper into the half-split point where $\eta_1 = 2\, \eta_{s_1}$. Curiously, we see that the weak conditions are enough to ensure $\cQ>0$ at this dangerous point as well and avoid the conjecture to be violated. Furthermore, if we take the derivative of $\cQ$ with the dimensionless parameter, $\frac{\eta_1}{\eta_{s_1}}$, we see that the derivative is always positive along edge (\Rmnum{1}) (actually, it is still positive up to the half-split point). This means that $\cQ$ monotonically increases with $\eta_1$ for this whole edge.

\subparagraph{Edge (\Rmnum{2}):}

We can do a similar expansion of $\cQ$ with $\frac{\eta_2}{\eta_{s_2}}$ for this edge. The coefficients we get are:
\bea
\bq_2^{(1)} ~&=&~ \frac{4\,\sum_I\,p^I}{p^1 p^2 p^3\,(1+3s^2)^3}\,\Big(({\cal P}^2)\, \xi^{(1)}(s) ~+~ ({\cal P}^{IJ})\, \xi^{(2)}(s)\Big)\,\eta_{s_2}\,, \\
\bq_2^{(2)} ~&=&~ \Big(({\cal P}^2)\, \xi^{(3)}(s) ~+~ ({\cal P}^{IJ})\, \xi^{(4)}(s)\Big)\,\eta_{s_2}^2\,, \\
\bq_2^{(3)} ~&=&~ \frac{p^1 p^2 p^3\,\sum_I\,p^I\,(s^3+3s^2-s-3)}{(1+3s^2)}\,\eta_{s_2}^3 ~=~ (p^1 p^2 p^3)^2\,\eta_{s_2}^4\,, \\
\bq_2^{(4)} ~&=&~ -\frac{(p^1 p^2 p^3)^2}{4}\,\eta_{s_2}^4\,,
\eea 
where $\cP^2$ and $\cP^{IJ}$ are defined as
\be
\cP^2 ~\equiv~ p^1 p^1 + p^2 p^2 + p^3 p^3\,\,, ~\qquad~ \cP^{IJ} ~\equiv~ p^1 p^2 + p^2 p^3 + p^3 p^1\,\,,
\ee 
and $\xi(s)$ are some polynomials of $s$:
\bea
\xi^{(1)}(s) ~&=&~ ~-~ 5 ~+~ 38\,s^2 ~+~ 48\,s^3  ~+~ 15\,s^4 \,\,, \\
\xi^{(2)}(s) ~&=&~ ~-~ 8 ~+~ s ~+~ 88\,s^2 ~+~ 102\,s^3 ~+~ 48\,s^4 ~+~ 9\,s^5 \,\,, \\
\xi^{(3)}(s) ~&=&~ ~-~ 13 ~-~ 6\,s ~+~ 34\,s^2 ~+~ 30\,s^3 ~+~ 3\,s^4 \,\,, \\
\xi^{(4)}(s) ~&=&~ ~-~ 25 ~-~ 12\,s ~+~ 74\,s^2  ~+~ 60\,s^3 ~+~ 15\,s^4  \,\,.
\eea
One can easily show that these four polynomials are all positive when $s>1$. Therefore, $\bq_2^{(1)}$ and $\bq_2^{(2)}$ are positive. Furthermore, one notices that in this edge, $\bq_2^{(3)}$ is exactly equal to  $4\,|\bq_2^{(4)}|$. This tells us that not only is $\cQ$ positive but also that it increases monotonically along this edge. Additionally, it also shows that the dangerous half-split point (i.e. $\eta_2=2\,\eta_{s_2}$) for this edge is also safe with positive $\cQ$. 

\subparagraph{Edges (\Rmnum{3}-\Rmnum{5}):}

All of these edges are single-center flows that terminate at some GH centers. As we have mentioned, we only need to check the conditions at the start and end points for these kinds of flows. All of the starting points should have been verified in the previous two edges and therefore, we only need to check the conditions (\ref{scf_CTC}) to ensure the end points are also well-behaved. 

For edge (\Rmnum{3}), we have already required the conditions in  (\ref{g0vz1}-\ref{g0vz3}). 
For the remaining two edges, we start with $h_{s_2}$:
\be
h_{s_2} ~=~ \frac{\Gamma_a + \Gamma_b}{r_{s_2}} ~+~ h_{s_1} \,\,.
\ee
%
%
To apply the condition to $\Gamma_a$ and $\Gamma_b$, we change the gauge respectively such that $\Gamma_a$ or $\Gamma_b$ explicitly become pure a positive/negative unit GH center and then check the electric part of $h_{s_2}$. The conditions for $\Gamma_a$ are:
%
%
%
\be
\frac{s\,(1+s^2)}{\mu_I\,(1 + 3\, s^2)} ~+~ \frac{1+5\,s^2}{1+3\,s^2} ~ >~  0 ~\quad~ I=1,2,3\,,
\label{Ga_scc}
\ee
and for $\Gamma_b$ are:
%
%
\be
\frac{2 + s + 4\,s^2 + s^3}{\mu_I\,(1+3\,s^2)} ~-~ 1  ~>~0 ~\quad~ I=1,2,3\,,
\label{Gb_scc}
\ee
where $\mu_I \equiv \frac{p^I}{\sum_I\,p^I}$ and since $p^I$ are all positive or all negative, $\mu_I$ are between $0$ and $1$. It is clear that these conditions are satisfied for $s>1$. Therefore, for the parallel charges, we conclude that $s>1$ is the only necessary and sufficient condition in order for the trees to exist and it defines the physical region.

\subsubsection{General charges}

In this subsection, we remove the restriction on $s_I$. Just like the parallel charge case, we take $p^I$ as some arbitrary constants which are either all positive or all negative and check the constraints imposed on $s_I$ by the sufficient conditions. At first, we divide the three dimensional $s_I$-space into eight regions according to the signs of $s_I$ (i.e., $(\pm,\pm,\pm)$) and look at the conditions for each region. Interestingly, we find that no matter what the values of $p^I$ are, the physical region is always confined in the all-positive region (i.e., $(+,+,+)$).

\begin{figure}[t]
\centering
\includegraphics[width=9cm]{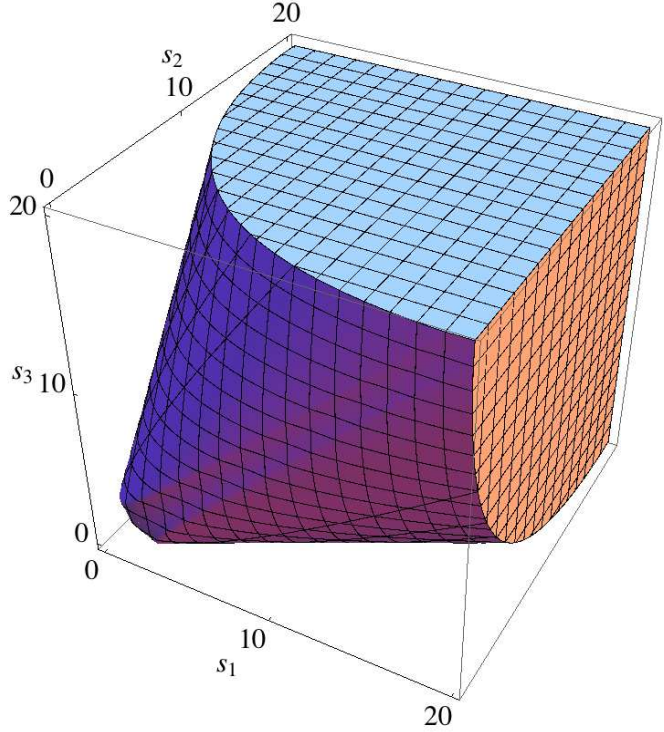}
 \caption{\it\small The physical region defined by $\Delta_s>0$ and $s_m>0$  in the positive $s_I$ region with $\mu_1=\mu_2=\mu_3=\frac{1}{3}$.}
\label{sI_space}
\end{figure}

In the following, we will assume $s_1 \geq s_2 \geq s_3$ but a similar argument can apply to the other cases by taking advantage of the cyclic nature of the conditions. To start, we express $\Delta_s$ as the following:
\be
\Delta_s ~=~ \overline{\Delta}_s ~+~ 2\,\left(\,(s_3-s_1)\,\mu_1+(s_3-s_2)\,\mu_2\,\right)\,s_m\,,
\label{Deltas_mu}
\ee
where $\overline{\Delta}_s$ is independent of $\mu_I$ and is defined as:
\be
\overline{\Delta}_s ~\equiv~ 2\,s_m\,(1-s_3)~-~ \,(1-s_1\,s_2\,s_3)\,,
\ee
also we have used $\mu_3=1-\mu_1-\mu_2$ and $s_m$ is defined in (\ref{smdef}). At first, we check the regions except the all-postive region:

\begin{itemize}
\item{$(s_1,s_2,s_3)\in (+,+,-)$: Because $s_m$ must be positive, the term that depends on $\mu_I$ in (\ref{Deltas_mu}) is always negative. Therefore the maximum of $\Delta_s$ is simply $\overline{\Delta}_s$. However, one can show that $\overline{\Delta}_s$ is negative in this region. Therefore, it is clear that $\Delta_s$ and $s_m$ cannot be simultaneously positive in this region as long as $p^I$ are all positive or all negative. 
}
\item{$(s_1,s_2,s_3)\in (+,-,-)$: To exclude this region, we need to include the other three conditions (\ref{g0vz1}-\ref{g0vz3}).  By combing  (\ref{g0vz2}) and  (\ref{g0vz3}), we have:
\be
2\,s_m+s_2+s_3 ~>~ 2\,\mu_1\,s_m\,.
\ee
Because $\mu_1$ and $s_m$ are positive values, this implies:
\be
2\,s_m+s_2+s_3 ~>~0\,.
\label{sm_cnd_pmm}
\ee
On the other hand, to maximize $\Delta_s$, we can set $\mu_1$ to zero but not $\mu_2$. Because $s_2$ is now negative, we can only take the minimum value of $\mu_2$ allowed by the condition (\ref{g0vz2}). Therefore the maximum value of $\Delta_s$ is:
\be
\Delta_s\,(\mu_1=0,\,\mu_2=-\frac{s_2}{2\,s_m})~=~ \overline{\Delta}_s ~+~ \,s_2\,(s_2-s_3)\,.
\label{Deltas_pmm}
\ee
One can show this maximum value of $\Delta_s$ is always negative under the condition (\ref{sm_cnd_pmm}) in this region. Therefore, this region is also excluded.
}
\item{$(s_1,s_2,s_3)\in (-,-,-)$: At first, by combing all of the three conditions (\ref{g0vz1}-\ref{g0vz3}), we have the following:
\be
s_1+s_2+s_3+2\,s_m ~>~ 0\,.
\label{sm_cnd_mmm}
\ee
To maximize $\Delta_s$, we need to take the minimum values of $\mu_1$ and $\mu_2$ allowed by the conditions (\ref{g0vz1}) and (\ref{g0vz2}) and we have:
\be
\Delta_s\,(\mu_1=-\frac{s_1}{2\,s_m},\,\mu_2=-\frac{s_2}{2\,s_m})~=~ \overline{\Delta}_s ~+~ s_2\,(s_2-s_3) + s_1\,(s_1-s_3)\,.
\label{Deltas_mmm}
\ee
One can show this maximum value of $\Delta_s$ is always negative under the condition (\ref{sm_cnd_mmm}) in this region. Therefore, this region is also excluded.
}
\end{itemize}
Through this analysis, one can conclude that the physical region must be confined in the all-positive region. Since the conditions (\ref{g0vz1}-\ref{g0vz3}) are trivially satisfied in this region, the precise physical region is defined only by the positivity of $\Delta_s$, $s_m$ and $s_I$. Of course, the exact physical region in $s_I$ space will depend on the value of $p^I$ or more precisely $\mu_I$. However, the general feature of this region is roughly the same with the boundary changing a little with $\mu_I$. For the special case with the equal dipole charges, the physical region is shown in Figure \ref{sI_space}.

%


It is quite interesting how these conditions interplay and the constraint that $s_I$ must be all positive emerges. Similarly, this implies the harmonic functions, $L_I$, are all positive along edges (\Rmnum{1}-\Rmnum{3}) and $V Z_I$ are also positive along these edges. In the following, we will check the remaining CTC's conditions along all edges in the physical region defined by the positivity of $s_m$, $\Delta_s$ and $s_I$.

\subparagraph{Edge (\Rmnum{1}):}

The coefficients of the expansion of $\cQ$ with $\frac{\eta_1}{\eta_{s_1}}$ in the general case are:
\bea
&& \bq_1^{(1)} ~=~ \eta_{s_1}\,, ~\quad~ \bq_1^{(2)} ~=~ \left (\,p_2\,p_3\,(1+s_1)+p_1\,p_2\,(1+s_2)+p_1\,p_2\,(1+s_3)\,\right)\,\eta_{s_1}^2\,, \nonumber \\
&& \bq_1^{(3)} ~=~ p^1 p^2 p^3\,\left(p^1 (1+s_2) (1+s_3)+p^2 (1+s_1)(1+s_3)+p^3(1+s_1)(1+s_2)\right)\,\eta_{s_1}^3\,, \nonumber \\
&& \bq_1^{(4)} ~=~ (p^1 p^2 p^3)^2\,\left((1+s_1)(1+s_2)(1+s_3) - \frac{1}{4}\,(2+2\,s_m+s_1+s_2+s_3)^2\right)\,\eta_{s_1}^4\,.
\eea
For the positive $s_I$, only $\bq_1^{(4)}$ can be negative. However, we will again show that $\bq_1^{(3)}+4\,\bq_1^{(4)}$ is always positive in the physical region. 

In the following, we will assume $s_3 \geq s_2 \geq s_1$ but a similar argument can apply to the other cases by using the cyclic nature of the conditions. At first, we isolate the $\mu$ dependent terms from $\bq_1^{(3)}+4\,\bq_1^{(4)}$ as:
\be
\bq_1^{(3)}+4\,\bq_1^{(4)} ~=~ (p^1 p^2 p^3)^2\,\eta_{s_1}^4\,\overline{\bq_1}\,,
\ee
with $\overline{\bq_1}$ is defined as:
\be
\overline{\bq_1} ~\equiv~ \widetilde{\bq_1}~+~\mu_1\,(1+s_2)(s_3-s_1) ~+~ \mu_2\,(1+s_1)(s_3-s_2)\,,
\label{bq1_def}
\ee
where $\widetilde{\bq_1}$:
\be
\widetilde{\bq_1} ~\equiv~ 4\,(1+s_1)(1+s_2)(1+s_3) - \,(2+2\,s_m+s_1+s_2+s_3)^2 +2\,s_m\,(1+s_1)(1+s_2)\,,
\ee
where $\mu_3=1-\mu_1-\mu_2$ is used. To verify $\bq_1^{(3)}+4\bq_1^{(4)}>0$, we will show that the minimum of  $\overline{\bq_1}$ is positive in the physical region. As we assume $s_3$ is larger than the other two parameters, we need to set $\mu_1$ and $\mu_2$ to their smallest allowed values to minimize $\overline{\bq_1}$. On the other hand, the parameters, $\mu_1$ and $\mu_2$, are constrained by the positivity of $\Delta_s$:
\be
\mu_1\,(s_3-s_1)\,~+~\mu_2\,(s_3-s_2) ~>~ \frac{1-s_1\,s_2\,s_3}{2\,s_m}-1+s_3\,.
\label{mu_constraint}
\ee
The minimum of $\overline{\bq_1}$ is either at $\mu_1=0$ or $\mu_2=0$ or $\mu_1=\mu_2=0$ depending on the value of $s_I$. For the case when both of them are zero, one can show that $\widetilde{\bq_1}$ is positive in the physical region. For the case with $\mu_1=0$, we can take $\mu_2$ to its minimum allowed value by (\ref{mu_constraint}):
\be
\mu_{2min} ~=~  \frac{1-s_1\,s_2\,s_3}{2\,s_m\,(s_3-s_2)}-\frac{1-s_3}{s_3-s_2}\,,
\ee
and in order for this to be a meaningful minimum, it must be between zero and one. Under this constraint, one can show that the minimum of $\overline{\bq_1}$ is still positive in the physical region for this value of $\mu_2$. Similarly, this is also true for the case where $\mu_2=0$. Therefore, this concludes that $\overline{\bq_1}$ is always positive in the physical region. As we have explained in section $\ref{parallel}$, this also implies that $\cQ$ increases monotonically with $\eta_1$ along the whole edge (\Rmnum{1}).

\subparagraph{Edge (\Rmnum{2}):}

The coefficients of the expansion of $\cQ$ with $\frac{\eta_2}{\eta_{s_2}}$ in this edge are:
\bea
\bq_2^{(1)} ~&=&~\frac{(p_1+p_2+p_3)^3}{2\,s_m\,p_1\,p_2\,p_3}\,\Big(\,\frac{1}{4\,s_m^2}\,(3\,\Delta_s+2) + \frac{\xi(s_I, \mu_I)}{4\,s_m}\, \nonumber \\
&& ~~\qquad~~ + \mu_1\,\mu_2\,(s_3+2) + \mu_2\,\mu_3\,(s_1+2) + \mu_1\,\mu_3\,(s_2+2)+1\, \Big)\,\eta_{s_2}\,,   \nonumber \\
\bq_2^{(2)} ~&=&~  \frac{(p_1+p_2+p_3)^2}{8\,s_m}\,\left(\,\frac{3}{s_m}\,(2\,\Delta_s+1) + \xi(s_I,\mu_I) + 8\,s_m\,( \mu_1\,\mu_2 + \mu_2\,\mu_3 + \mu_1\,\mu_3)\,\right)\,\eta_{s_2}^2 \,, \nonumber \\
 \bq_2^{(3)} ~&=&~ (p^1 p^2 p^3)^2\,\eta_{s_2}^4\,,  ~\qquad~ \bq_2^{(4)} ~=~-\frac{1}{4}\, (p^1 p^2 p^3)^2\,\eta_{s_2}^4\,,
 \label{qc2_gc}
\eea
where $\xi(s_I,\mu_I)$ is:
\be
\xi(s_I,\mu_I) ~\equiv~  2\,(s_1+s_2+s_3) +  4\,(s_2\,s_3\,\mu_1 + s_1\, s_3\, \mu_2 + s_1\, s_2 \,\mu_3) - 8\,s_m\,.
\ee
Notice that the positivity of $\bq_2^{(0)}$ should already be verified at edge (\Rmnum{1}) and we do not need to check it here. From (\ref{qc2_gc}), we immediately see that $\bq_2^{(4)}$ is always negative. However, it is also clear that $\bq_2^{(3)}+4\,\bq_2^{(4)}=0$. This means that $\bq_2^{(3)}$ is large enough to overcome $\bq_2^{(4)}$ before this flow reaches the split point at $\eta_2=\,\eta_{s_2}$ or even deeper to the half-split point at $\eta_2=2\,\eta_{s_2}$. For the positivity of the remaining two coefficients, we only need to check $\xi(s_I,\mu_I)$. With similar arguments as we have been using, one can show $\xi(s_I,\mu_I)$ is always positive in the physical region. Therefore, we conclude that not only is $\cQ$ positive along this edge but it also increases monotonically with $\eta_2$. 

\subparagraph{Edges (\Rmnum{3}-\Rmnum{5}):}

For these three edges, we need to check the conditions for single center flow. The condition for edge (\Rmnum{3}) has been included in the definition of the physical sector and is automatically satisfied. For the remaining two edges, we just need to generalize the conditions in (\ref{Ga_scc}, \ref{Gb_scc}). The conditions for $\Gamma_a$ are:
\bea
1+\frac{1}{8\,\mu_1\,s_m}\,\left(\,4\,\Delta_s + (s_2-s_3)^2-s_1^2+2\,s_1+3\,\right) ~&>&~ 0\,, \\
1+\frac{1}{8\,\mu_2\,s_m}\,\left(\,4\,\Delta_s + (s_1-s_3)^2-s_2^2+2\,s_2+3\,\right) ~&>&~ 0\,, \\
1+\frac{1}{8\,\mu_3\,s_m}\,\left(\,4\,\Delta_s + (s_1-s_2)^2-s_3^2+2\,s_3+3\,\right) ~&>&~ 0\,,
\eea
and for $\Gamma_b$:
\bea
-1+\frac{1}{8\,\mu_1\,s_m}\,\left(\,4\,\Delta_s - (s_2-s_3)^2+s_1^2+2\,s_1+5\,\right) ~&>&~ 0\,, \\
-1+\frac{1}{8\,\mu_2\,s_m}\,\left(\,4\,\Delta_s - (s_1-s_3)^2+s_2^2+2\,s_2+5\,\right) ~&>&~ 0\,, \\
-1+\frac{1}{8\,\mu_3\,s_m}\,\left(\,4\,\Delta_s - (s_1-s_2)^2+s_3^2+2\,s_3+5\,\right) ~&>&~ 0\,.
\eea
Using similar arguments, one can show that all of these conditions are satisfied in the physical region.

This concludes the verification that the strong existence conditions of the trees are satisfied in the physical region for this sector. According to the conjecture, every existing tree should correspond to some well-behaved supergravity solution. In other words, there exists  a well-behaved solution for every point in the physical region (e.g. Figure \ref{sI_space}) with arbitrary all-positive or all-negative dipole charges. Before we can claim these solutions are microstate geometries of some zero-entropy black rings, we need to verify the following ratio:
\be
\label{nudef}
\nu ~\equiv~ \left|\frac{\la \Gamma_a+\Gamma_b\,,\,\Gamma_0 \ra}{\la \Gamma_a\,,\,\Gamma_b \ra}\right| ~=~ 2\,s_m\,.
\ee 
This ratio measures the relative size of the bubble that supports the ring's radius and the bubble that nucleated inside the ring. For a three-center solution in bubbled geometries to resemble a classical zero entropy black ring, this ratio should be much larger than one (i.e., $\nu \gg {\cal O}(1)$). Since this ratio is just proportional to  $s_m$, one can see that it is not bounded above in the physical region (Figure \ref{sI_space}). This can also be easily seen in the parallel charge case. The only dangerous part is near the boundary where $s_m$ goes to zero. When the solution is close to such a boundary, the ratio can become comparable or even small than order one and it will no longer resemble a black ring. However, the majority of the solutions in the physical region still have a large ratio and can be regarded as the simplest bubbled supertubes.


\subsection{Sector (\Rmnum{2})}

For this sector, one or two of the dipole charges of the ring are negative. Naively, it seems unlikely to have well-behaved solutions in this sector. However, we will show that the trees still exist in some limited region.


To simplify the analysis, we reduce one degree of freedom and take $s_1=s_2=s$. We will use the following parameterization:
\be
\Gamma_r^{(m)} ~=~ (0,\, p^1,\, p^2,\, p^3) ~=~ (0,\, p+q,\, p-q,\, -y\,p)\,, 
\ee
where $y$ is a dimensionless quantity. For this sector, we will focus on the region where $p^1$ and $p^2$ are positive while $p^3$ is negative and without loss of generality, we assume $p^1>p^2$ (i.e., $p>q>0$ and $y>0$). Notice that all of CTC's conditions are invariant if we flip the signs of all $k^I$ simultaneously (of course, the asymptotic vector must change accordingly). Therefore, the case where two $p^I$ are negative should be the mirror of  the case we analyze here. 

We will follow a similar procedure to the previous section. Due to the simplification, the $s_I$-space now is only two dimensional and we divide it into four regions based on the signs of $s$ and $s_3$. By carefully checking the weak existence condition (particularly, CTC's conditions for $\Gamma_a$, $\Gamma_b$ edges and the positivity of the two separations.), we find that the physical region is confined in $(s, s_3) \in (-,+)$. For this region,  the charges of the harmonic functions, $L_I$, are again all positive. Furthermore, the  only conditions we need for this region are the ones  that require two separations are positive:
\bea
\label{rs_cond}
\eta_{s_1} ~&=&~ \frac{4\,(y-2)}{(1 + 4\,s\,s_3 - s_3^2)\,(p^2-q^2)\,y }~>~0 ~\Longrightarrow~ \frac{(y-2)}{(1 + 4\,s\,s_3 - s_3^2)} ~>~ 0\,, \\
\eta_{s_2} ~&=&~\frac{\xi(y,s,s_3)}{(1 + 4\,s\,s_3 - s_3^2)\,(p^2-q^2)\,y }~>~0 ~\Longrightarrow~ \frac{\xi(y,s,s_3)}{(1 + 4\,s\,s_3 - s_3^2) } ~>~ 0\,,
\label{rss_cond}
\eea
where $\xi$ is:
\be
\xi(y,s,s_3) ~=~ (1+4\,s\,s_3-s_3^2)\,\Big(\,(y-2)\,(1-s_3)+2\,(s-s_3)\,\Big) ~+~ 4\,(s^2\,s_3-1)\,(y-2)\,.
\ee
Notice that when $y=2$, the three dipole charges sum to zero and the separation of $\Gamma_r$ and $\Gamma_0$ diverge. This means that the $h_0$ is already at the split point and consequently, the tree cannot exist. For $0<y<2$ and $y>2$, the physical region where the tree exists is bounded by two lines in ($s$, $s_3$)-space. These two lines are defined by the following conditions respectively:
\be
1+4\,s\,s_3-s_3^2 ~=~ 0\,, ~~\quad~~\xi(y,s,s_3) ~=~ 0\,.
\label{twolines}
\ee
For $0<y<2$, the first line provides the lower bound of $s_3$ while the second line gives the upper bound. For $y>2$, the upper and lower bounds are reversed. These two lines always cross with each other at a particular point, $(s,s_3)\approx (-4.01545, 0.062)$, independent of $y$. At this point, the ratio between two separations diverge (i.e., $\frac{\eta_{s_1}}{\eta_{s_2}}$) and it is not possible to satisfy both conditions, (\ref{rs_cond}) and (\ref{rss_cond}), simultaneously for $-4.01545 \lesssim s < 0$. Furthermore, it can be easily checked that the upper bound of $s_3$ is always smaller than the maximum value, $-2+\sqrt{5}$, which is reached at $y=0$ and $s \rightarrow -\infty$. Therefore, we can see the physical region is confined in a very narrow semi-infinite band\footnote{In fact, even if we allow $s_1$ and $s_2$ to be different, their difference is still quite limited (i.e., $|s_1-s_2| \lesssim 1$) no matter what  the choice of $p^I$ is. Therefore, in general, the physical region is confined in a very narrow semi-inifite box.} in $(s, s_3)$-space. Also, we carefully checked that the strong existence conditions are indeed satisfied in the physical region by using a similar procedure to the previous subsection.


To understand the physical interpretation for this sector, we verify the ratio, $\nu$, in (\ref{nudef}). Interestingly, unlike sector (\Rmnum{1}), the ratio in the physical region is actually bounded above for $0<y<2$:
\be
\nu ~=~ \frac{1}{4}\,(-1-4\, s\, s_3 + s_3^2) ~<~ \frac{1}{4}\,(-1+\frac{2}{y})\,. 
\ee
In order to have a large ratio, $y$ must be close to zero. Since $y$ controls how large the negative $p^3$ is compared to the remaining $p^I$, this implies this negative $p^3$ must be very small compared to the sum of the three $p^I$ in order to have the solution resemble the classical zero entropy black ring. On the other hand, for $y>2$, the maximum value of $\nu$ is $\frac{1}{4}$ and moreover, the sum of the three $p^I$ becomes negative. In this case, even though the solutions exist, they are no longer similar to classical black rings. 

\begin{figure}[t]
\centering
 \begin{tabular}{cc}
\includegraphics[width=0.4\linewidth]{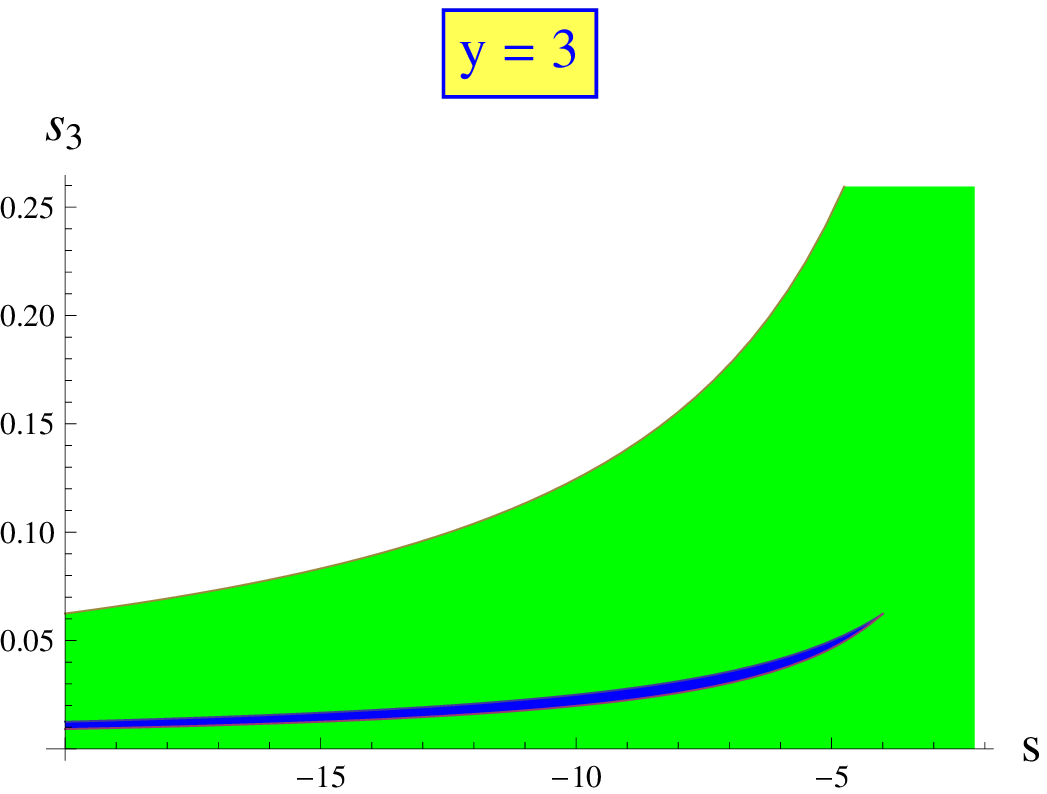} &
\includegraphics[width=0.4\linewidth]{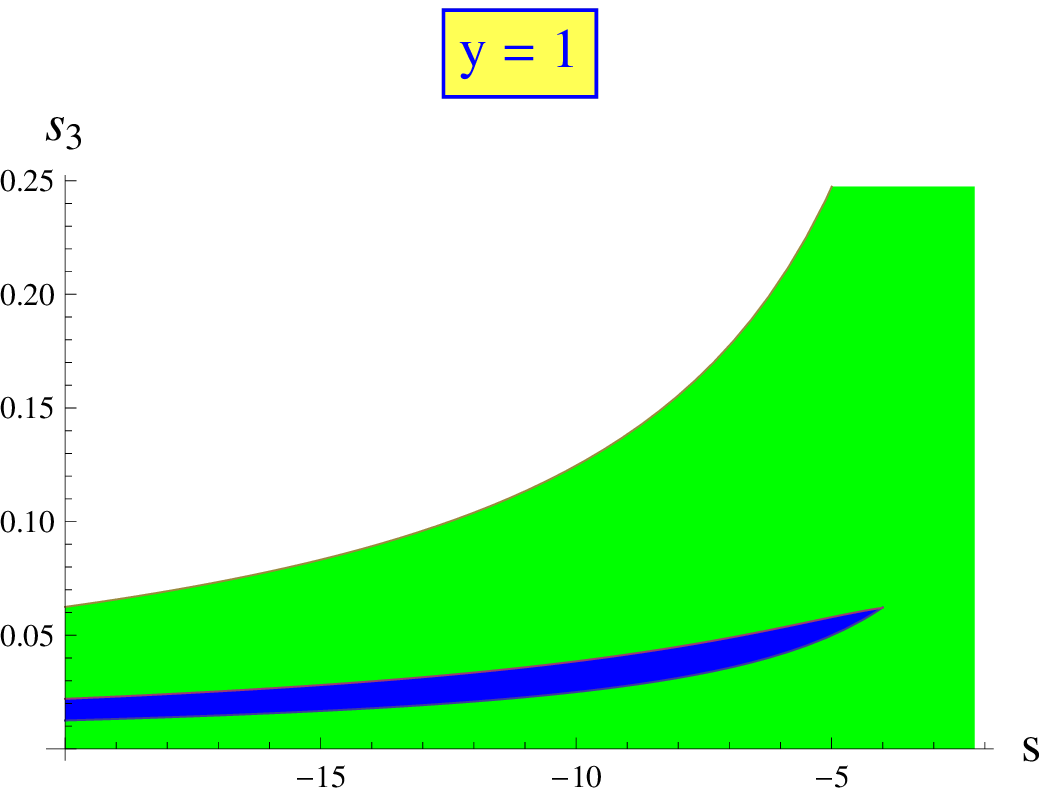} \\
\includegraphics[width=0.4\linewidth]{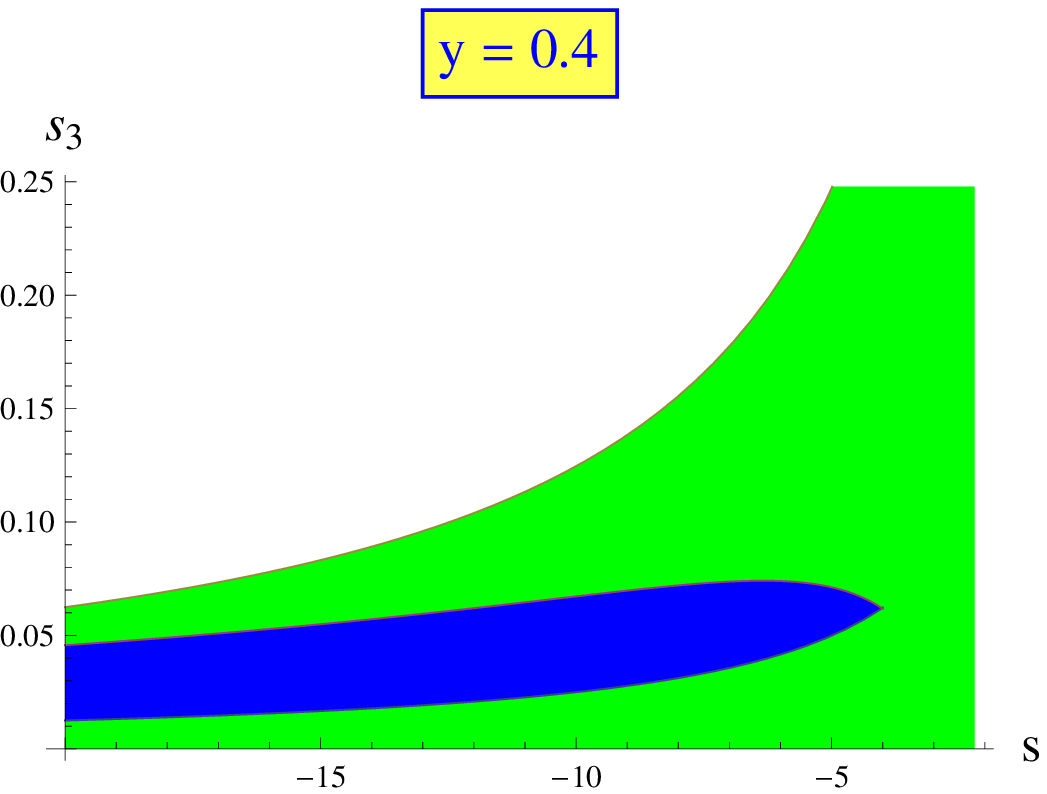} &
\includegraphics[width=0.4\linewidth]{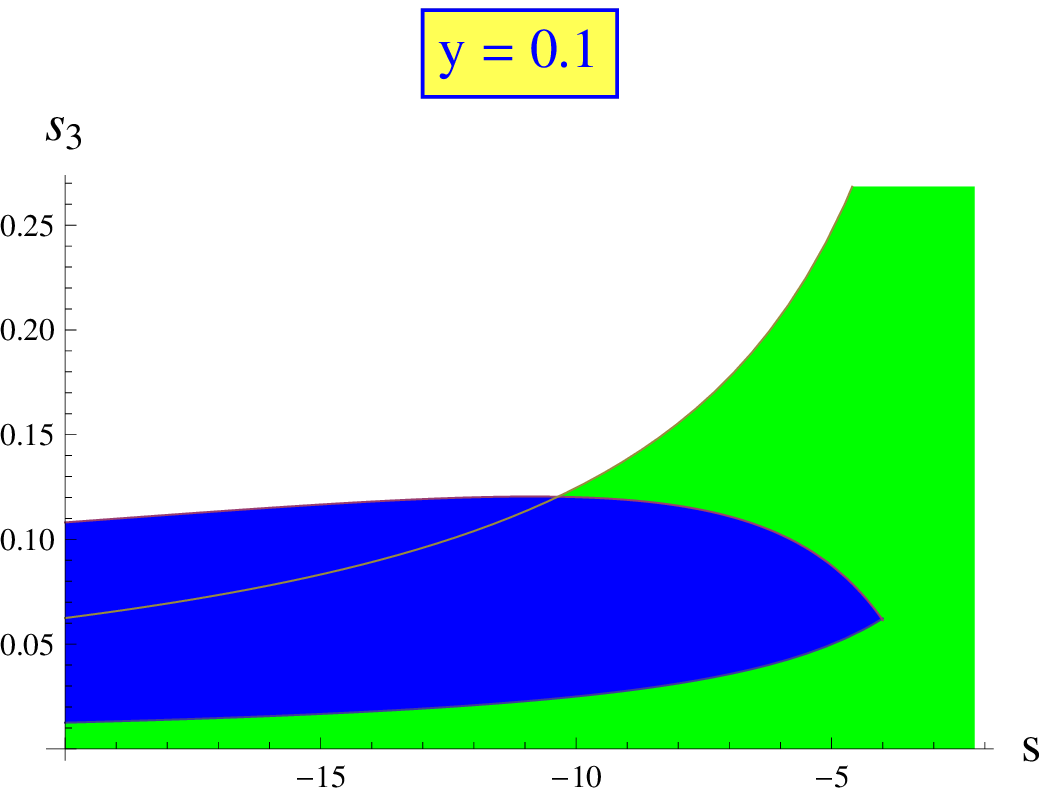} \\
 \end{tabular}
 \caption{\it\small These graphics compare two regions in $(s,s_3)$-space for four values of $y$. The light blue region is the physical region of the tree while the light green one is the region that satisfies triangle scaling conditions.}
\label{phscalreg}
\end{figure}

To understand the physical interpretation of the solutions when $y$ is not so small, we investigate another possibility. Specifically, we find that three centers are very likely to form scaling solutions in this sector. To study this possibility, we investigate the conditions for the  formation of the triangular scaling solutions:
\bea
\la \Gamma_a\,, \Gamma_b \ra\, \la \Gamma_b\,, \Gamma_0 \ra ~>~0\,, ~\quad~  \mbox{and two other cyclic ones}\,, \\
|\la \Gamma_a\,, \Gamma_b \ra  + \la \Gamma_b\,, \Gamma_0 \ra |  ~>~  |\la \Gamma_0\,, \Gamma_a \ra |\,, ~\quad~  \mbox{and another two cyclic ones}\,,
\eea
where the first three conditions require the directions of the fluxes to form a closed loop and the next three conditions require them to satisfy triangle inequalities. These conditions can be reduced to ones that involve only $s$ and $s_3$. If we look at the region, $s<-1$ and $s_3>0$, the triangle scaling conditions are reduced to:
\be
\frac{-5+s_3^2}{4\,s_3} ~<~ s ~<~ \frac{-4-s_3+s_3^3}{2\,(1+s_3^2)}\,, ~~\quad~~0<s_3<1\,.
\ee
Unlike the physical region of the split flow component, this triangle scaling region defined by the above conditions does not depend on the dipole charges. Therefore, for any positive $y$, the scaling region remains the same. On the other hand, the physical region changes with $y$ and we compare the two regions for several values of $y$ in Figure \ref{phscalreg}.
%

We find that for $y\geq 0.4$, the physical region is completely submerged in the triangle scaling region. For a larger value of $y$, the scaling region dominates and this implies that the three centers are more likely to form a scaling solution than become a three-center tree. Additionally, we randomly pick several solutions in the region where the physical and scaling regions overlap, we find that those solutions have bifurcation points similar to the example in section \ref{bifurcation}. This means that the split flow component only appears from the scaling component at the bifurcation point which is usually close to the split point. If the solution becomes scaling and is well-behaved, it is in fact more similar to a five dimensional black hole rather than a black ring. Since a triangular scaling solution is continuously connected to a scaling point, based on the dynamic version of the conjecture, we can check $\cQ$ and $V Z_I$ along the single center flow of the total charge to ensure it is well-behaved. We find that as long as the asymptotic charges are positive (i.e., $s<-1$ and $s_3>-1$), every coefficient of the radial expansion of every CTC's condition is positive. Therefore, the solutions in the triangle scaling region are well-behaved and should be regarded as the microstate geometries of some five dimensional black holes.

From the above analysis, we conclude that the majority of sector (\Rmnum{2}), especially for the solutions with larger $y$, should be regarded as the microstate geometries of some black holes while the solutions with smaller $y$ can be regarded as the transition phase from black rings to black holes.

\subsection{Sector (\Rmnum{3})}

For this sector, we cannot find an efficient way to parametrize the charge parameters. However, through some numerical studies, we know the allowed region in the charge space is also very limited. For example, if one takes equal charge case (i.e., $k_{a,b}^1=k_{a,b}^2=k_{a,b}^3=k_{a,b}$), one finds that no tree can exist. However, one can still find some examples in which the tree does exist. A particular simple example is:
\be
\Gamma^{(m)}_0=(-1,0,0,0)\,, ~\quad~ \Gamma_a^{(m)} =  (\,1,\, 150,\, 40,\, 20)\,, ~\quad~ \Gamma_b^{(m)} = (\,1,\, 10,\, 50,\, 30\,)\,.
\ee
We find that the tree exists in this setting and the corresponding solution are well-behaved. The asymptotic charges for this solution are all positive but the physical interpretation is unclear.


\section{Conclusions}
\label{conclusion}

In this paper, we have proposed a procedure to clarify part of CTC's free sector in bubble geometries based on ideas similar to the split attractor flow conjecture. Using this procedure and starting from any particular set of centers including the asymptotic vector, one can derive every existing tree without referring to the moduli space. The structure of the tree is determined by the simple thin graph that we called a ``skeleton" and the existence conditions are based purely on checking CTC's conditions at several points or lines on this skeleton. If the conjecture is true, every existing tree should correspond to a split flow component in which every solution is well-behaved. 

By studying several numerical examples, we have shown how this correspondence explicitly works through two pictures with one involving locating the ``image" of the flows in space and the other involving disassembling and assembling of the solutions. We also find that the conjecture works well for the split flow solutions and we cannot find any counter example. Although, this procedure seems to be quite effective even with some complicated trees, it is also clear that this is not the full story. The scaling components in which some or all of the centers form the scaling solutions cannot be captured by this procedure. Through the numerical studies, we show explicitly that the scaling solutions behave like some single-center solutions in the split flow pictures even though they are composed of several GH centers. On the other hand, in some numerical examples, we also find that the ability to distinguish between the scaling and split flow components may depend on the asymptotic vector (moduli). Specifically, when the split flow and scaling components coexist, the two components may exist as a whole component and only split into two when the asymptotic vector is tuned to the bifurcation point. It is still unclear if this is a general behavior or just some exotic phenomena. Also, this bifurcation point may have some physical meaning which is unknown at this moment.

We also associate the simplest bubbled supertubes with some three-center trees and see how the existence conditions of the trees constrain their charge parameter space. The resulting region yields the CTC's free sector of the simplest bubbled supertubes based on the conjecture. Inside this region, the majority of the solutions indeed look like classical zero-entropy black rings. However, for solutions near the boundary of the region, the flux that supports the radius of the ring becomes too small and makes the solutions become more like black holes rather than rings. In fact, near the boundary or even some region beyond it, the charge parameter space becomes dominated by the scaling region and the three centers are more likely to form a scaling solution than a split flow solution. If they do form the scaling solution, this system has the total GH charge equal to one and should be considered as microstate geometries of some black holes according to \cite{Bena:2006kb}. There exists an overlap region for which the physical interpretation is ambiguous. A simple possibility is to consider this region as the transition phase between black holes and rings.

Through several numerical examples, it is clear that the scaling solution plays an important role in microstate geometries of large black rings/holes. Specifically, it seems that the only way to produce black-hole like attractor points in bubble geometries is through the scaling solutions. However, the procedure in this paper cannot capture these solutions and a similar systematic method to analyzing them is needed. If the bifurcation point we have discovered is quite general, it may provide a hint to search the scaling solutions by using the split flow trees.



Furthermore, it is also interesting to understand what exactly happens in five dimensional moduli space. In \cite{Kraus:2005gh}, it was found that the attractors of the black rings and the black holes are determined from the extremization of the two different central charge functions, $Z_m$ and $Z_e$. These functions are governed by the dipole charges and asymptotic electric charges respectively. Furthermore, for the attractor flow of a black ring, neither function is monotonic. While $Z_e$ decreases for a large radius, it is the function $Z_m$ which decreases for a small radius as one approaches the ring from infinity. This strange behavior is actually quite natural from the perspective of the split flow picture. The attractor flows simply ``switch" from the flow with GH charge one to the flow with GH charge zero after passing the split point and the crossover region is naturally near the split point. From this picture, one may be able to find a unique expression for the attractor point and central charge for a single flow and understand the corresponding split attractor flow picture in five dimensional moduli space. We leave this issue and the problem related with the scaling solutions to a future study.

\section*{Acknowledgment}

We would like to thank Sheer El-Showk for helpful discussions.

\bibliographystyle{JHEP}
\bibliography{refslist-proc}

\end{document}